\documentclass[preprint,aps,showpacs,preprintnumbers,amsmath,amssymb,nofootinbib]{revtex4}
\usepackage{epstopdf}
\usepackage{epsfig}
\usepackage{graphicx}
\usepackage{braket}

\usepackage{color}

\begin{document}

\newcommand{\be}{\begin{equation}}
\newcommand{\ee}{\end{equation}}
\newcommand{\bea}{\begin{eqnarray}}
\newcommand{\eea}{\end{eqnarray}}

\newcommand{\nn}{\nonumber}
\def\CP{{\it CP}~}
\def\cp{{\it CP}}

\title{\large Implications of  lepton flavour violation on long baseline neutrino oscillation experiments}
\author{Soumya C. and R. Mohanta }
\affiliation{School of Physics, University of Hyderabad, Hyderabad - 500 046, India }

\begin{abstract}

Non-standard neutrino interactions (NSIs), the sub-leading effects in the flavour transitions of neutrinos, 
play a crucial role in the determination of the various unknowns in neutrino oscillations, such as neutrino mass hierarchy, 
Dirac CP violating phase and the octant of atmospheric mixing angle. In this work, we focus on the possible  implications of lepton flavour violating (LFV) 
NSIs, which  generally affect the neutrino propagation,   on the determination of the these unknown oscillation parameters.  
We study the effect of these NSIs on the physics potential of the currently running and upcoming long-baseline experiments, i.e.,  T2K, NO$\nu$A and DUNE. We also check  the allowed  oscillation parameter space in presence of LFV NSIs. 

\end{abstract}

\maketitle

\section{Introduction}

Neutrino oscillation \cite{SK,SNO,KL,T2K,DC,DB,RENO}, the phenomenon of flavour transition of neutrinos, 
provides strong evidences for neutrino mass and mixing. Further, the three flavour neutrino oscillation model 
has become a very successful theoretical framework,  which could accommodate  almost all neutrino oscillation experimental 
data except some results in very short baseline experiments. However, some of the oscillation parameters \cite{PMNS1,PMNS2} 
(Dirac CP violating phase, neutrino mass hierarchy and the  octant of atmospheric mixing angle) in the standard paradigm  are still 
not known. Recently, Daya Bay \cite{DB1-13,DB2-13}, RENO \cite{RENO-13} and Double CHOOZ \cite{DC-13} experiments have observed that 
the value of reactor mixing angle is significantly large (close to its upper bound), which improves the sensitivities to determine these 
unknowns by enhancing 
the matter effect. Therefore, a well understanding of sub-leading contributions to neutrino oscillation,  coming from various  
new physics scenarios, may lead to the enhancement of physics potentials of long-baseline neutrino oscillation experiments.

Non-standard neutrino interactions (NSIs) \cite{NSI-1,NSI-2} can be considered as sub-leading effects  in the neutrino oscillations, 
which  arise from various new physics scenarios beyond the  standard model. The NSIs, which come from Neutral Current (NC) 
interactions can affect the propagation of neutrino, whereas NSIs  coming from the Charged Current (CC) interactions of neutrinos 
with quarks and leptons can affect the production and detection processes of neutrinos. 
However, in this work, we consider only the NSIs which affect the propagation of neutrinos. The Lagrangian  
corresponds to NSIs during the propagation  is given by \cite{NSI-L}, 
\begin{equation}
{\cal L}_{\rm NSI} = -2\sqrt{2}G_F\varepsilon_{\alpha\beta}^{fC}(\overline{\nu}_\alpha \gamma^\mu P_L\nu_\beta)(\overline{f} \gamma_\mu P_C f) \,, 
\end{equation}
where $G_F$ is the Fermi coupling constant,  $\varepsilon_{\alpha\beta}^{fC}$ are the  new coupling constants, so called NSI parameters, $f$ is  
fermion and $P_C=(1\pm\gamma_5)/2$ are the  right ($C = R$) and left ($C = L$)  chiral projection  operators. The NSI contributions which are 
relevant as neutrino propagate through the earth are those coming from the interaction of neutrino with $e$, $u$ and $d$ because the 
earth matter is made up of these fermions only. Therefore, the effective NSI parameter is given by
\begin{equation}
\varepsilon_{\alpha\beta} = \sum_{f=e,u,d}\frac{n_f}{n_e} \varepsilon_{\alpha\beta}^{f} \,,
\end{equation}
where $\varepsilon^f_{\alpha\beta}= \varepsilon_{\alpha\beta}^{fL} + \varepsilon_{\alpha\beta} ^{fR}$, $n_f$ is the number density of the fermion $f$ and $n_e$ the 
number density of electrons in earth. For earth matter, we can assume that the number densities of electrons, protons and neutrons are equal, i.e, 
$n_n \approx n_p =n_e$, which implies that $n_u \approx n_d = 3 n_e$.

NSIs and their consequences have been studied  quite extensively in the literature both in model dependent (mass models) and independent ways. Furthermore,
there are studies, which have been done to investigate the effect of NSIs on  atmospheric neutrinos \cite{nsiatm-1,nsiatm-2,nsiatm-3},
 solar neutrinos \cite{nsisolar-1, nsisolar-2, nsisolar-3, nsisolar-4, nsisolar-5}, accelerator neutrinos 
\cite{Nacc-1, Nacc-2, Nacc-3, Nacc-4, Nacc-5, Nacc-6, Nacc-7, Nacc-8, Nacc-9, Nacc-10, Nacc-11} and supernova neutrinos 
\cite{supernsi-1, supernsi-2, supernsi-3}. However, it is very crucial to understand the implications of new physics effects  at
the long baseline experiments like T2K, NO$\nu$A and DUNE. In this regard, there are many recent works which 
have been discussed the various aspects of NSIs at long baseline experiments  \cite{1511.05562,1511.06357,1601.03736}, 
for instance in \cite{1601.07730}, the authors have obtained the constrain on NSI parameters using long baseline experiments and in 
\cite{1601.00927} authors have discussed the degeneracies among the oscillation parameters in presence of NSIs. However, in this paper, we focus on the effect of the lepton flavour violating NSIs on the determination of various unknowns at long baseline experiments.

We have discussed  the physics potential of long baseline experiments in our previous papers \cite{soumya,deepthi}. 
As neutrino oscillation physics already entered into its precision era, one should take care of various sub-leading effects such as 
NSIs in the oscillation physics. In this regard, we would like to study the effect of the lepton flavor violating NSIs on the determination 
of oscillation parameters. This paper is organized as follows. In section II, we discuss the basic formalism of neutrino oscillation including NSI effects. 
 In section  III, we study the effect of NSI parameters on $\nu_{e}$ appearance oscillation  probability. 
The effect of LFV NSI on Physics potential of long baseline  experiments are discussed in section IV. 
In section V, we discuss the parameter degeneracies among the oscillation parameters in presence of NSIs. Section VI contains the summary and conclusions.
\section{Neutrino oscillation with NSIs}

In the standard oscillation (SO) paradigm, the propagation of neutrino through matter is described by the Hamiltonian
\begin{eqnarray}
H_{SO} &=&  H_{0} + H_{matter} \nonumber\\
&=&\frac{1}{2E} U\cdot {\rm diag}(0,\Delta m^2_{21},\Delta m^2_{31}) \cdot U^{\dagger} +  {\rm diag}(V_{CC},0,0)\;,
\end{eqnarray}
where the $H_{0}$ is the  Hamiltonian in vacuum, $\Delta m^2_{ji} = m^2_{j}-m^2_{i}$ is neutrino mass squared difference, $H_{matter}$ 
is the Hamiltonian responsible for matter effect, $V_{CC}=\sqrt{2}G_{F}n_{e}$ is the matter potential and $U$ is the PMNS mixing matrix which is described 
by three mixing angles ($\theta_{12},\theta_{13},\theta_{23}$) and one phase ($\delta_{CP}$) and is given by
    \begin{equation}
    U_{PMNS}= \left( \begin{array}{ccc} c^{}_{12} c^{}_{13} & s^{}_{12}
    c^{}_{13} & s^{}_{13} e^{-i\delta} \\ -s^{}_{12} c^{}_{23} -
   c^{}_{12} s^{}_{13} s^{}_{23} e^{i\delta} & c^{}_{12} c^{}_{23} -
   s^{}_{12} s^{}_{13} s^{}_{23} e^{i\delta} & c^{}_{13} s^{}_{23} \\
   s^{}_{12} s^{}_{23} - c^{}_{12} s^{}_{13} c^{}_{23} e^{i\delta} &
    -c^{}_{12} s^{}_{23} - s^{}_{12} s^{}_{13} c^{}_{23} e^{i\delta} &
   c^{}_{13} c^{}_{23} \end{array} \right),
    \label{standpara}
    \end{equation}
with  $c_{ij} = \cos\theta_{ij}$ and $s_{ij} = \sin\theta_{ij}$.
The  NSI Hamiltonian, which is coming from  the interactions of neutrinos  as they propagate through matter is given by
\begin{equation}
H_{NSI} = V_{CC}
 \begin{pmatrix}
  \varepsilon_{ee} & \varepsilon_{e\mu} & \varepsilon_{e\tau} \\
  \varepsilon^*_{e\mu} & \varepsilon_{\mu\mu} & \varepsilon_{\mu\tau} \\
  \varepsilon^*_{e\tau} & \varepsilon^*_{\mu\tau} & \varepsilon_{\tau\tau}
 \end{pmatrix},
 \label{nsi}
\end{equation}
where $\varepsilon_{\alpha\beta} =|\varepsilon_{\alpha\beta}|e^{i\delta_{\alpha \beta}}$ are the complex NSI parameters, 
which give the coupling strength of non-standard interactions. The off-diagonal elements of the NSI Hamiltonian 
($\varepsilon_{e\mu},\varepsilon_{e\tau}$ and $\varepsilon_{\mu\tau}$) are the lepton flavor violating NSI parameters, which are our subject of interest.

Almost all current neutrino oscillation data are consistent with the standard oscillation paradigm. Therefore,  the effect of NSI on the oscillation 
phenomena is expected to be very small.  Moreover, some neutrino mass models for instance, triplet seesaw model \cite{0811.3346}, Zee Babu model \cite{0909.0455} 
predict the value of NSI parameters of the order of $10^{-4}-10^{-3}$, which depend on the scale of new physics and the neutrino mass ordering.  
The strong constraints on NSI parameters make them very difficult to be observed in the long baseline experiments. 
Therefore, we use phenomenological way to study the effect of NSIs on the physics potential of such experiments. 
The model independent current  upper bounds of NSI parameters  at 90 \% C.L. are given as  \cite{epsiloneff-1, nsibound-1}
\begin{eqnarray}
 |\varepsilon_{\alpha\beta}|
 \;<\;
  \left( \begin{array}{ccc}
4.2  & 0.3 &  0.5 \\
0.3 & 0.068 & 0.04 \\
0.5  & 0.04 & 0.15 \\
  \end{array} \right).\label{nsibound}
\end{eqnarray} 
From the above equation, it should be noted that the bound on LFV-NSI parameters as $|\varepsilon_{e\mu}|<0.3, |\varepsilon_{\mu\tau}|<0.04$ 
and $|\varepsilon_{e\tau}|<0.5$, therefore, in our analysis we use the
representative   values for $\varepsilon_{e\mu}, \varepsilon_{\mu\tau}$ and $\varepsilon_{e\tau}$  close to their upper bounds, i.e.,  as  
0.2, 0.03 and 0.3 respectively. It should  also be noted that each NSI parameter $\varepsilon_{\alpha \beta}$ has a CP phase $\delta_{\alpha \beta}$, which can vary between $-\pi$ to $\pi$.  

\section{LFV-NSI effect on $\nu_{e}$ appearance probability}

In general, the measurement of  branching ratios (BRs) and the CP violation parameters can be used to 
probe the  New Physics effects or non-standard interactions in the flavor sector. If any inconsistency  found between the 
experimental observed values and the corresponding SM predictions in these observables, it would imply
the presence of new physics. However, in the case of neutrinos one can not use branching ratio measurements to study  the new physics effects, 
since the mass difference between neutrinos is really small and also experiments detect neutrinos as flavour states (mixed state of mass eigenstates). 
The various issues regarding the BR measurement of neutrinos are discussed  in \cite{9507344}. Therefore, in the case of neutrinos,
 new physics effect can be studied by using the oscillation probabilities. The super-beam experiments like T2K, NO$\nu$A and DUNE use muon neutrino beams as neutrino source. Therefore, 
in this section, we discuss the consequences of LFV-NSI parameters on neutrino  appearance ($\nu_{\mu} \rightarrow \nu_{e} $) probability.

\begin{figure}[!htb]
\begin{center}
\includegraphics[width=5.4cm,height=4.5cm]{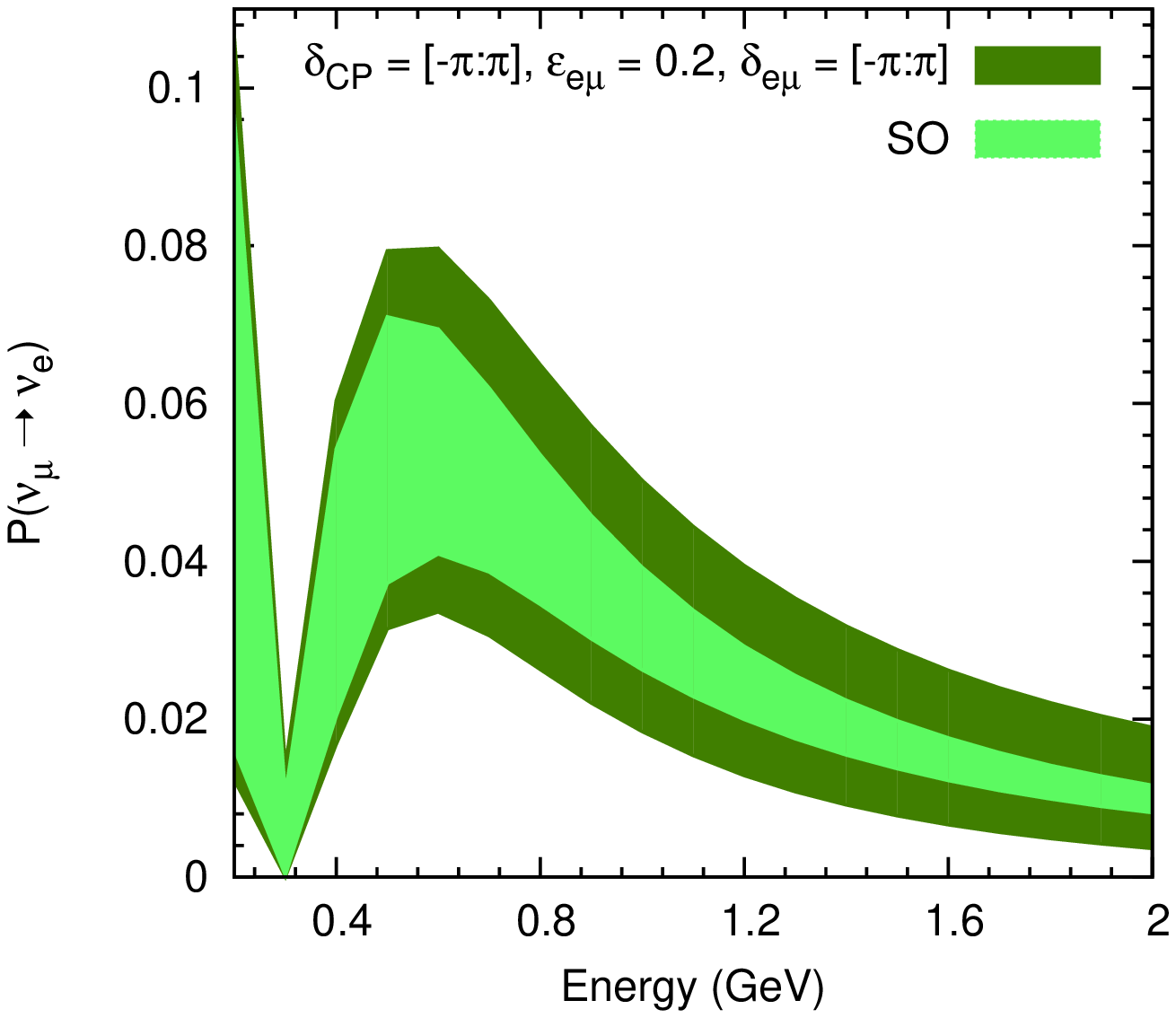}
\includegraphics[width=5.4cm,height=4.5cm]{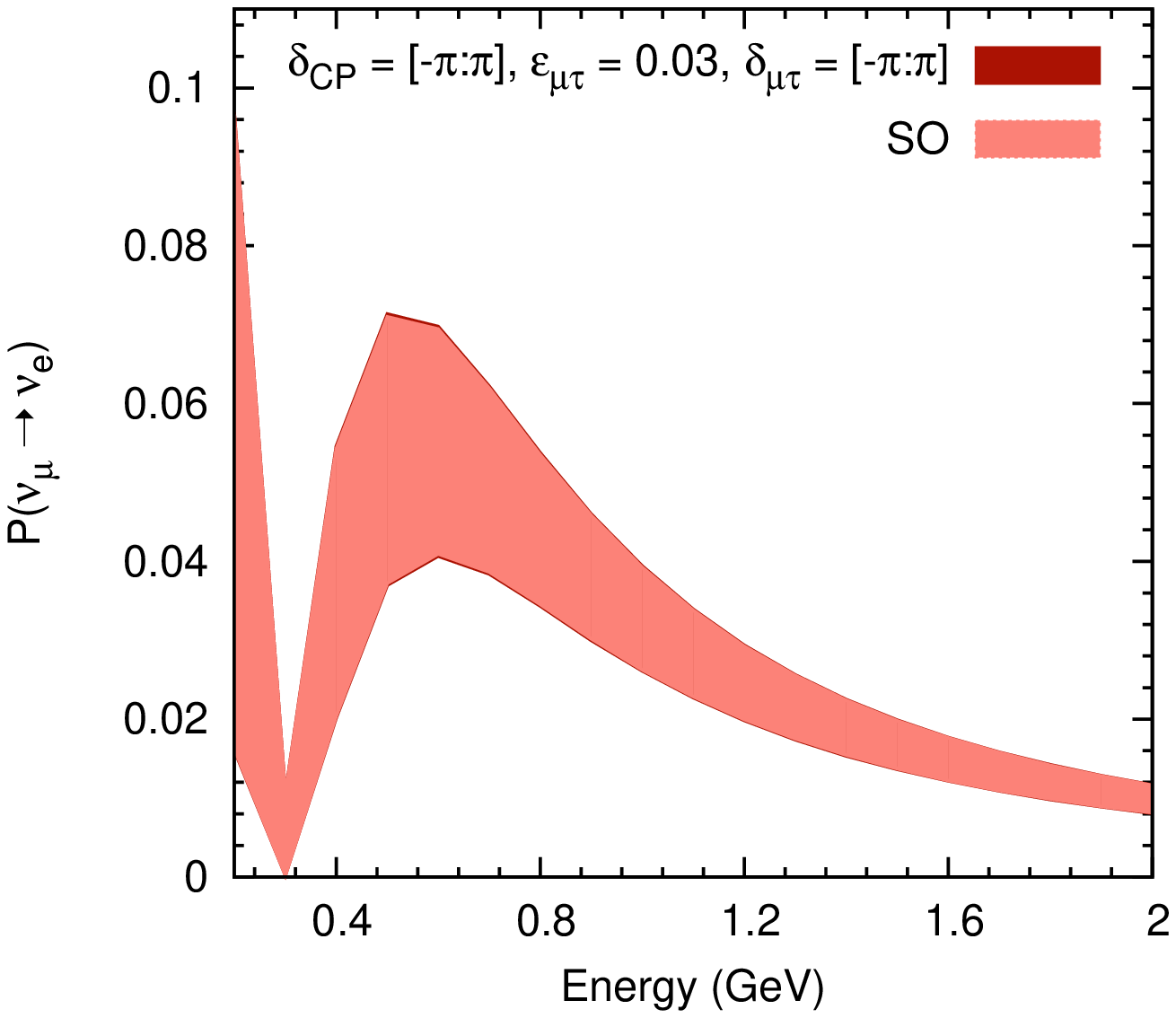}
\includegraphics[width=5.4cm,height=4.5cm]{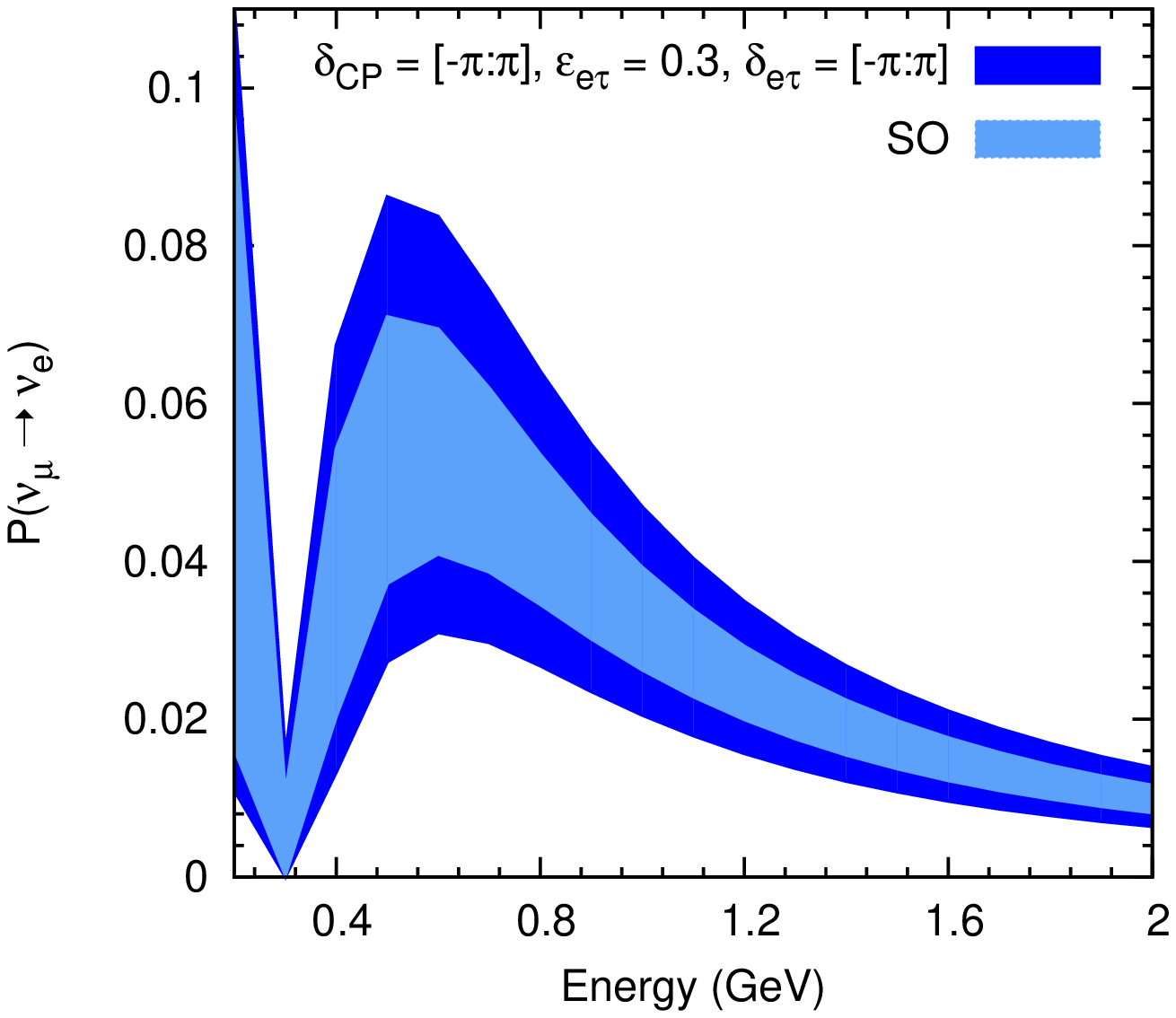}
\includegraphics[width=5.4cm,height=4.5cm]{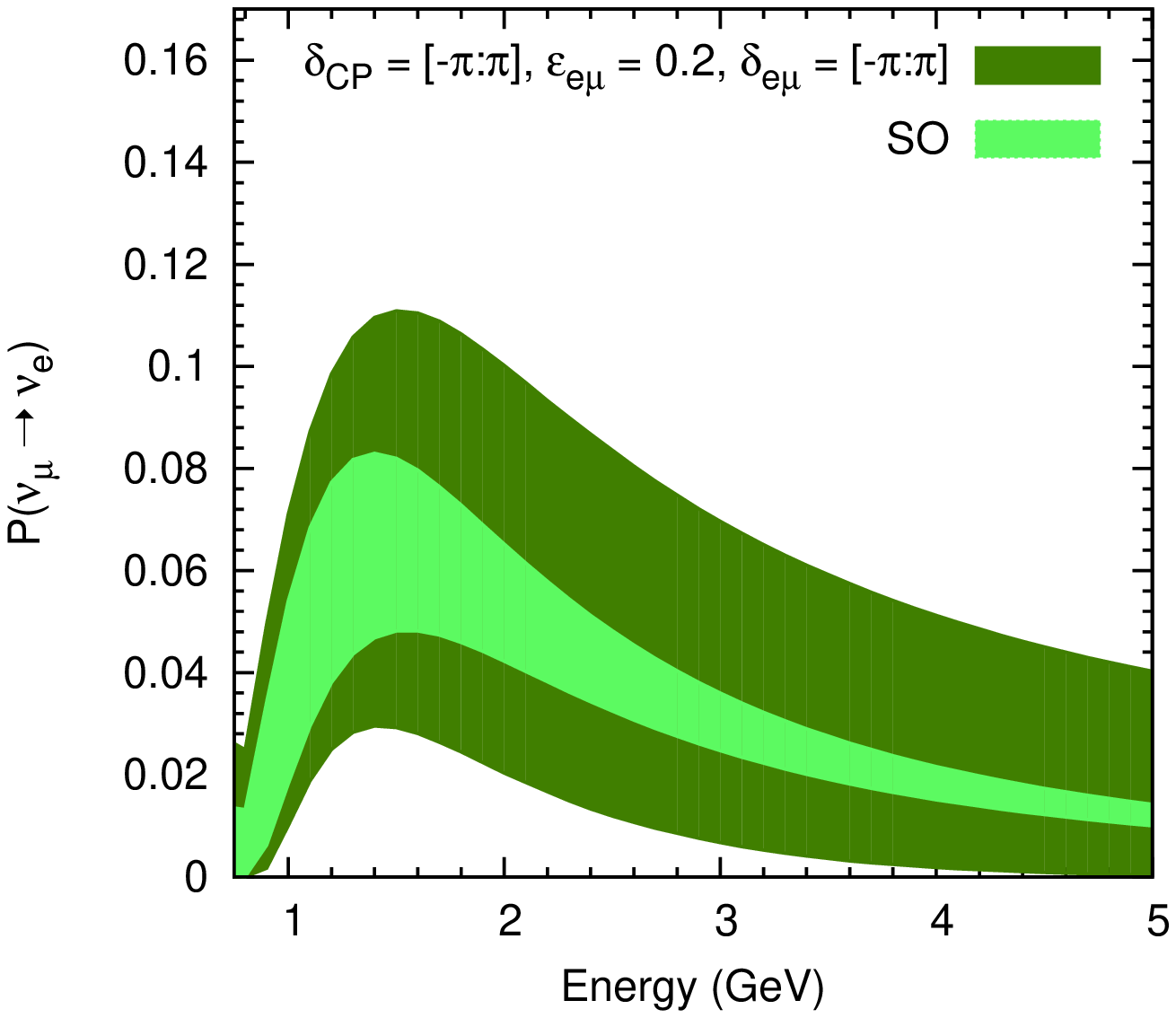}
\includegraphics[width=5.4cm,height=4.5cm]{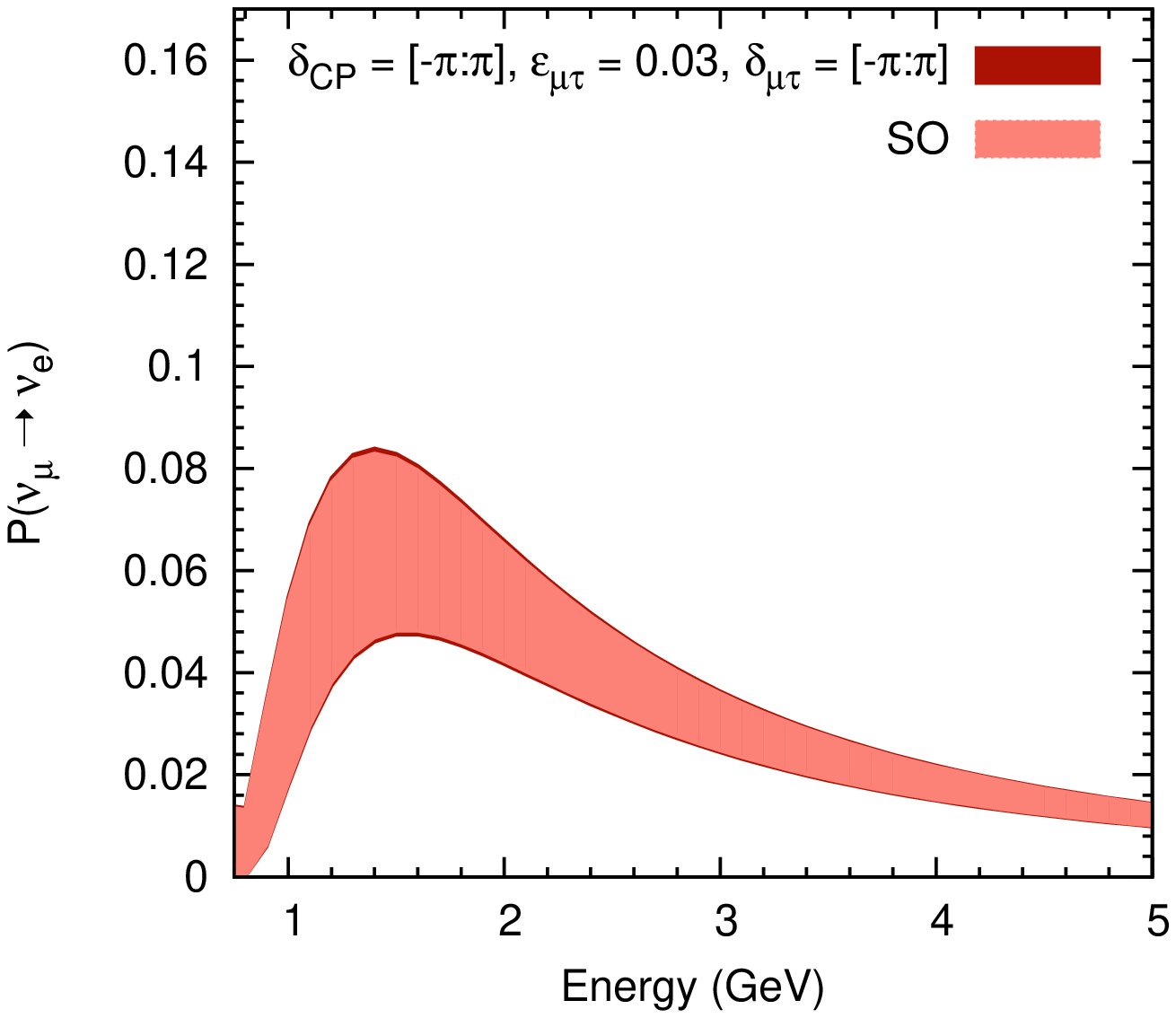}
\includegraphics[width=5.4cm,height=4.5cm]{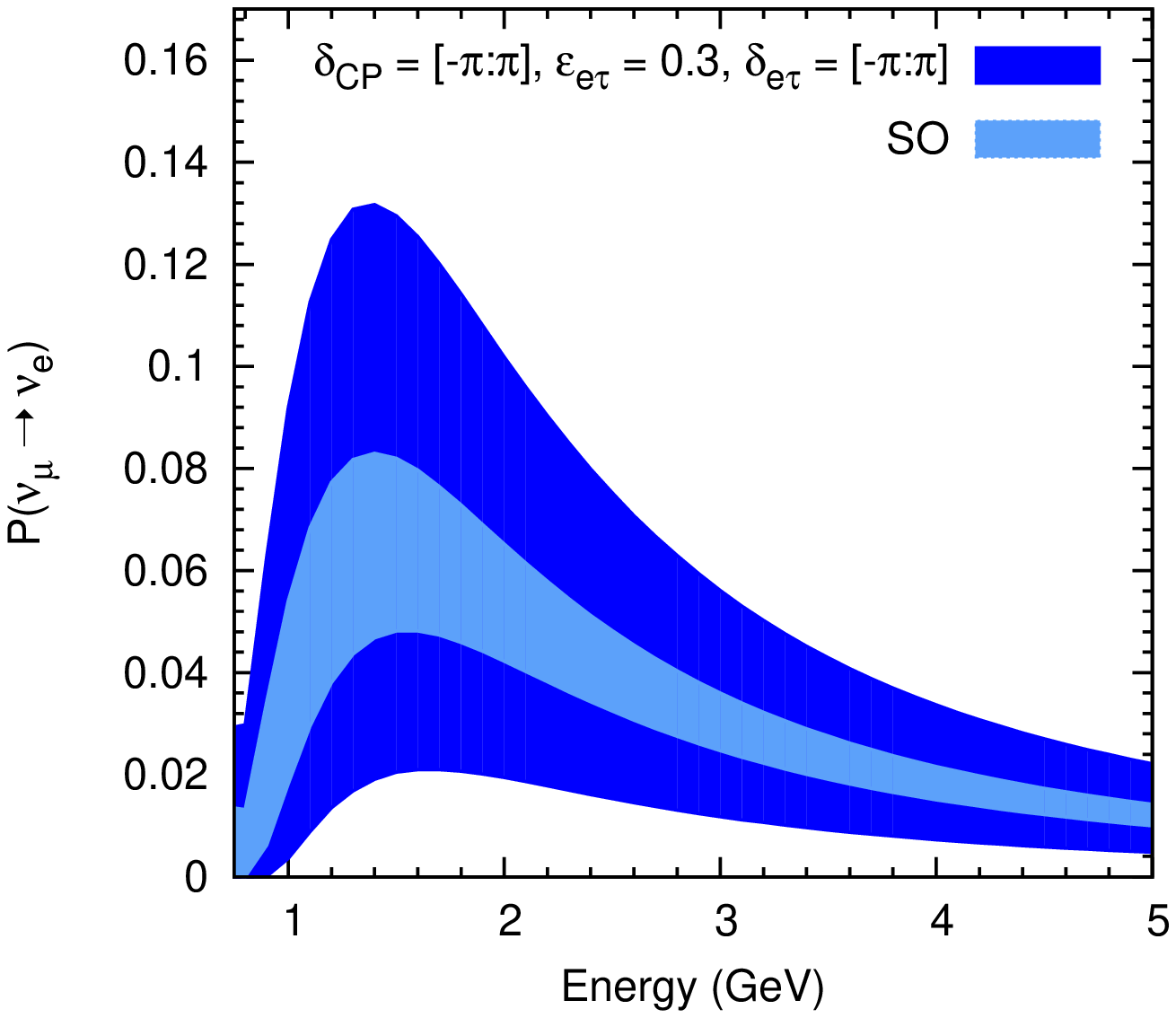}
\includegraphics[width=5.4cm,height=4.5cm]{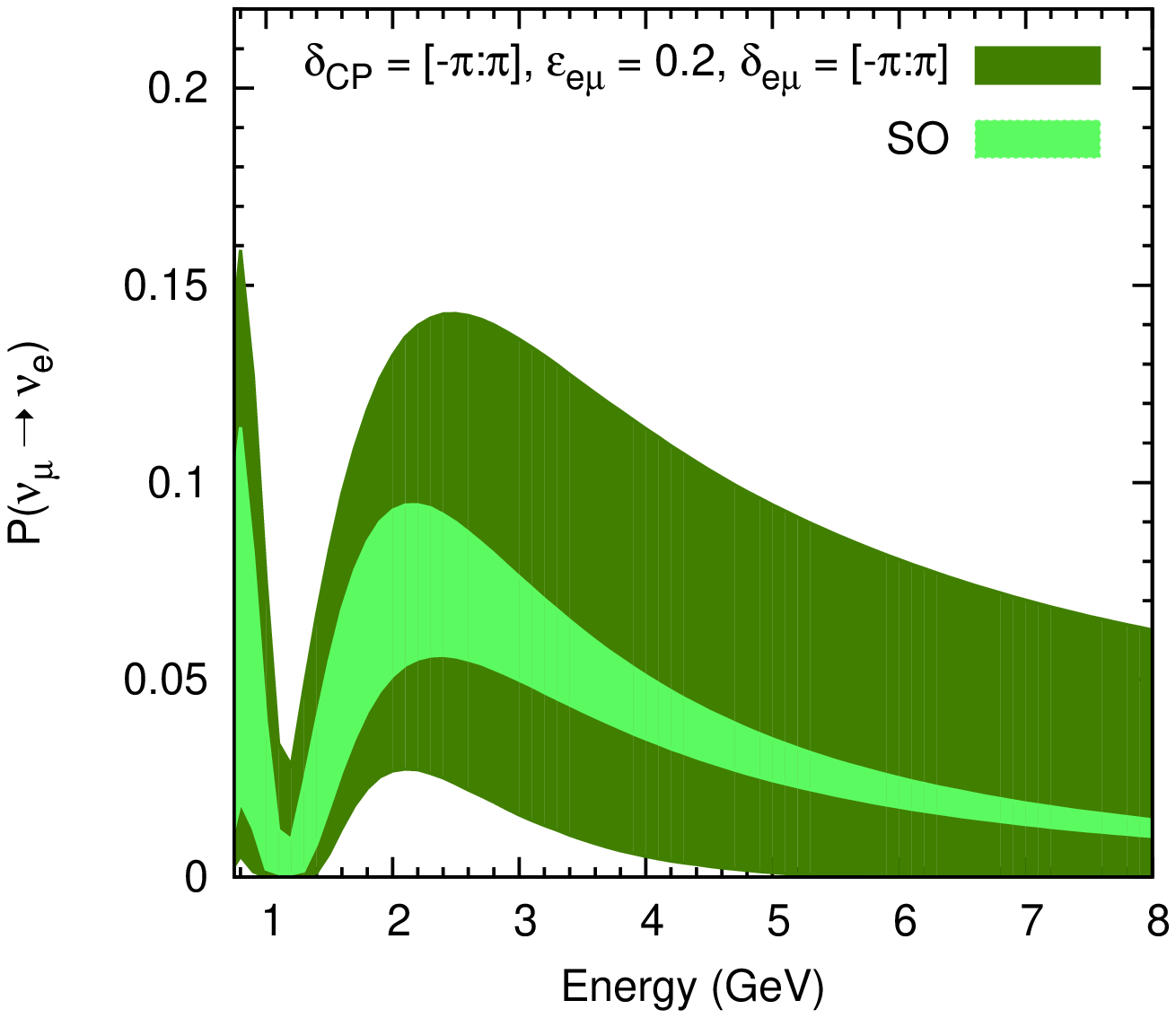}
\includegraphics[width=5.4cm,height=4.5cm]{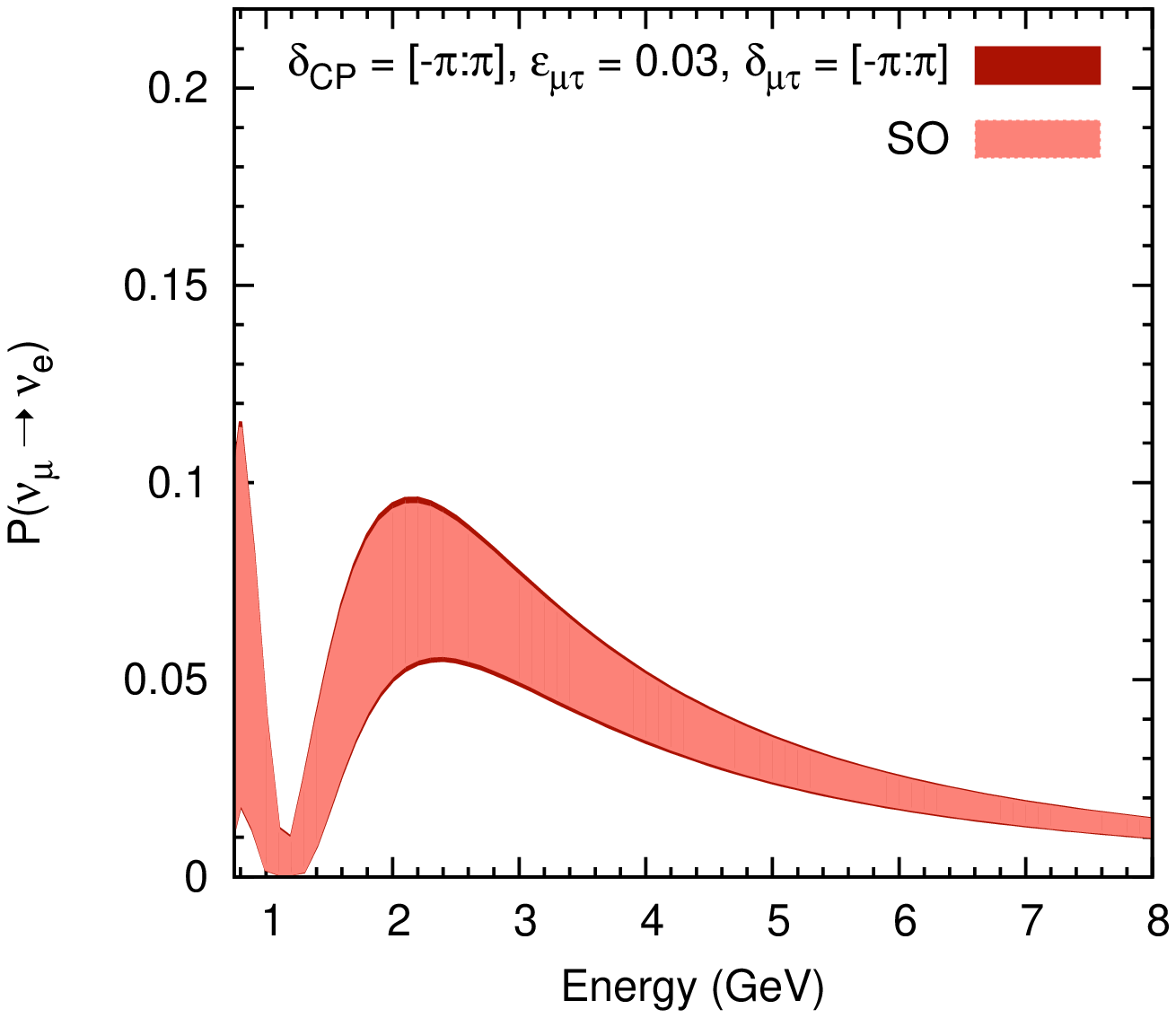}
\includegraphics[width=5.4cm,height=4.5cm]{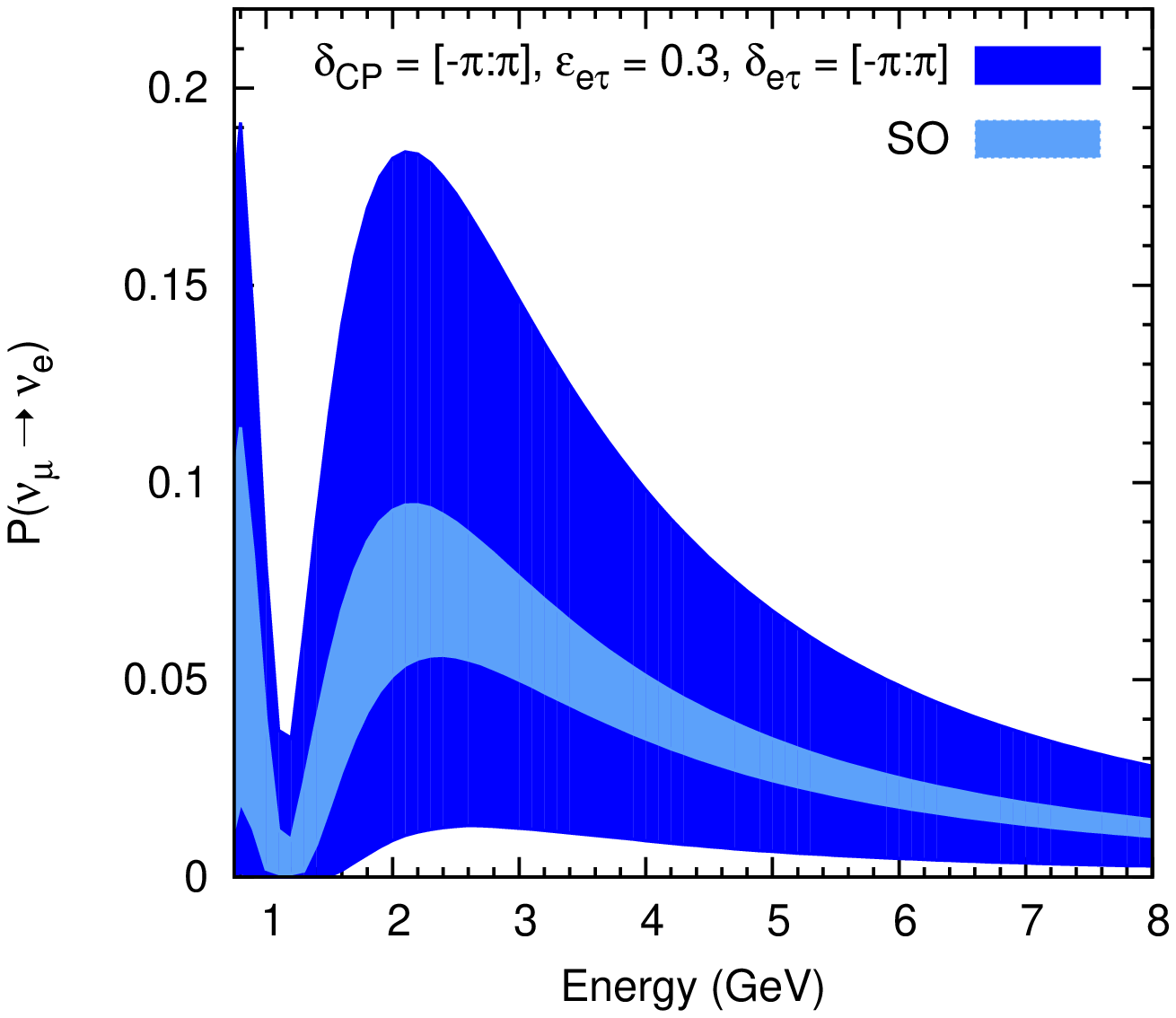}
\end{center}
\caption{Neutrino appearance probability for the $\nu_{\mu} \rightarrow \nu_{e}$ without NSI (light shaded region) and with NSI (dark shaded green, 
red and blue regions are correspond to  $\varepsilon_{e\mu}, \varepsilon_{\mu\tau}$ and $ \varepsilon_{e\tau}$ parameters contribution respectively) 
for T2K (top panel), NO$\nu$A (middle panel) and DUNE (bottom panel). The hierarchy is assumed to be NH}
\label{Pnu1}
\end{figure}

\begin{table}[h]
    \begin{center}
    \vspace*{0.1 true in} 
    \begin{tabular}{|l|c|c|c|}\hline
    Expt. setup& \text{T2K} & \text{NO$\nu$A} & \text{DUNE}\\
    & \cite{{t2k-2,t2k-3,t2k-4}}& \cite{{nova,nova-1,sanjib}}& \cite{{lbne,lbne-2}}\\\hline
    Detector &Water Cherenkov&Scintillator&Liquid Argon\\
    Beam Power(MW)& 0.75 & 0.77 & 0.7\\
    Fiducial mass(kt) & 22.5  & 14 & 35\\
    Baseline length(km)& 295 & 810 & 1300 \\
    Running time (yrs)&5 (3$\nu$+2$\bar{\nu}$)&6 (3$\nu$+3$\bar{\nu}$)&10 (5$\nu$+5$\bar{\nu}$) \\
    \hline
    \end{tabular} 
    \caption{The experimental specifications.}
    \label{experiments}
    \end{center} 
    \end{table}
     
We use GLoBES package \cite{Huber:2004-1, Huber:2009-2} along with snu plugin  \cite{snu1,snu2} for our analysis to study the implications of LFV-NSI on the 
propagation of 
neutrinos. The experimental details of T2K, NO$\nu$A and DUNE that we consider in this analysis are given in Table \ref{experiments}.  The values of standard oscillation parameters that we use in the analysis are given in the Table \ref{true-osc}.

   \begin{table}[h]
    \begin{center}
    \vspace*{0.1 true in}
    \begin{tabular}{|c|c|}\hline
    Oscillation Parameter & True Value\\ \hline
    $\sin^2\theta_{12}$ & 0.32 ~ \\\hline
    $\sin^2 2\theta_{13}$ & 0.1 ~ \\\hline
    $\sin^2 \theta_{23}$ & 0.5, 0.41 (LO), 0.59 (HO) ~ \\\hline
    $\Delta m_{atm}^2$ & $2.4 \times 10^{-3} ~{\rm eV}^2$ for NH ~ \\
    & $-2.4 \times 10^{-3} ~{\rm eV}^2$ for IH ~ \\\hline
    $\Delta m_{21}^2$ & $7.6 \times 10^{-5}~ {\rm eV}^2$ ~ \\\hline
    $\delta_{CP} $ & $0^\circ$  ~ \\\hline
    \end{tabular}
    \caption{The true values of oscillation parameters considered in the simulations are taken from \cite{T2K-oct}.}
    \label{true-osc}
    \end{center}
    \end{table} 
    
For an illustration, we show the   calculated transition probability  with and without NSI for T2K (top panel), NO$\nu$A (middle panel) and DUNE 
(bottom panel)  by assuming hierarchy as NH  in Fig.~\ref{Pnu1}, for neutrinos. In the figure, the light shaded regions correspond to probability 
in the standard oscillation (SO) paradigm, whereas the dark shaded green, red, and blue regions represent the additional contribution to the oscillation probability, which are coming from NSI parameters $\varepsilon_{e\mu}, \varepsilon_{\mu\tau}$ and $\varepsilon_{e\tau}$  respectively. From the figure, we can see that the NSI contribution to oscillation probability is noteworthy in presence of $\varepsilon_{e\tau}$ and $ \varepsilon_{e\mu}$ parameters, whereas the contribution from $\varepsilon_{\mu\tau}$ is negligible. 
It can also be seen  from the figure that there is  significant change in the oscillation probability  in the presence of NSIs for both 
NO$\nu$A and DUNE, whereas for T2K,  the effect is found to be  rather small, i.e., NO$\nu$A and DUNE are more sensitive to  NSI effects. 
We can also see that there is a substantial change in the oscillation probability of DUNE experiment in the  presence of NSI. Therefore, DUNE 
experiment  can be used to investigate various effect of NSI, which are expected to be observed in the  long baseline experiment. Moreover, 
NSI can even affect the results, which require much precision on their measurements for the  determination of the unknowns in neutrino sector, 
of  the currently running experiments like  T2K and NO$\nu$A.

\section{NSI effect on Physics potential of long baseline  experiments}
The primary objective  of long baseline experiments is  the  determination of the various unknowns (Neutrino mass ordering, 
CP violating phase and octant of atmospheric mixing angle) in the phenomenon of neutrino oscillation. In this section, we discuss the effect of 
LFV-NSI on the  determination of these unknowns. From the previous section, we found that the NSI parameter $\varepsilon_{e\tau}$ can significantly change the oscillation probability. Therefore, for simplicity, we focus  on the effect of $\varepsilon_{e\tau}$ on the determination of other unknowns in neutrino oscillation sector. We also compare the effect of  NSIs on physics potential of different experiments that have been considered in this paper. All the sensitivities are computed by using GLoBES.

\subsection{ Effect on the determination of neutrino mass ordering} 
 
\begin{figure}[!htb]
\begin{center}
\includegraphics[width=5.4cm,height=4.5cm]{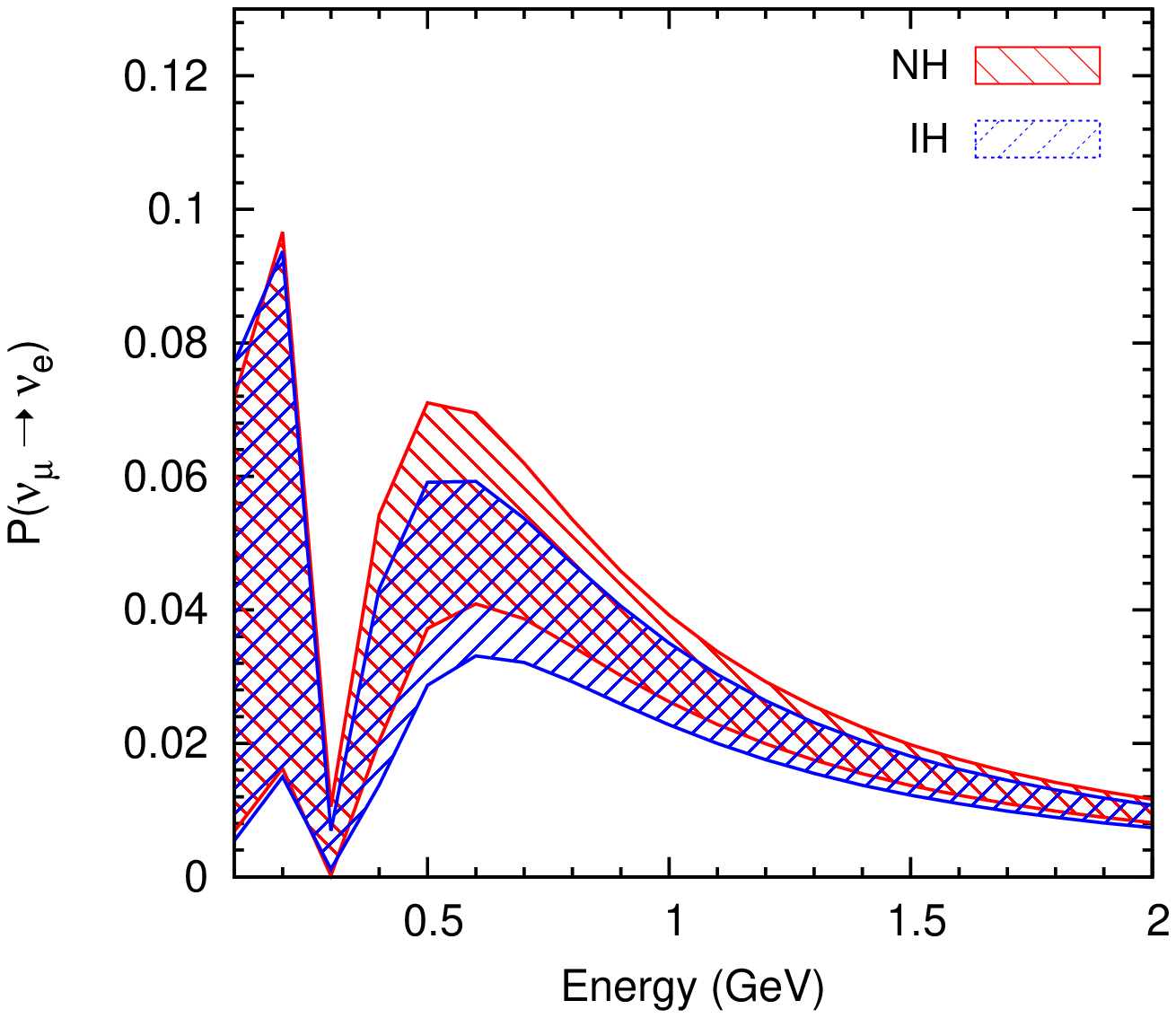}
\includegraphics[width=5.4cm,height=4.5cm]{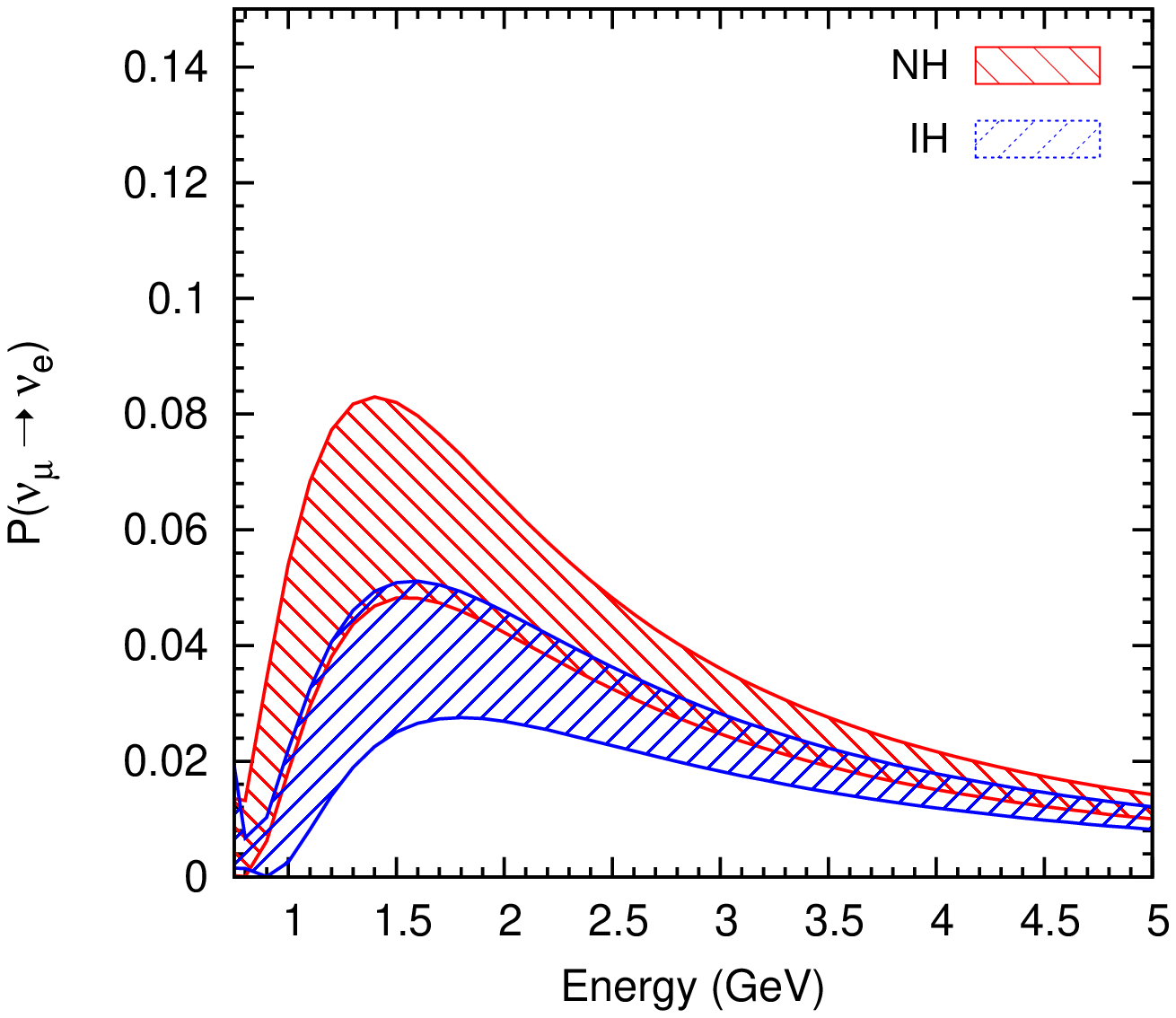}
\includegraphics[width=5.4cm,height=4.5cm]{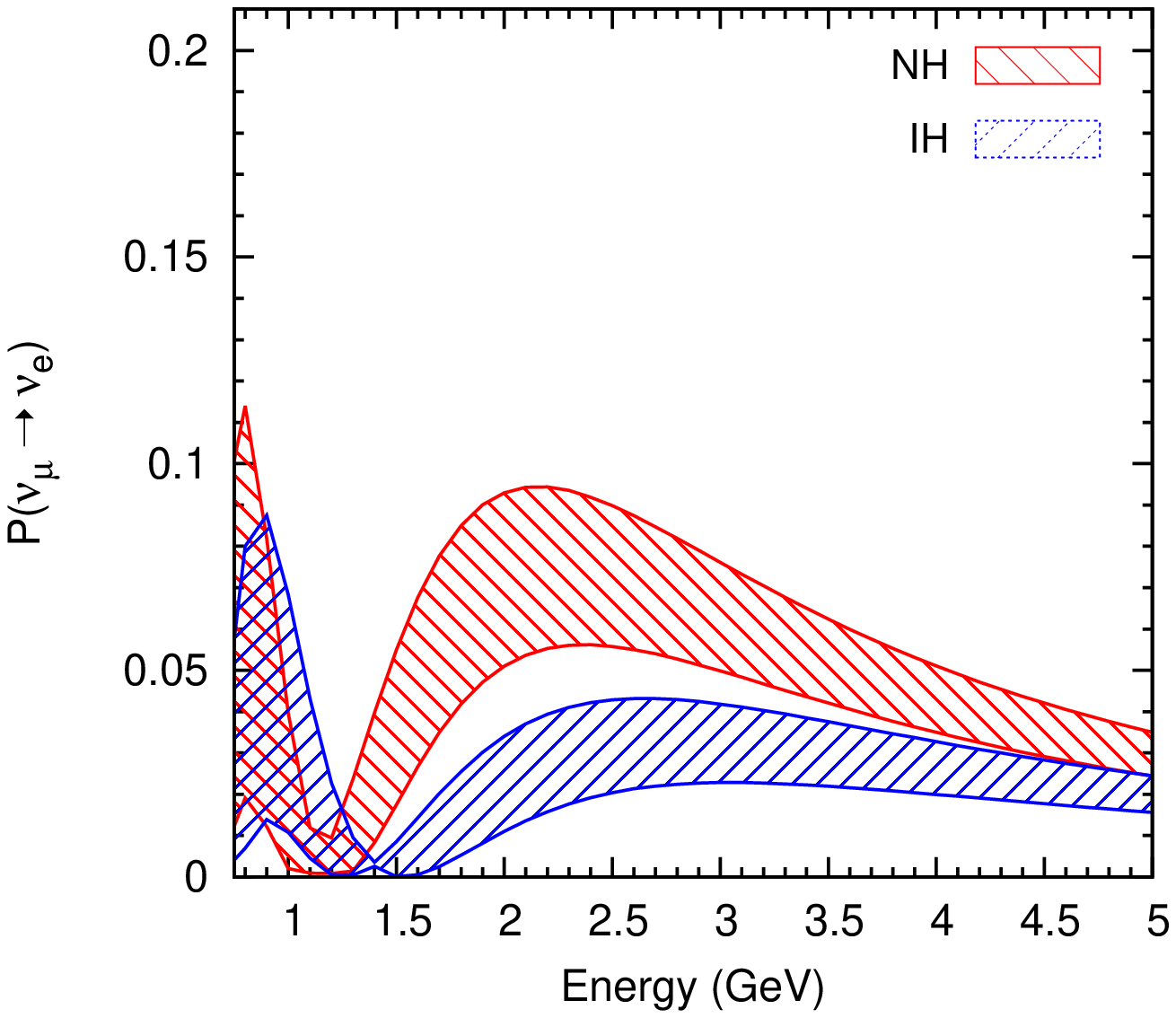}
\includegraphics[width=5.4cm,height=4.5cm]{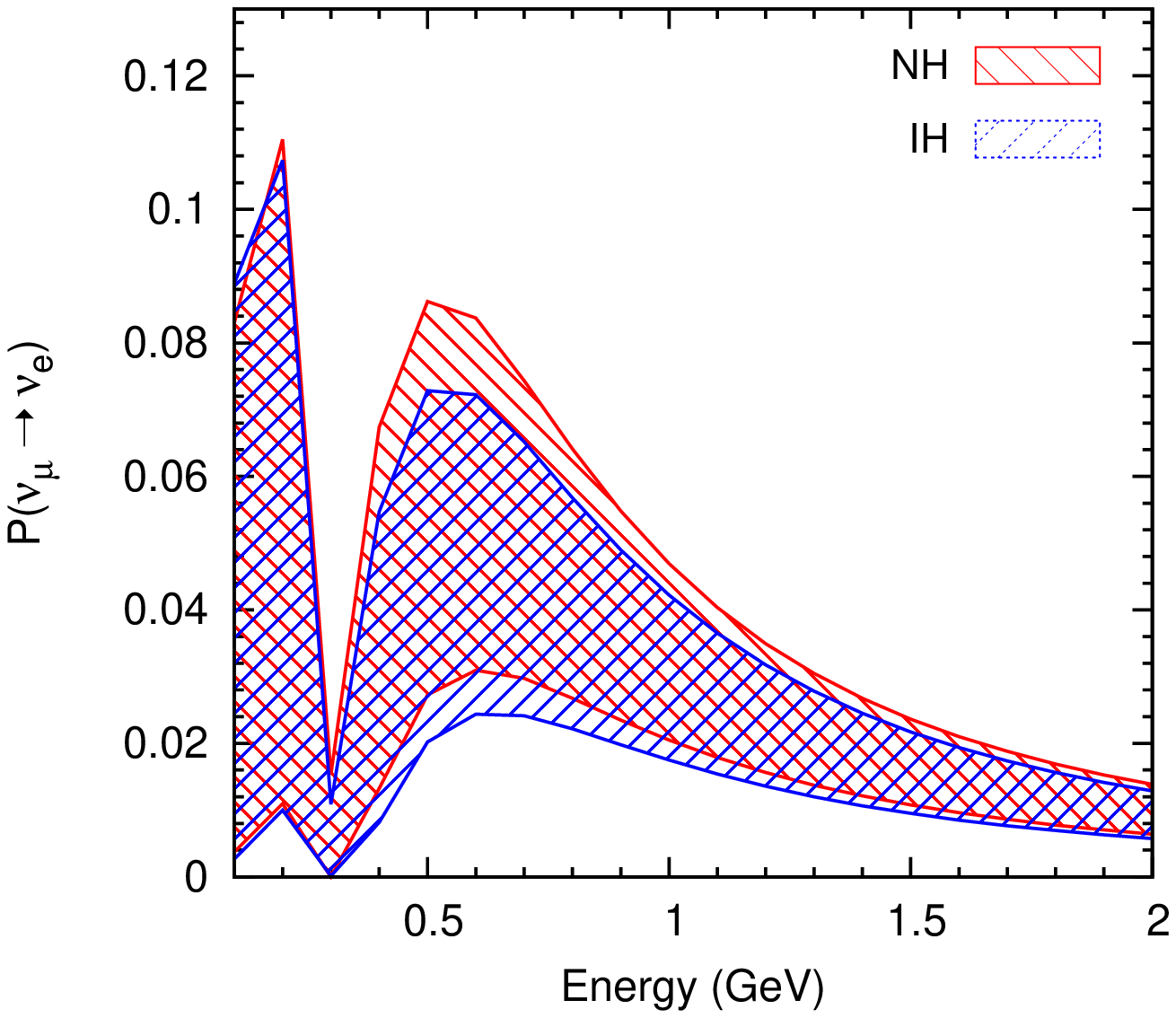}
\includegraphics[width=5.4cm,height=4.5cm]{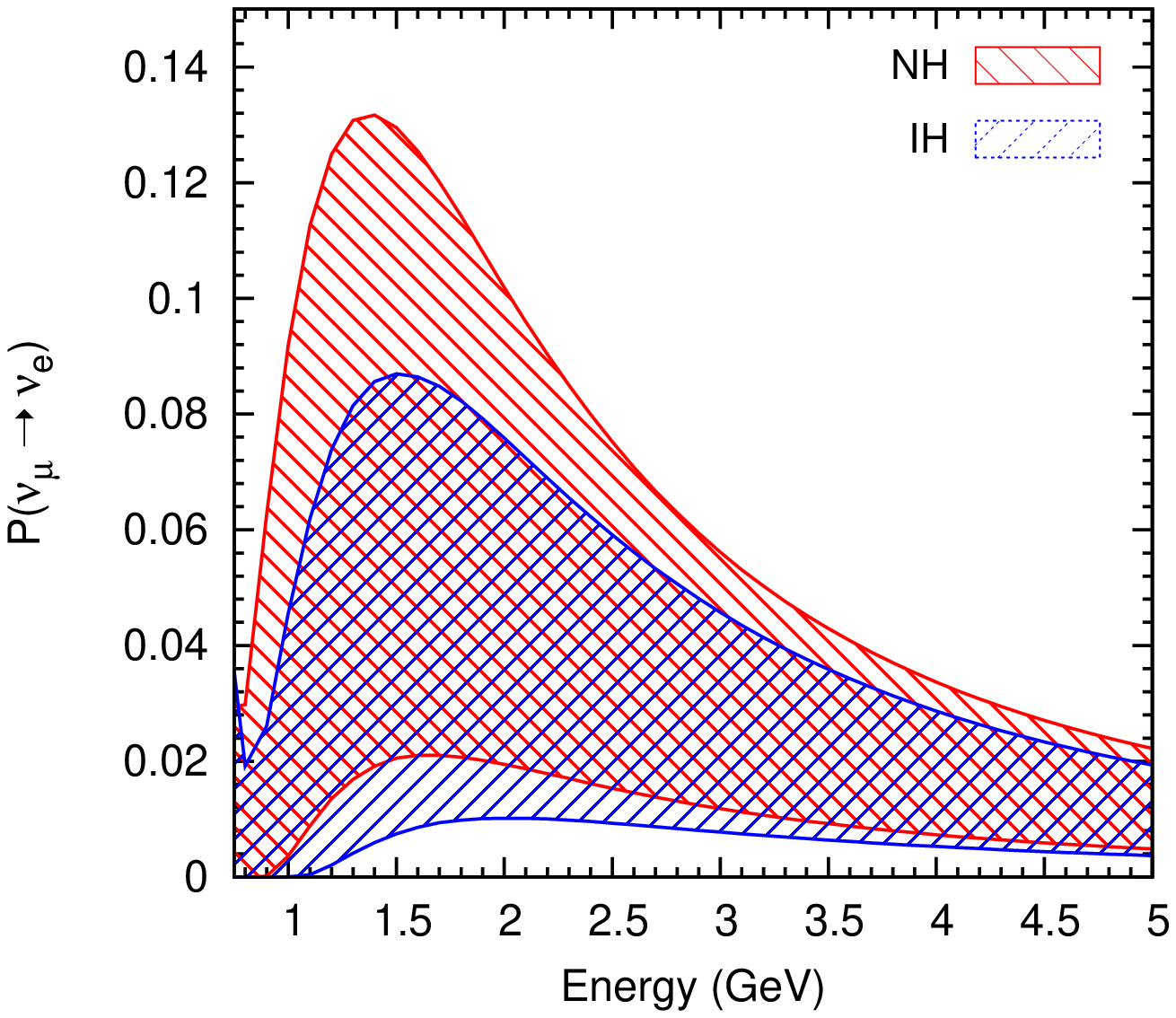}
\includegraphics[width=5.4cm,height=4.5cm]{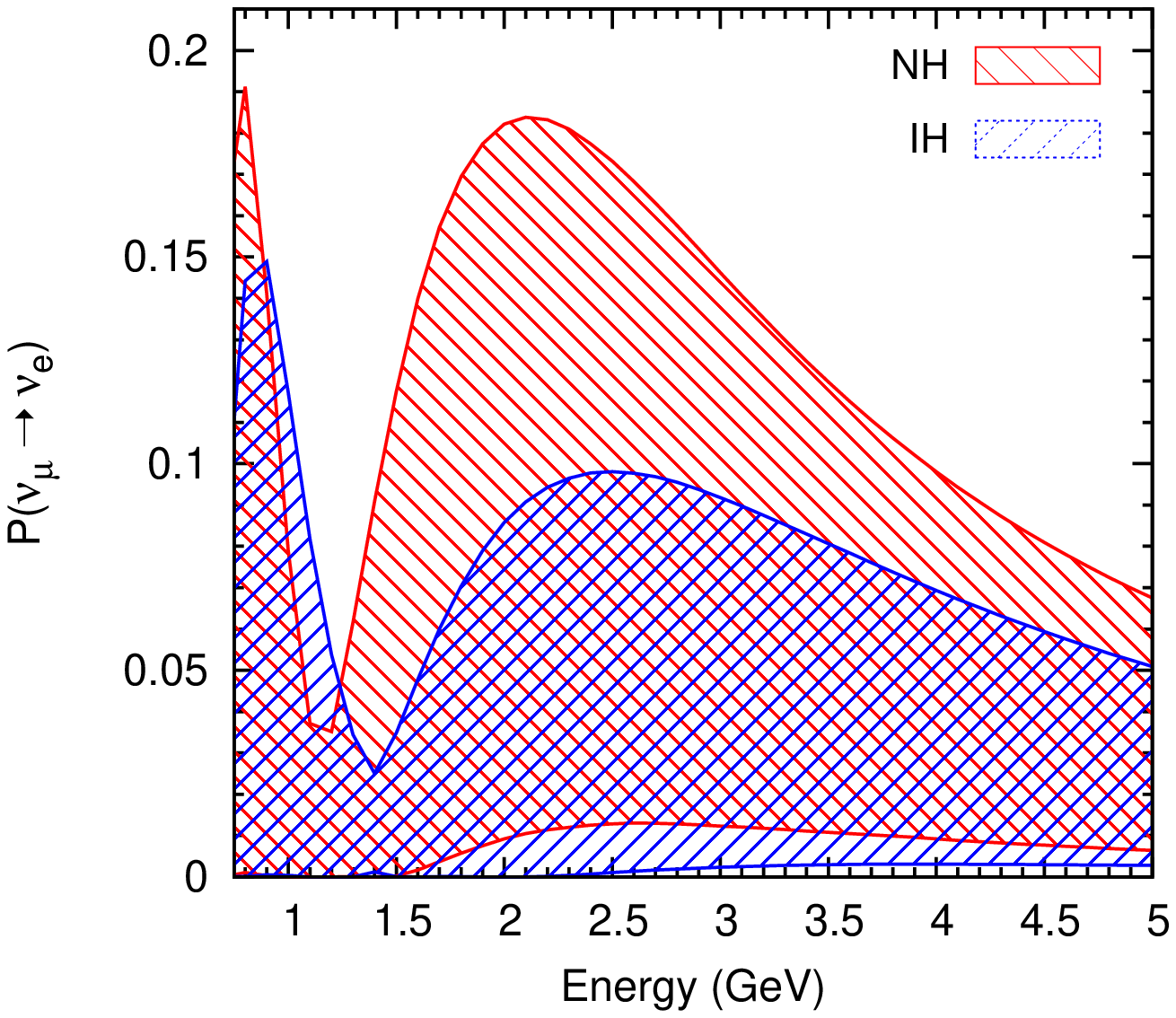}
\end{center}
\caption{Neutrino appearance probability for the $\nu_{\mu} \rightarrow \nu_{e}$ without NSI (top panel) and with NSI (bottom panel) by assuming both NH (red) and IH (blue) for T2K (left panel), NO$\nu$A (middle panel) and DUNE (right panel).}
\label{MH-OSC}
\end{figure}

So far, we do not know whether the hierarchy of neutrino mass is Normal ($m_1<m_2<<m_3$) or Inverted ($m_3<<m_1<m_2$). The MSW effect, the so called 
matter effect  plays a crucial role in the determination of neutrino mass hierarchy,  because unlike vacuum oscillation,  
they give different contributions to oscillation probability for NH and IH,  as one can see from the top panel of Fig.~\ref{MH-OSC}. 
Therefore, a thorough study of effect of NSIs on the determinations of MH is of great importance in oscillation physics.

\begin{figure}[!htb]
\begin{center}
\includegraphics[width=5.4cm,height=4.5cm]{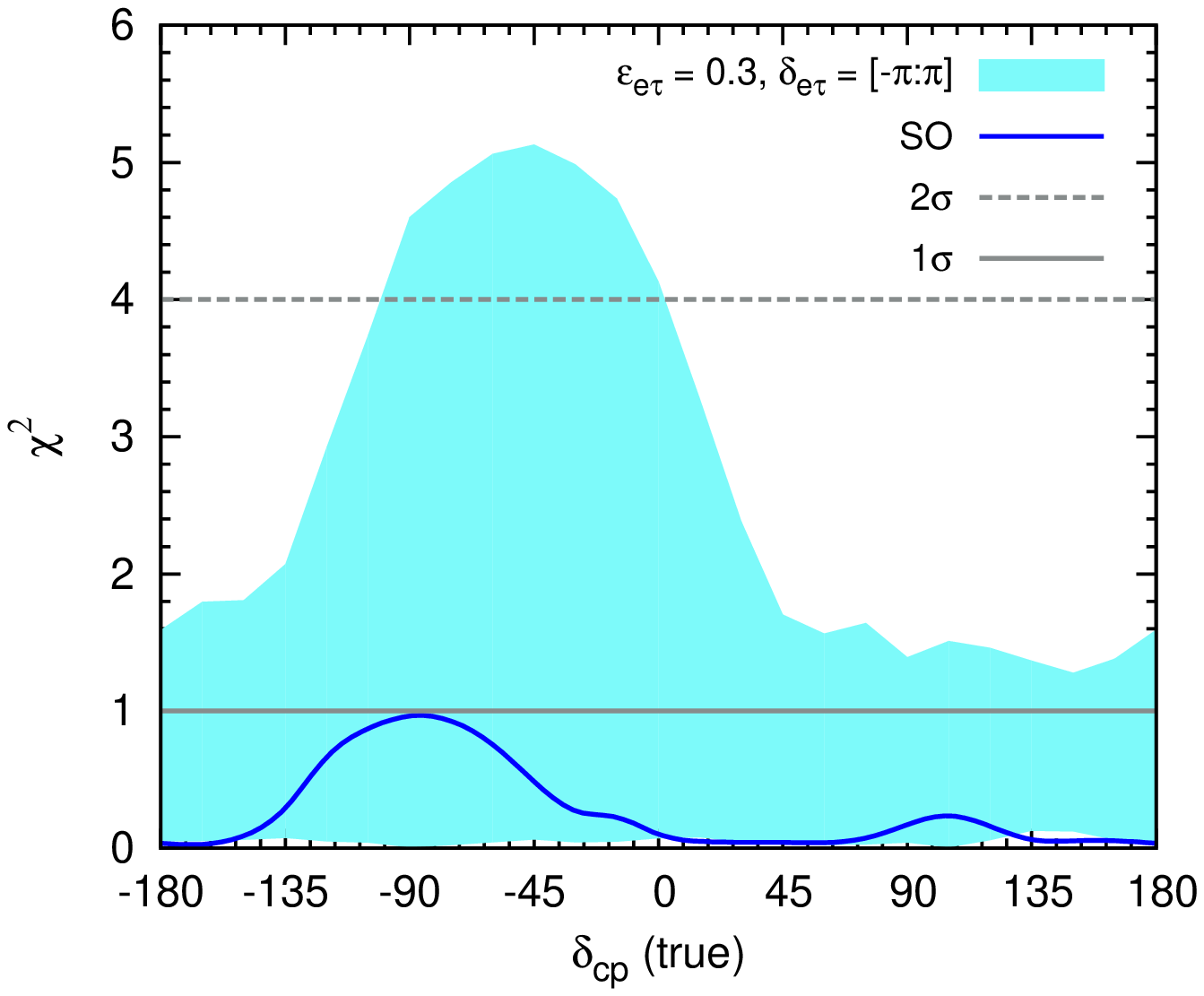}
\includegraphics[width=5.4cm,height=4.5cm]{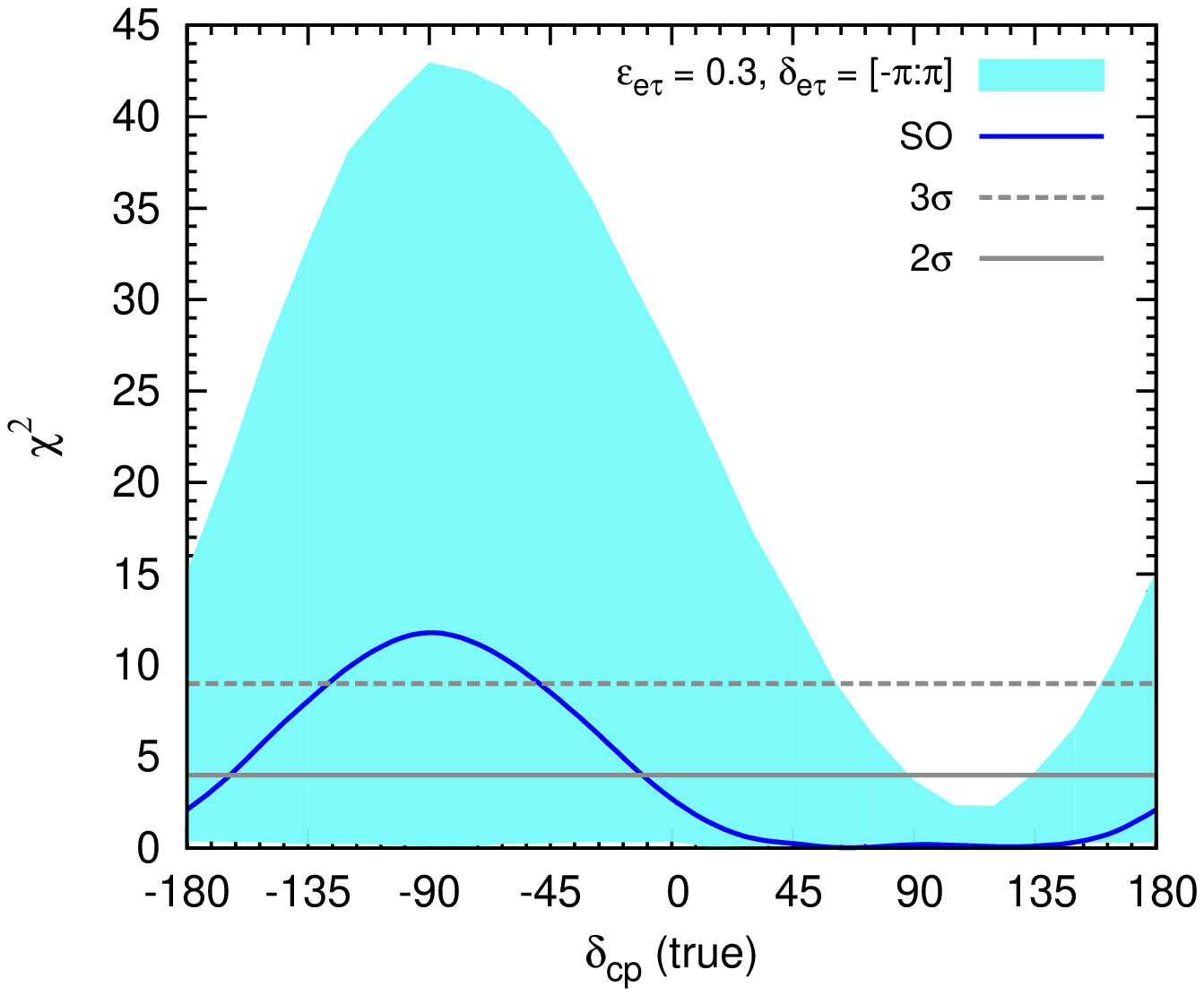}
\includegraphics[width=5.4cm,height=4.5cm]{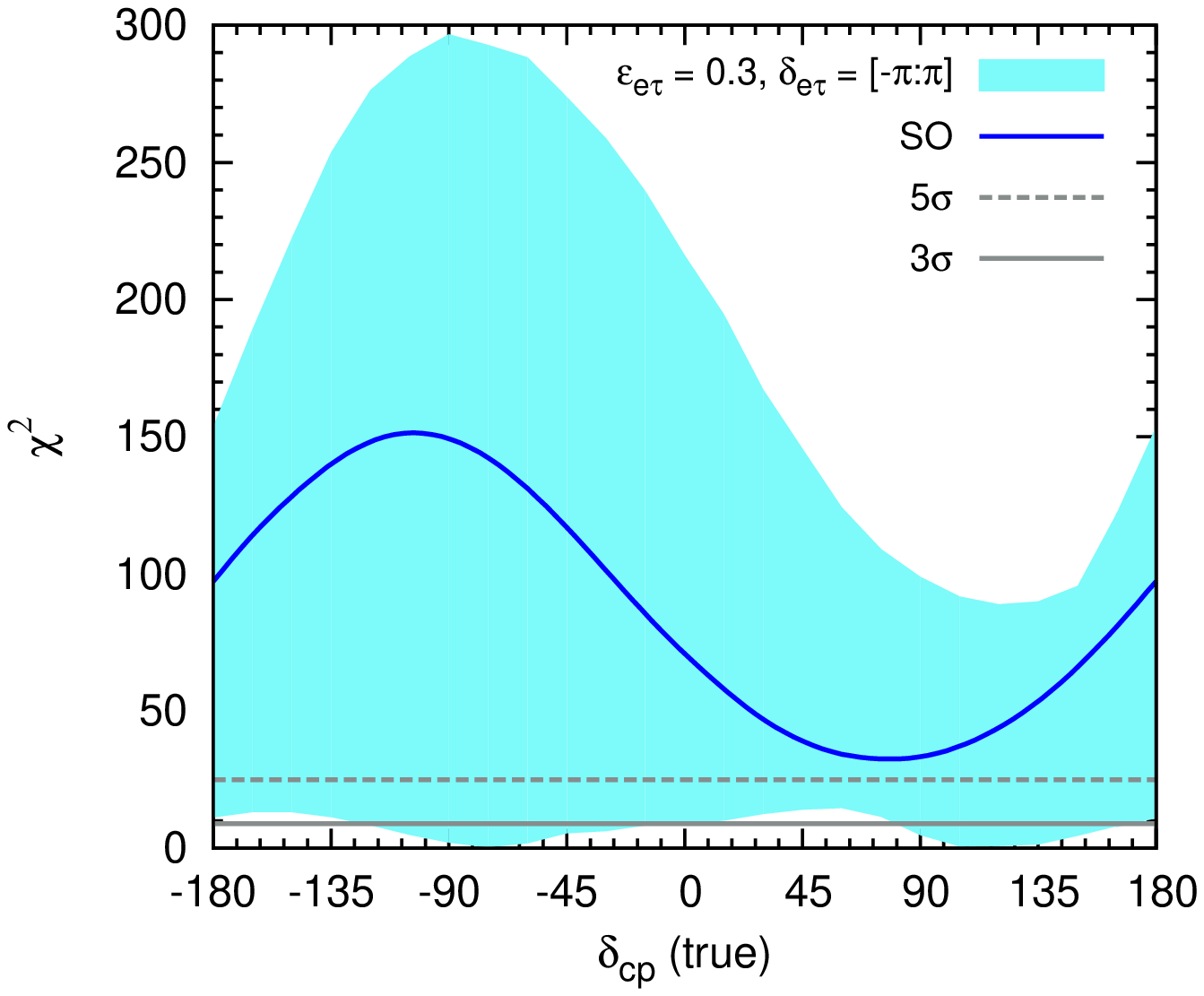}
\end{center}
\caption{ Mass hierarchy sensitivity as a function of true values of $\delta_{CP}$. The blue solid line in the figure corresponds to MH sensitivity without NSI, whereas blue band in the figure shows the MH sensitivity  in presents of NSI ($\varepsilon_{e\tau}$ =0.3) in the allowed range of $\delta_{e\tau}$ for T2K (left panel), NO$\nu$A (middle panel) and DUNE (right panel).}
\label{MH}
\end{figure}

 However, if we compare the  top and bottom panels of Fig.~\ref{MH-OSC}, we can see that there is considerable overlap between 
the hierarchies in the presence of NSIs and this overlap will worsen the hierarchy determination capability of long-baseline experiments. 
Further, the MH sensitivity as a function of  true values of Dirac CP phase $\delta_{CP}$ is shown in Fig.~\ref{MH}. In the figure,  
the solid blue line corresponds to the MH sensitivity in SO, which is obtained by comparing true event spectrum as NH and test event spectrum as IH. 
The blue band in the figure shows the variation in MH sensitivity for different values of $\delta_{e\tau}$ with $\varepsilon_{e\tau} =0.3$. In all cases,
 we do marginalization over SO parameters in their allowed parameter space and add a prior on $\sin^2 2\theta_{13}$. From the figure, it is clear that though the presence of NSI worsen MH sensitivity, there is a possibility to determine mass hierarchy for T2K (NO$\nu$A)  above  $2\sigma$ ($3\sigma$) for 30 \% (75\%) of parameter space of $\delta_{CP}$.  

\subsection{ Effect on the determination of octant of $\theta_{23}$}  

The precision measurements of atmospheric neutrino oscillation data by Super-Kamiokande experiment prefers a maximal 
mixing of $\theta_{23}$, i.e,  $\theta_{23} =\pi/4$. However, disappearance measurements of MINOS \cite{minos23}, point towards non-maximal mixing, 
which contradicts the measurements of Super-Kamiokande. Therefore, there are two possibilities, either $\theta_{23} < \pi/4$, so called 
Lower Octant (LO) or $\theta_{23} >\pi/4$, so called Higher Octant (HO). The T2K disappearance measurement, which provides the most
precise value of  $\theta_{23}$, indicates that $\theta_{23}$ is  near to maximal. However,  T2K data along with 
reactor data show that $\theta_{23}$ is in higher octant. The resolution of such tension between LO and HO of atmospheric mixing angle 
is one of the challenging goal of long baseline neutrino oscillation experiments. In this section, we discuss the 
effect of LFV-NSI on the resolution of octant of atmospheric mixing angle.

\begin{figure}[!htb]
\begin{center}
\includegraphics[width=5.4cm,height=4.5cm]{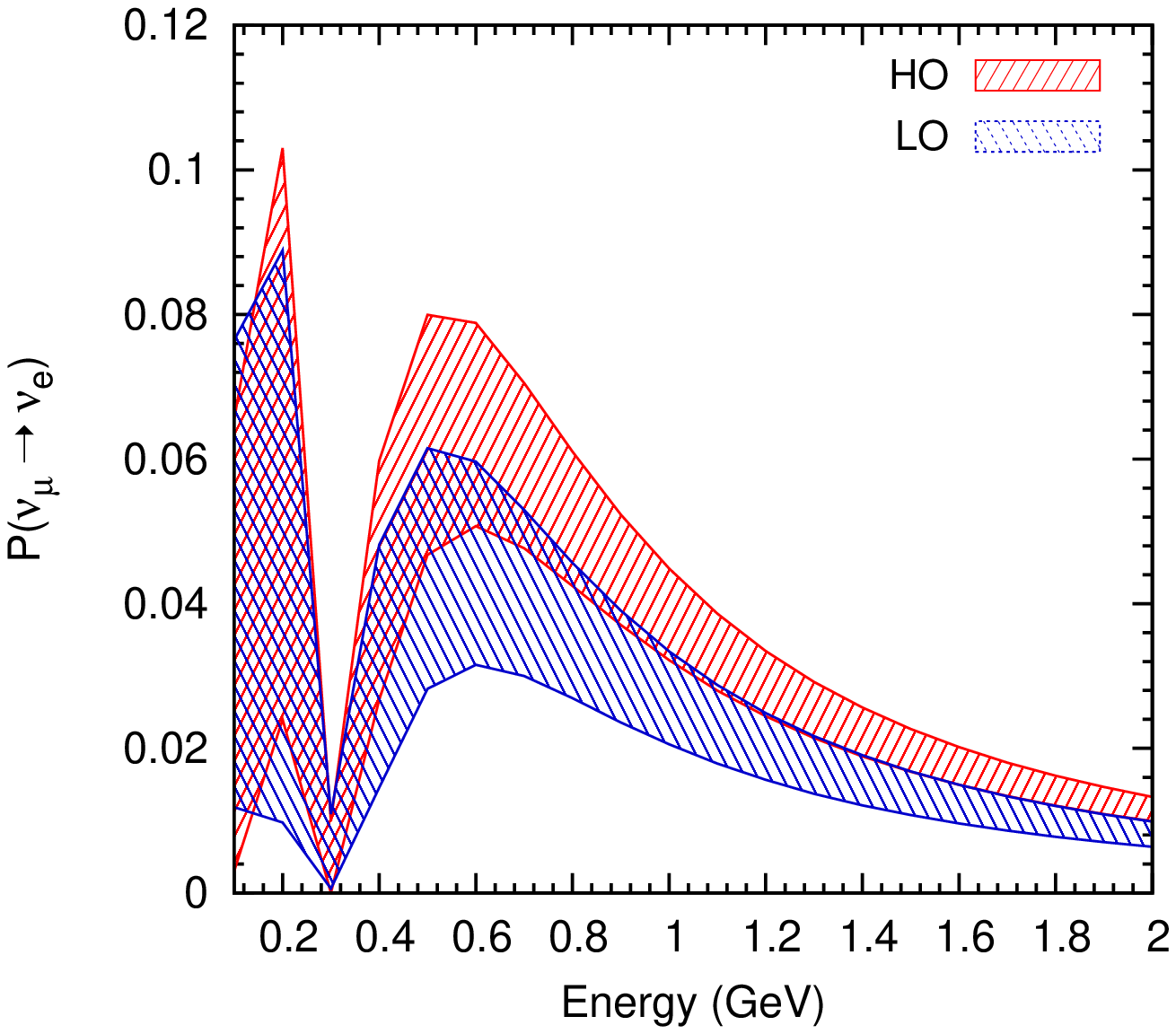}
\includegraphics[width=5.4cm,height=4.5cm]{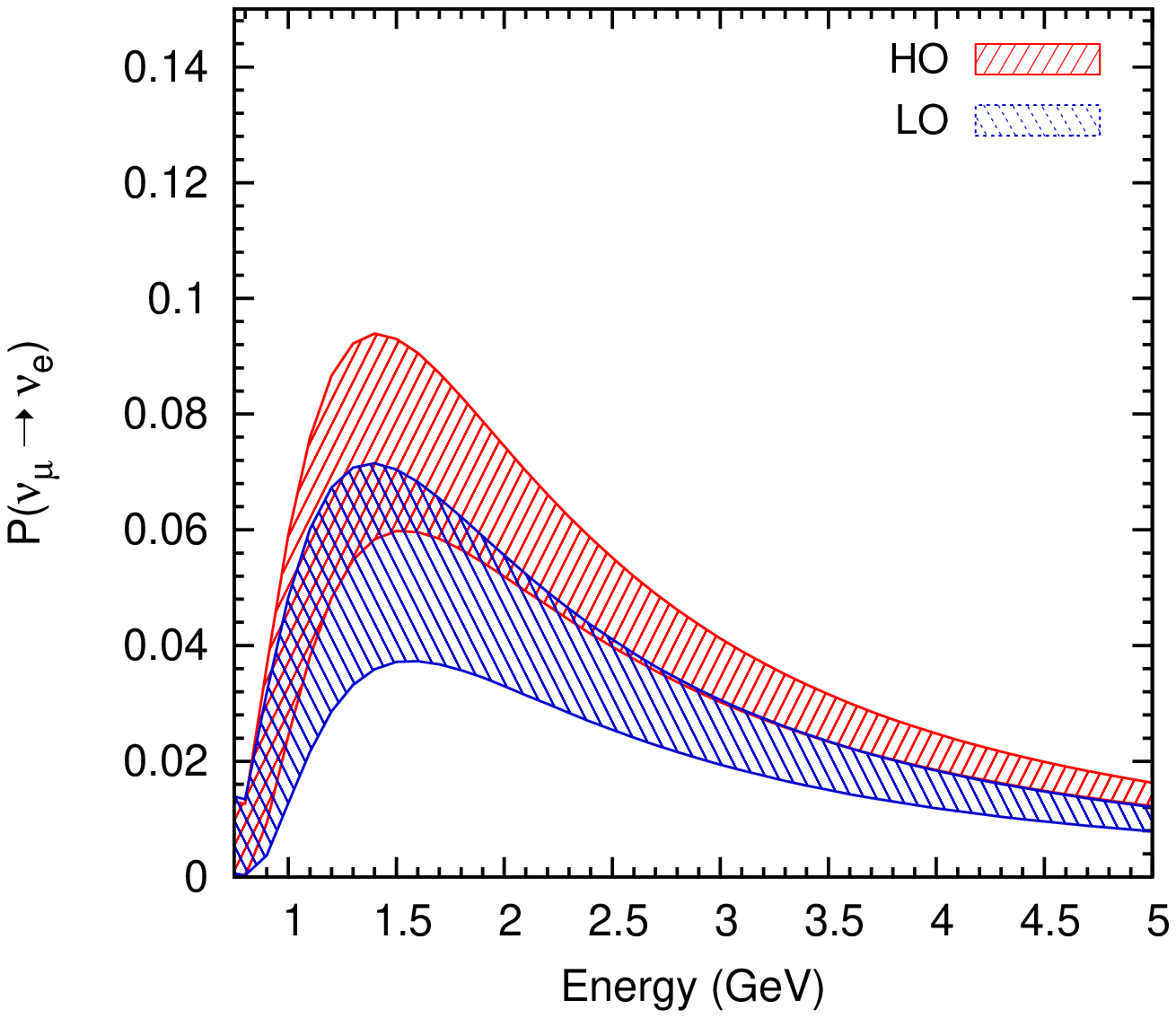}
\includegraphics[width=5.4cm,height=4.5cm]{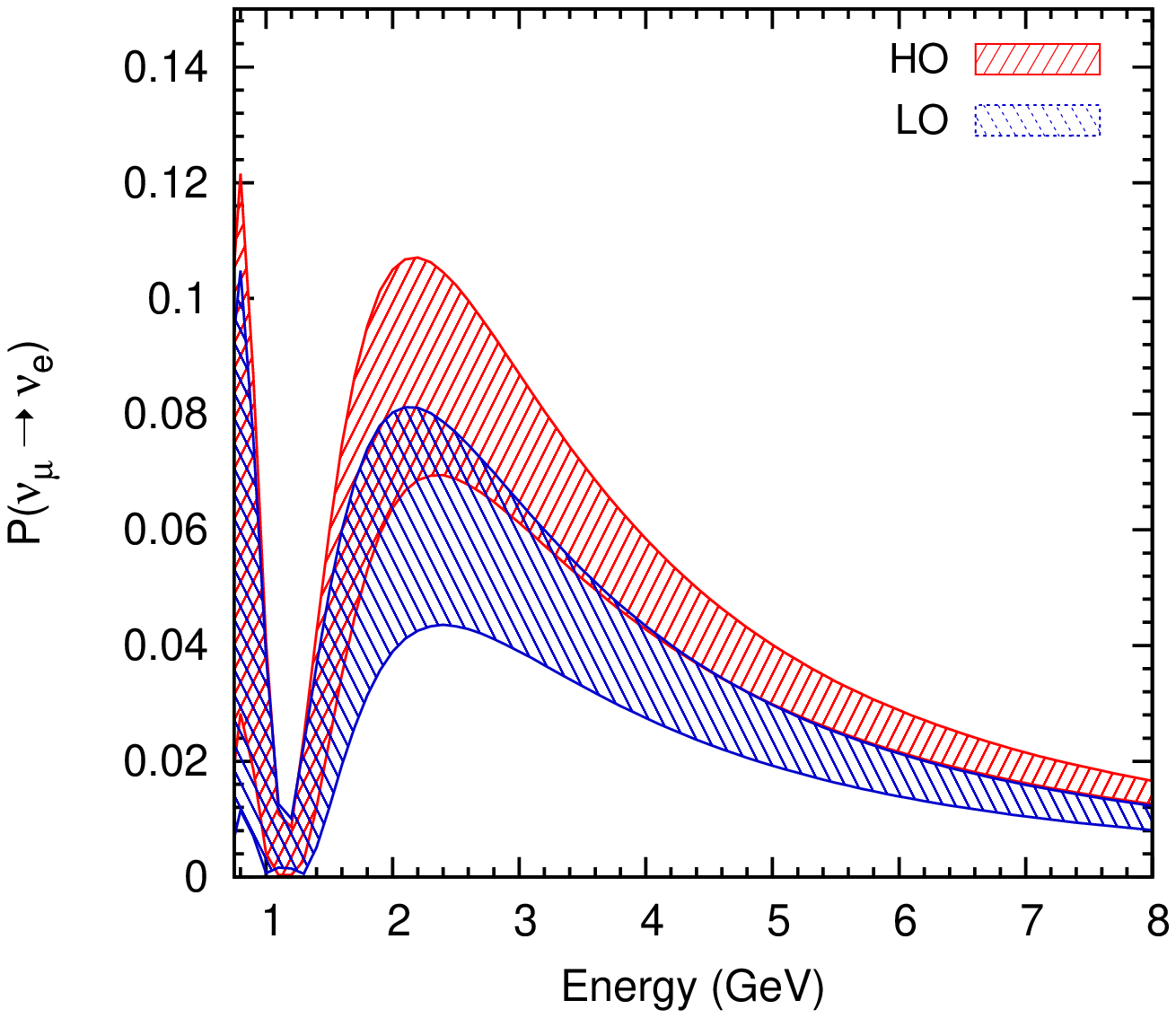}
\includegraphics[width=5.4cm,height=4.5cm]{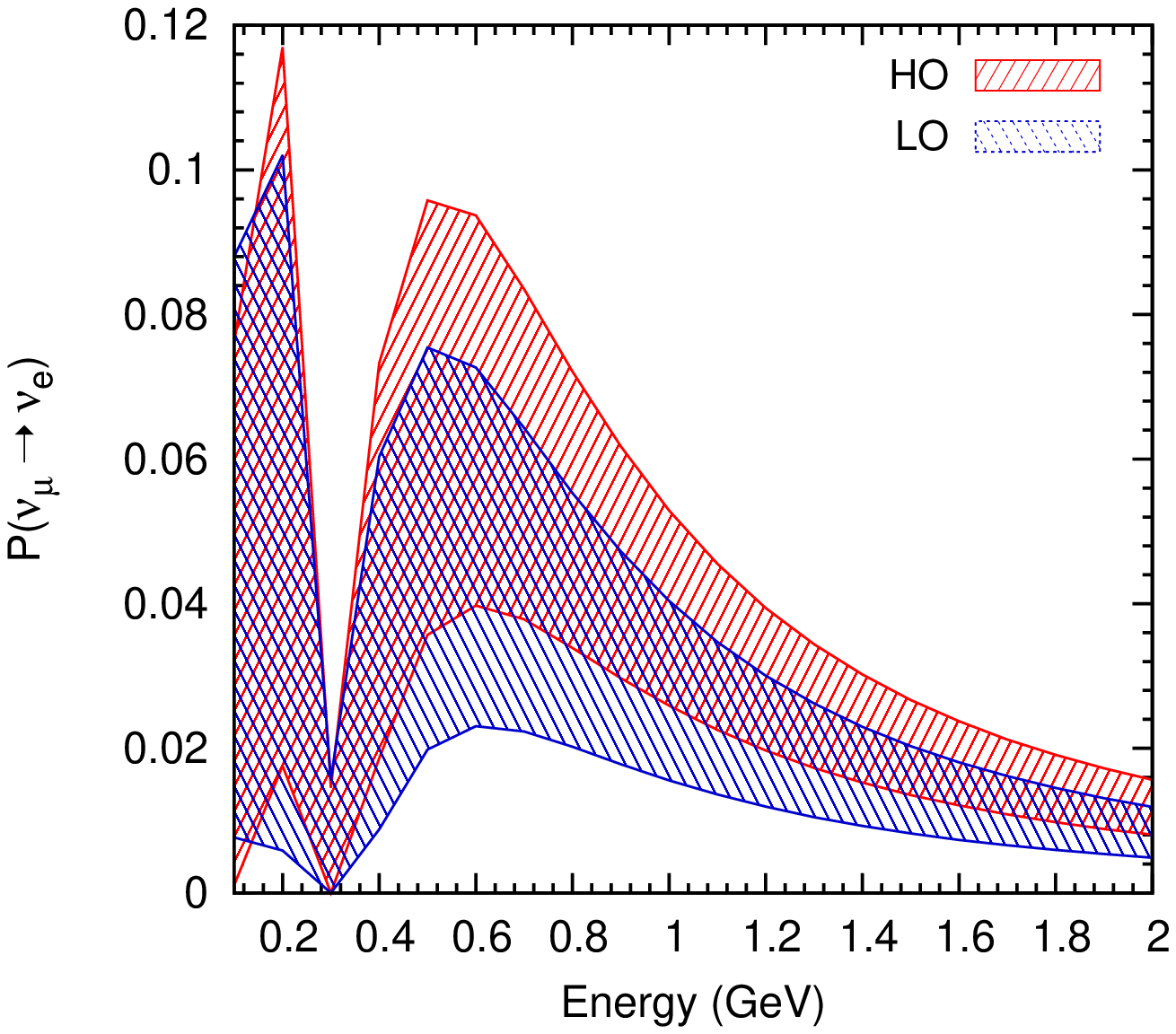}
\includegraphics[width=5.4cm,height=4.5cm]{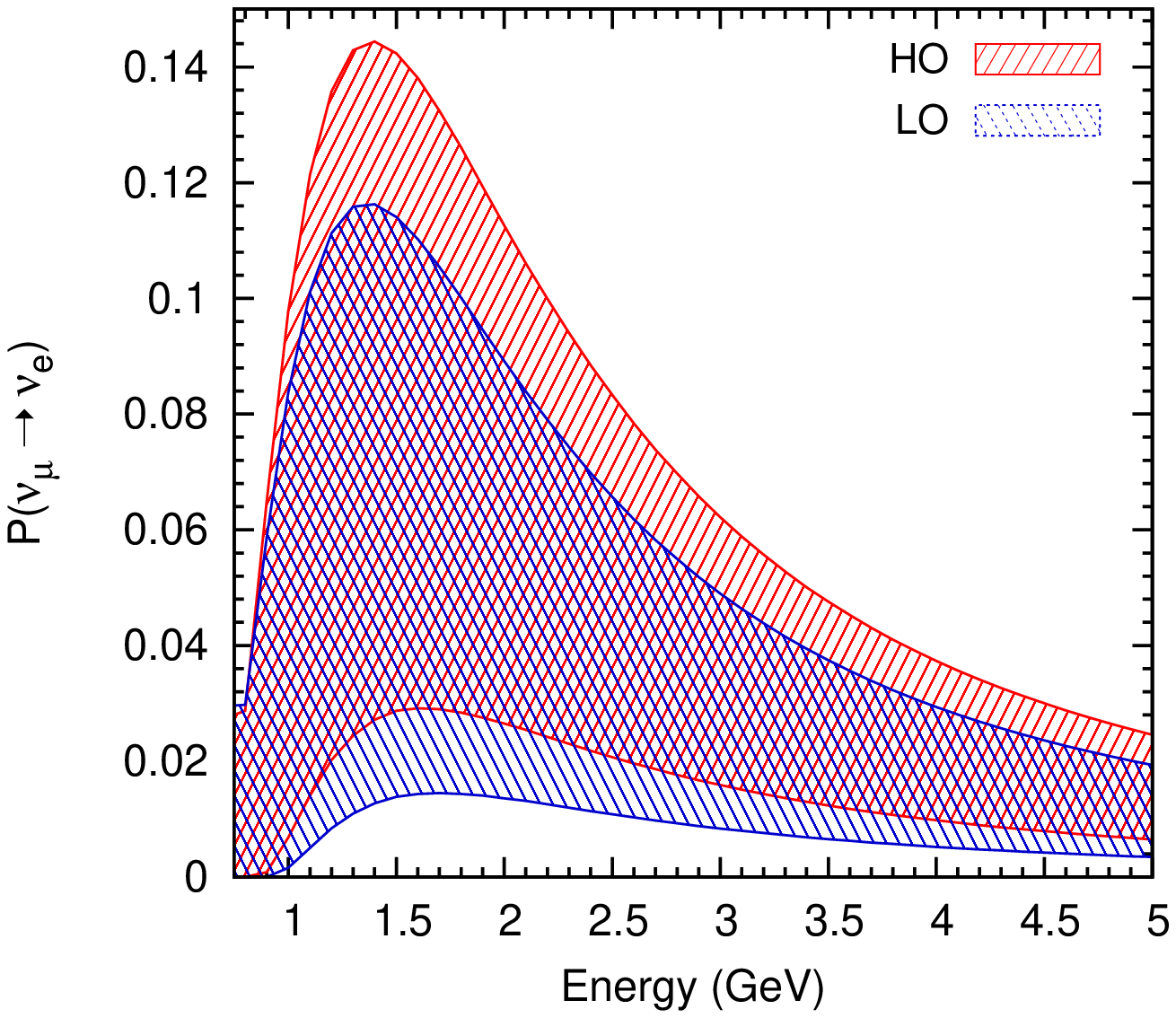}
\includegraphics[width=5.4cm,height=4.5cm]{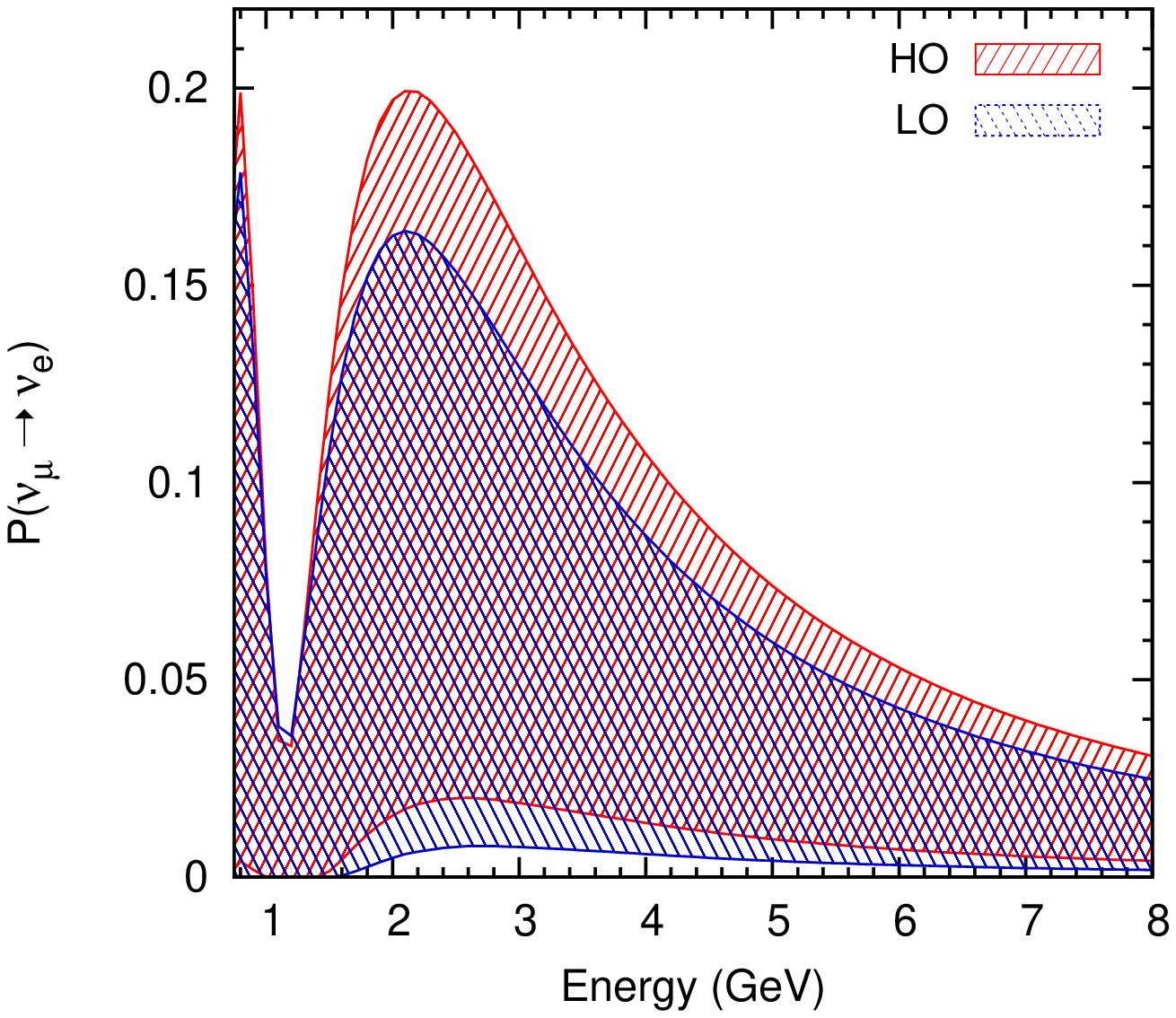}
\end{center}
\caption{Neutrino appearance probability for the $\nu_{\mu} \rightarrow \nu_{e}$ without NSI (top panel) and with 
NSI (bottom panel) by assuming both HO (red) and LO (blue) for T2K (left panel), NO$\nu$A (middle panel) and DUNE (right panel).}
\label{OCT-OSC}
\end{figure}

\begin{figure}[!htb]
\begin{center}
\includegraphics[width=5.4cm,height=4.5cm]{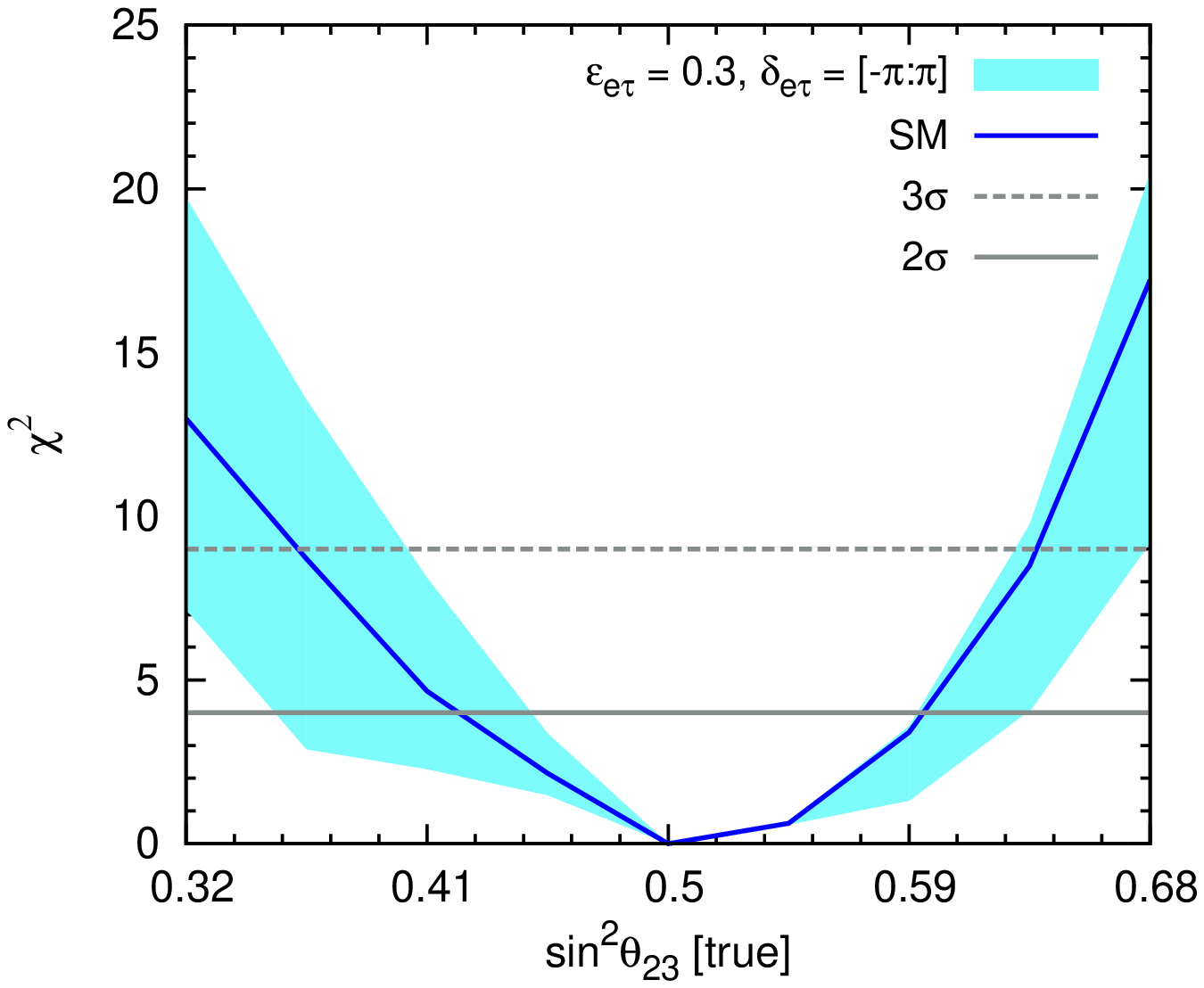}
\includegraphics[width=5.4cm,height=4.5cm]{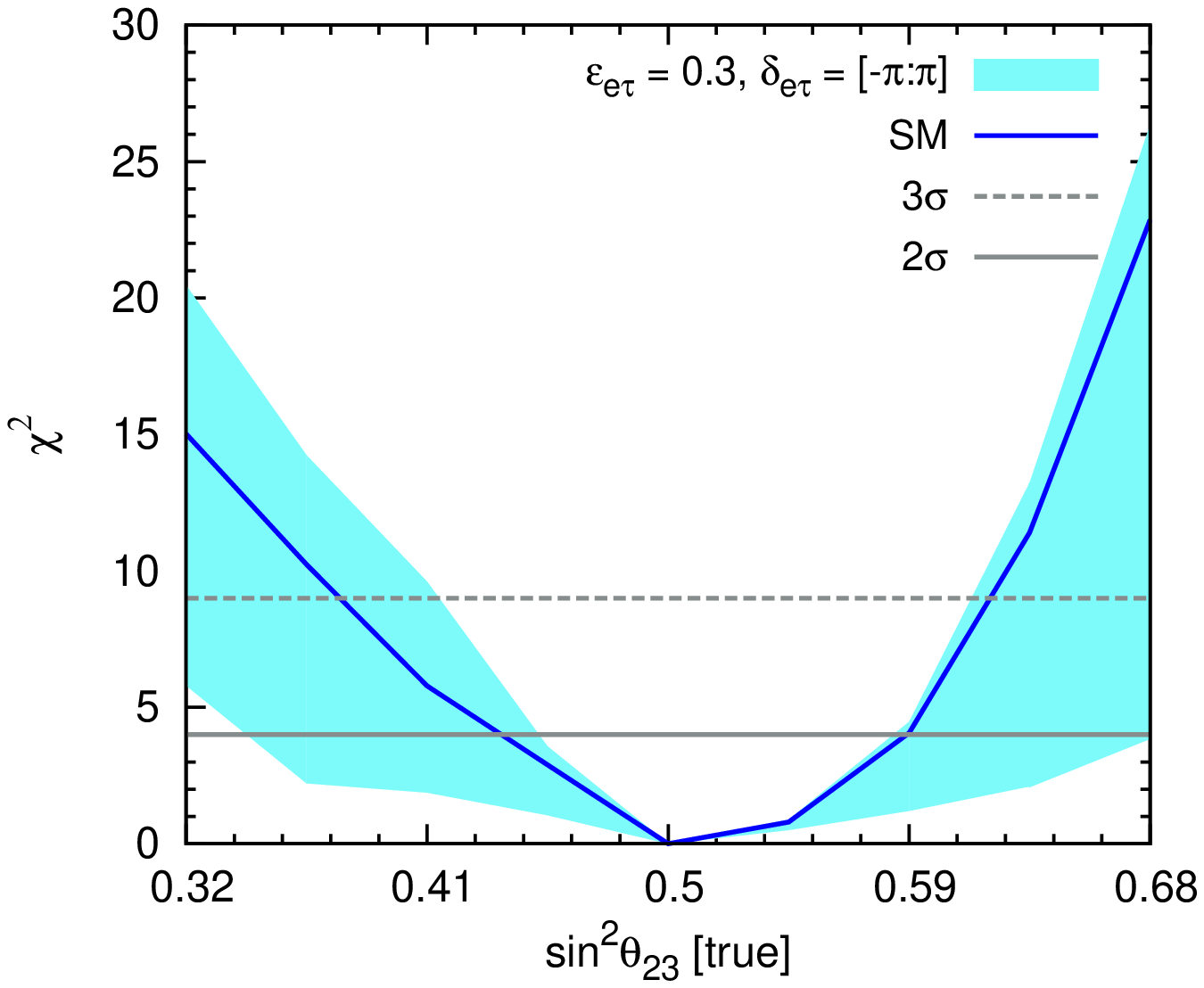}
\includegraphics[width=5.4cm,height=4.5cm]{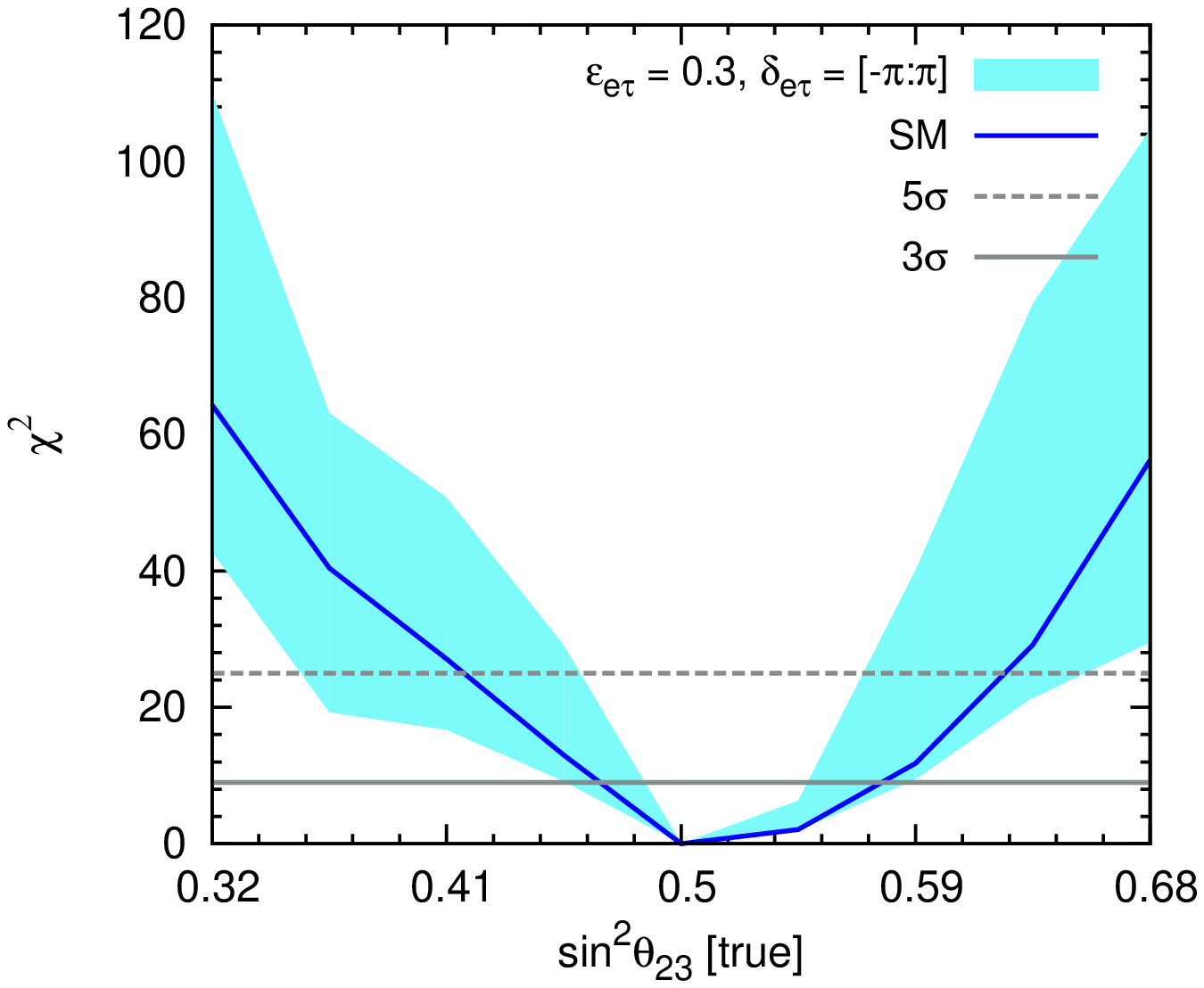}
\end{center}
\caption{ Octant sensitivity as a function of true values of $\sin^2\theta_{23}$. The blue line in the figure corresponds to octant without NSI, whereas light blue band in the figure shows the octant sensitivity  in presents of NSI ($\varepsilon_{e\tau}$ =0.3) in the allowed range of $\delta_{e\tau}$ for T2K (left panel), NO$\nu$A (middle panel) and DUNE (right panel). Neutrino MH is assumed to be Normal Hierarchy}
\label{oct-sensitivity}
\end{figure}

The octant degeneracy  is merely a consequence of inherent structure of three flavour neutrino oscillation probability, where a set of oscillation 
parameters gives disconnected regions in neutrino oscillation parameter space and it makes too difficult to find the true solution. However, 
the matter effect in long baseline experiments can help to resolve the octant of $\theta_{23}$ \cite{OCt-2}, since the oscillation probability 
gives different contributions to HO and LO as one can see from the upper panels of Fig.~\ref{OCT-OSC}. From the lower panels of the figure, it can be seen
 that there is considerable overlap between the lower and higher octants in the presence of LFV-NSI, which will worsen the sensitivity of long 
baseline experiments in the determination of octant of $\theta_{23}$. Moreover, the octant sensitivity  as a function of true value of 
$\sin^2\theta_{23}$ is given in Fig.~\ref{oct-sensitivity}.  The octant sensitivity is obtained by comparing true event spectrum  (HO/LO) 
with test event spectrum (LO/HO). While calculating the $\chi^2$, we do marginalization over SO parameter space in their allowed values 
and add a prior on $\sin^2 2\theta_{13}$. From the figure, we can see that there is a possibility of enhancement in the sensitivity of  
octant of atmospheric mixing angle in the presence of LFV-NSIs, though LFV-NSIs worsen the sensitivity.

\subsection{ Effect on the determination of CP violating phase $\delta_{CP}$} 

One of the main objectives of long-baseline neutrino oscillation experiments is the determination of the CP violation (CPV) in the leptonic sector. Therefore, it is crucial to study the effect of NSI on the determination of CPV at T2K, NO$\nu$A and DUNE experiments.  The direct measurement of CP violation can be obtained by looking at the difference in the  transition probability of CP conjugate channels i.e, 
by analyzing the $\nu_e$ appearance  and $\bar{\nu}_e$ appearance probabilities.

\begin{figure}
\begin{center}
\includegraphics[width=5.4cm,height=4.5cm]{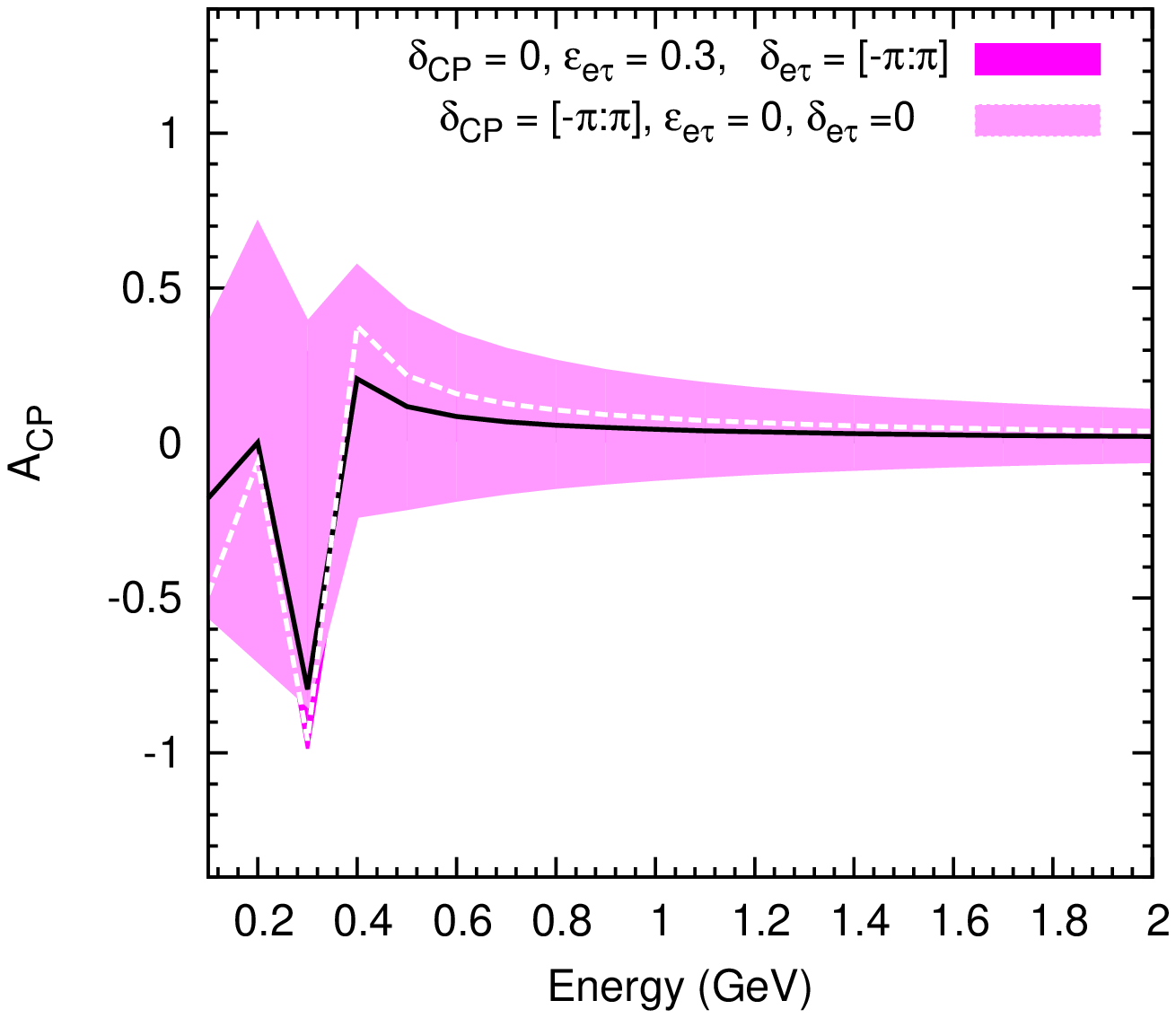}
\includegraphics[width=5.4cm,height=4.5cm]{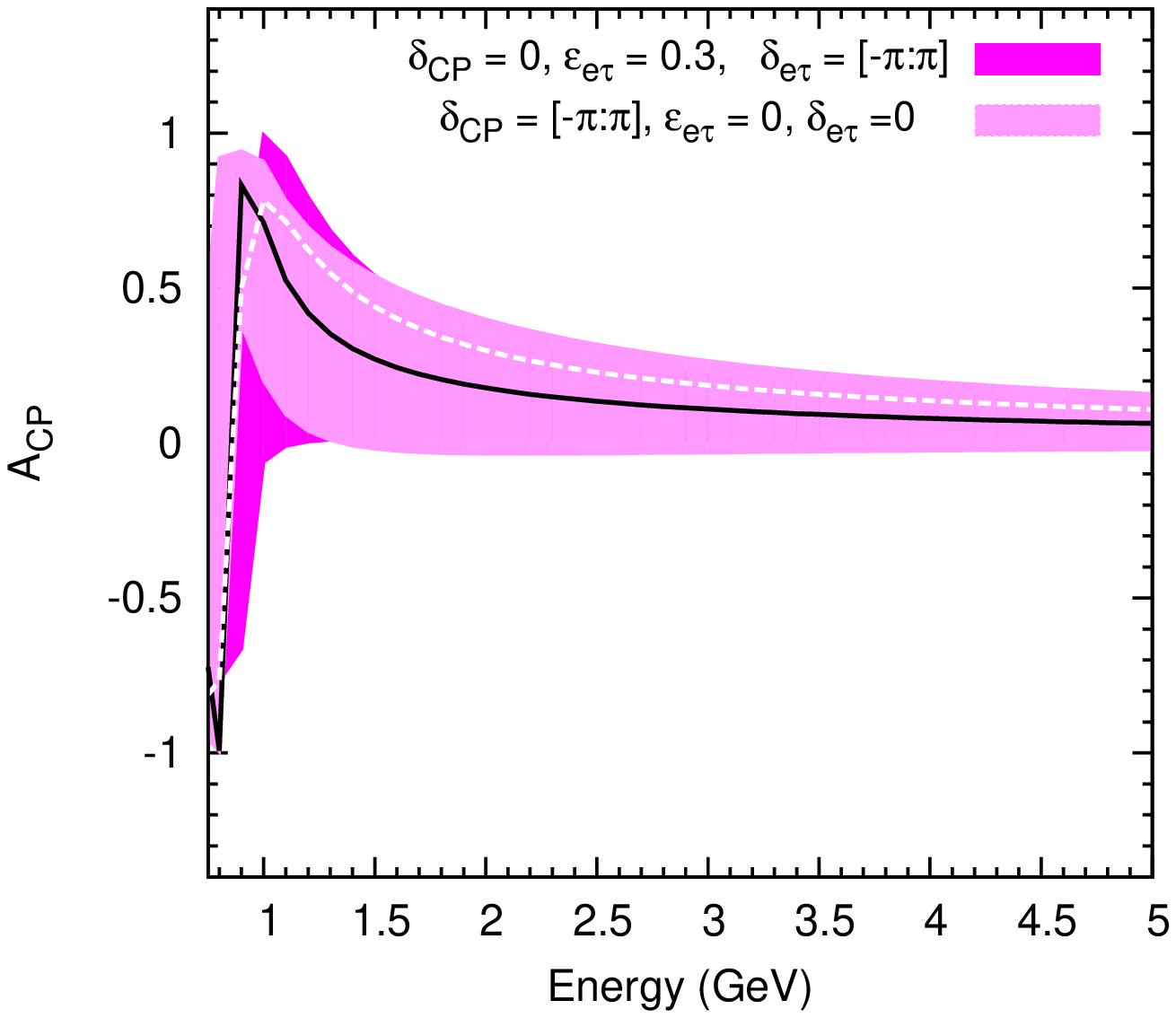}
\includegraphics[width=5.4cm,height=4.5cm]{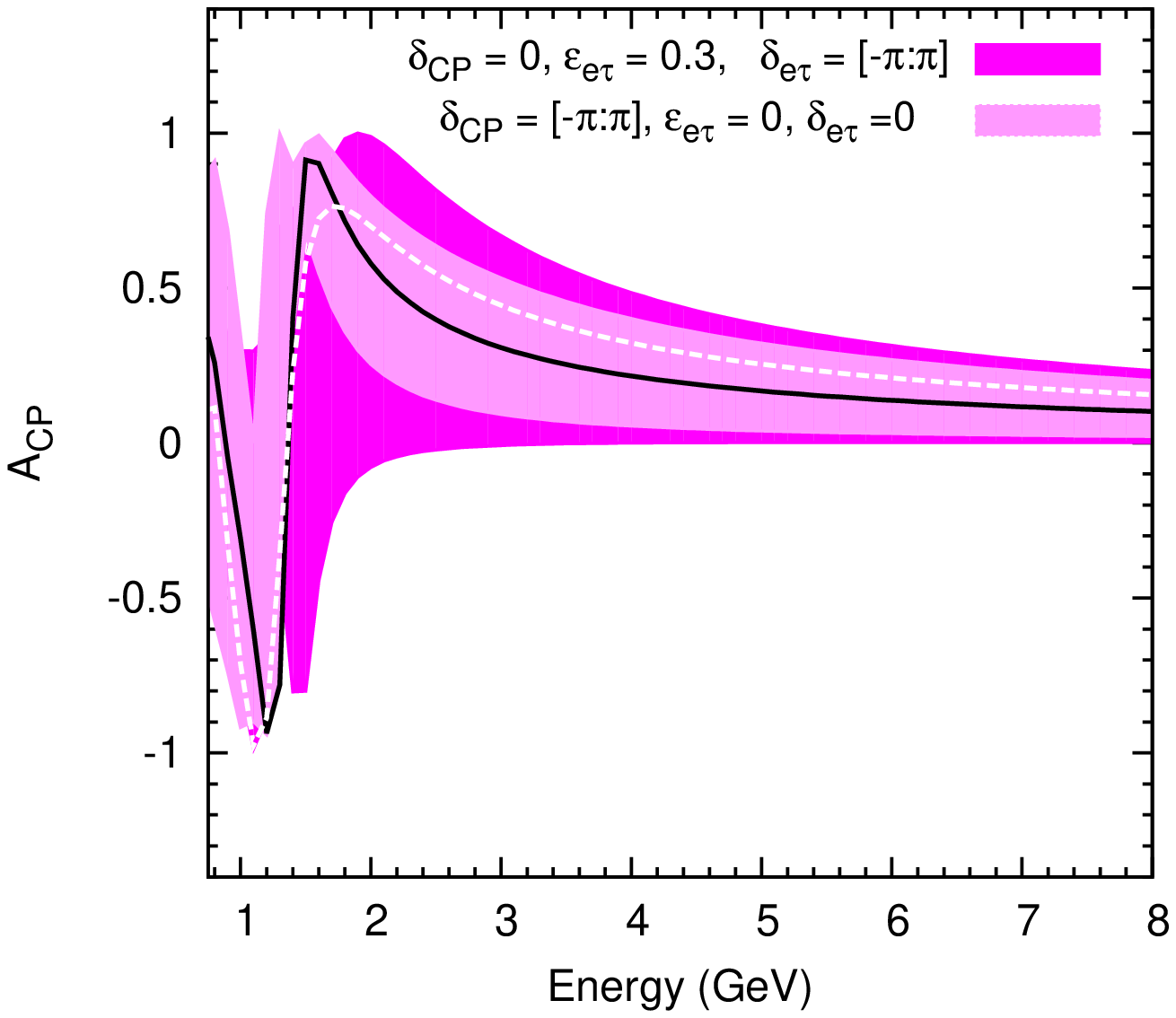}
\includegraphics[width=5.4cm,height=4.5cm]{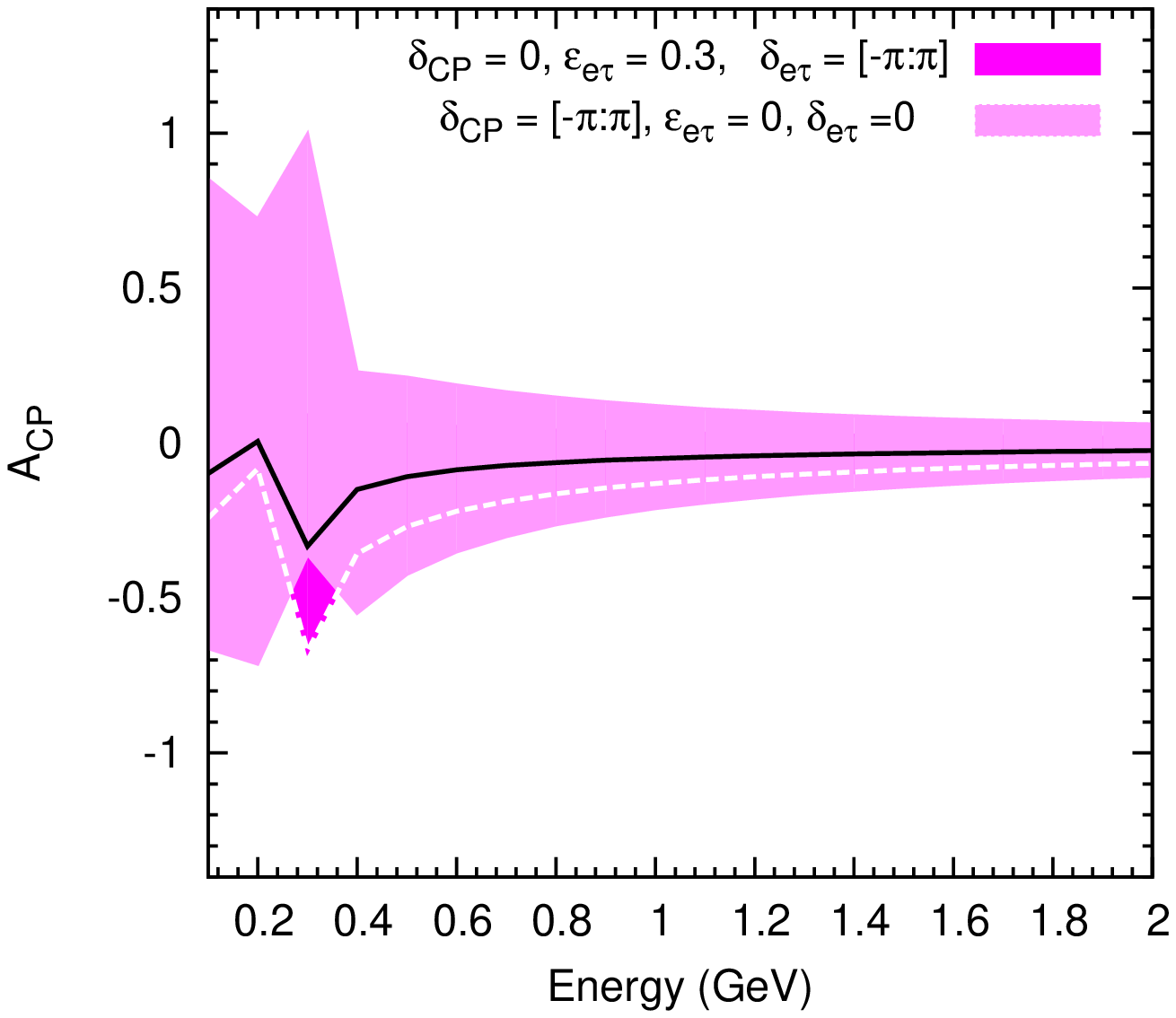}
\includegraphics[width=5.4cm,height=4.5cm]{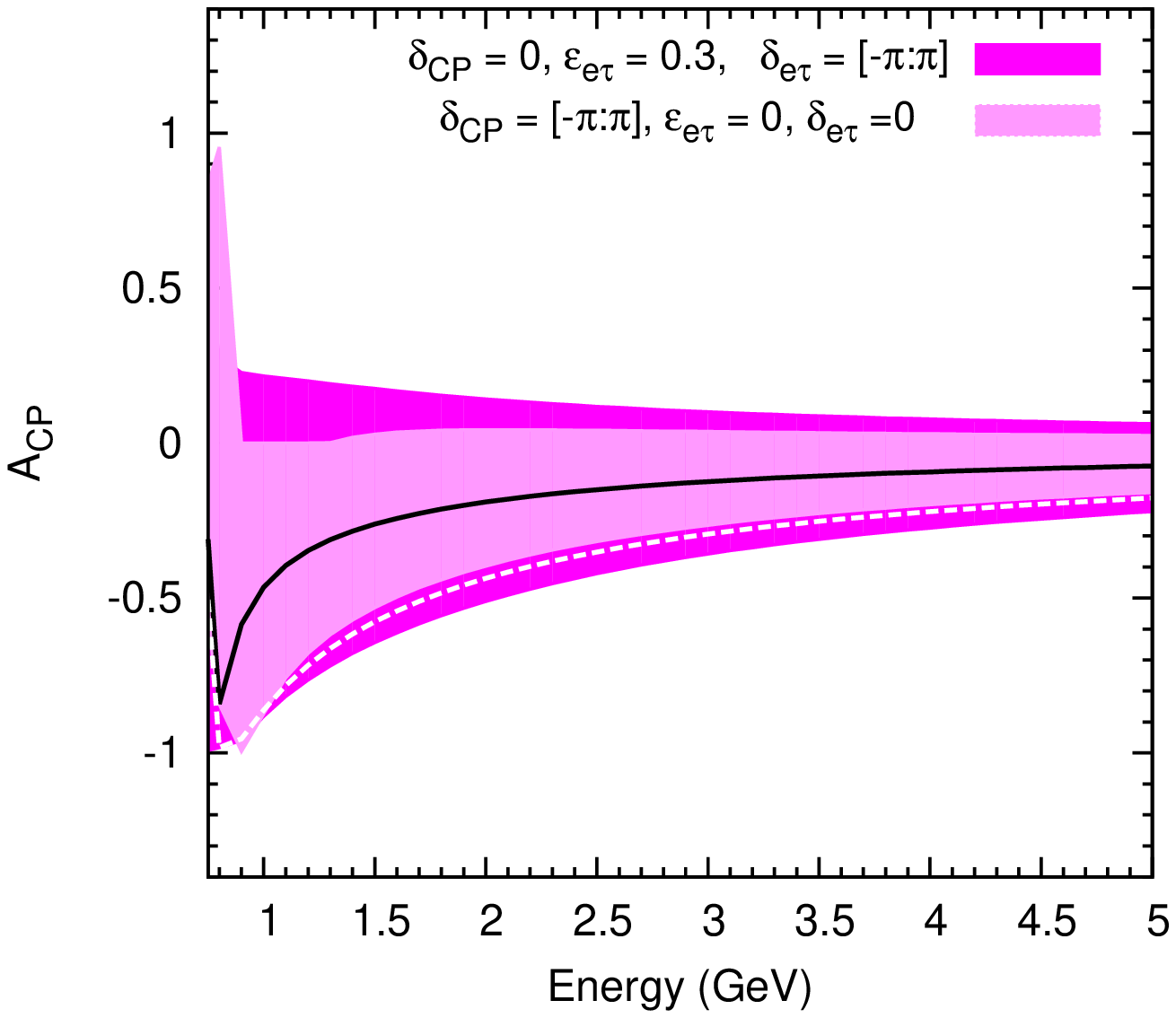}
\includegraphics[width=5.4cm,height=4.5cm]{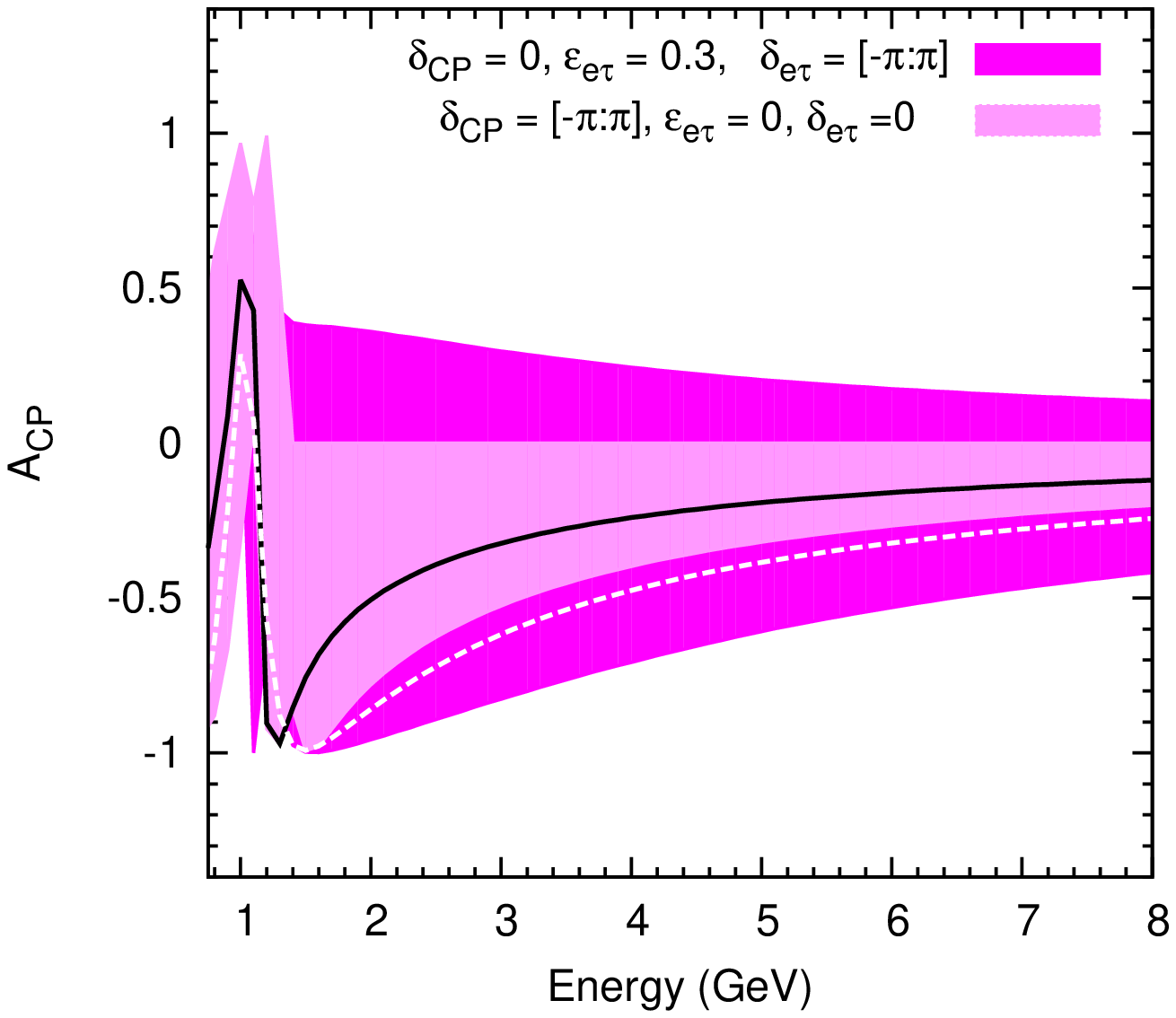}
\end{center}
\caption{The CP asymmetry bands for T2K (left panel), NO$\nu$A (middle panel) and DUNE (right panel) 
without NSI (light coloured band) and with NSI (dark coloured band) by assuming both NH (top panel) and IH (bottom panel). 
The solid black line corresponds to CP asymmetry for $\delta_{CP} = 0$ without NSI, where as the dashed white line corresponds 
to CP asymmetry for $\delta_{CP} = 0$ with NSI ($\varepsilon_{e\tau} =0.3$).}
\label{ACP}
\end{figure}

\begin{figure}
\begin{center}
\includegraphics[width=5.4cm,height=4.5cm]{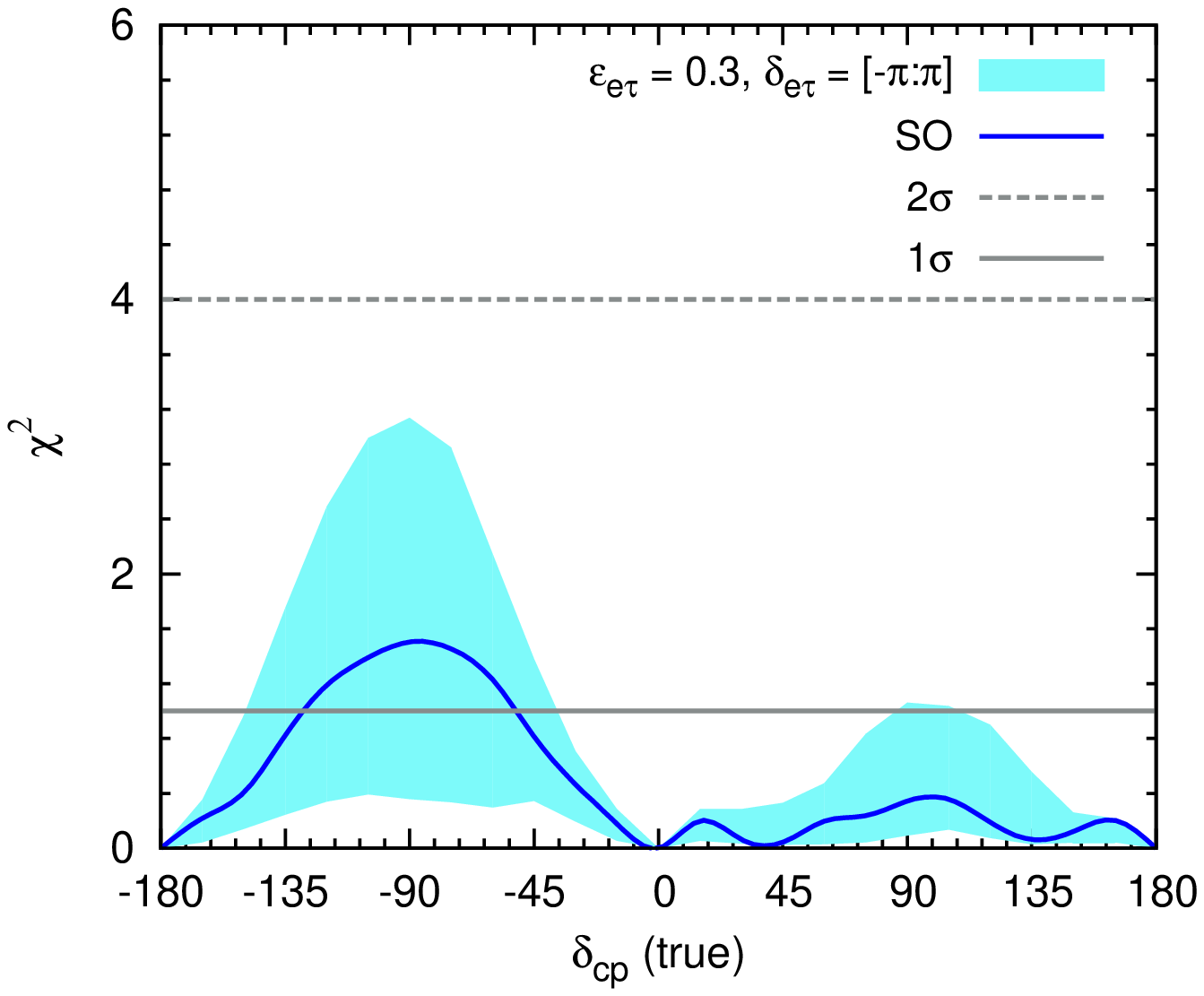}
\includegraphics[width=5.4cm,height=4.5cm]{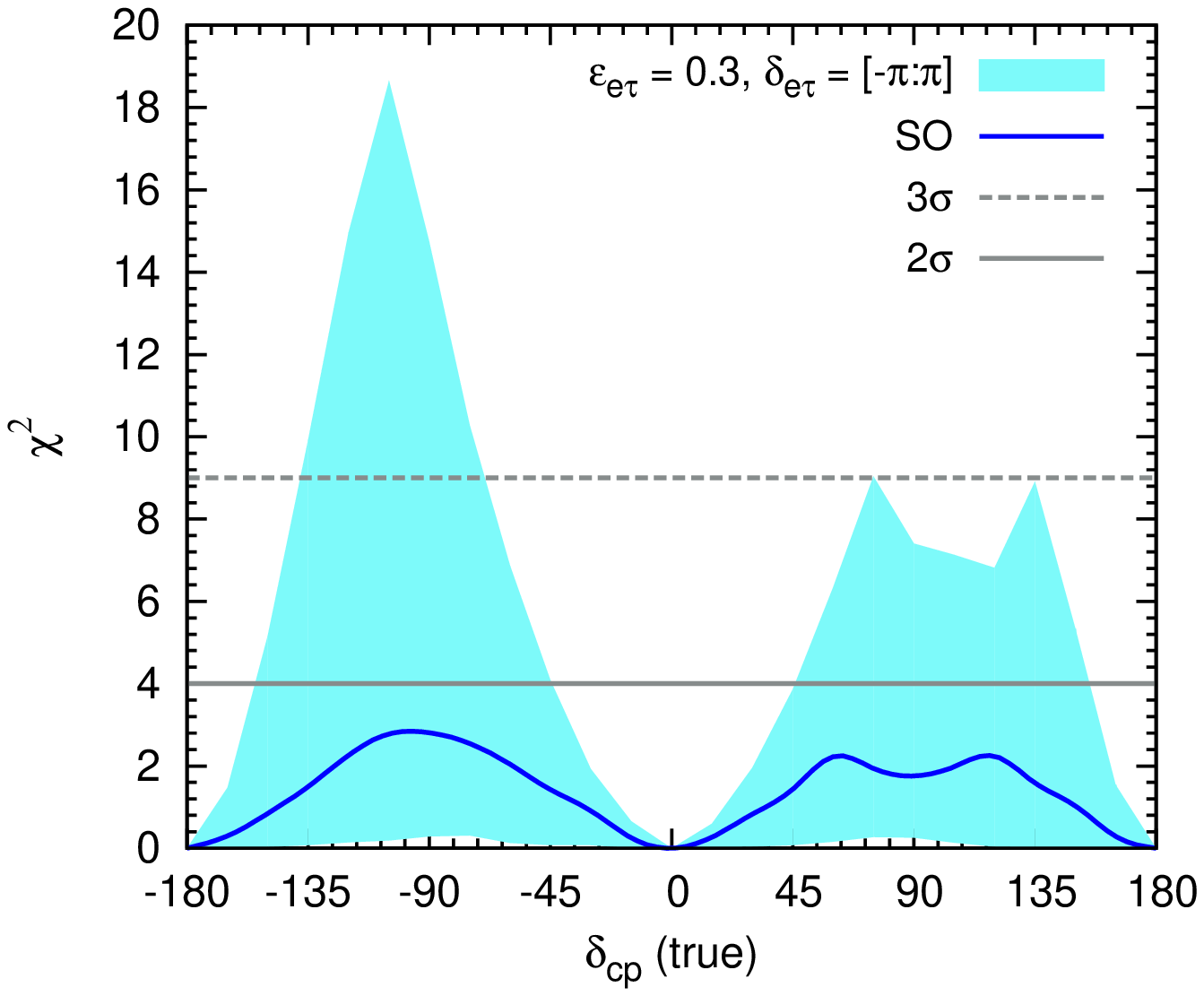}
\includegraphics[width=5.4cm,height=4.5cm]{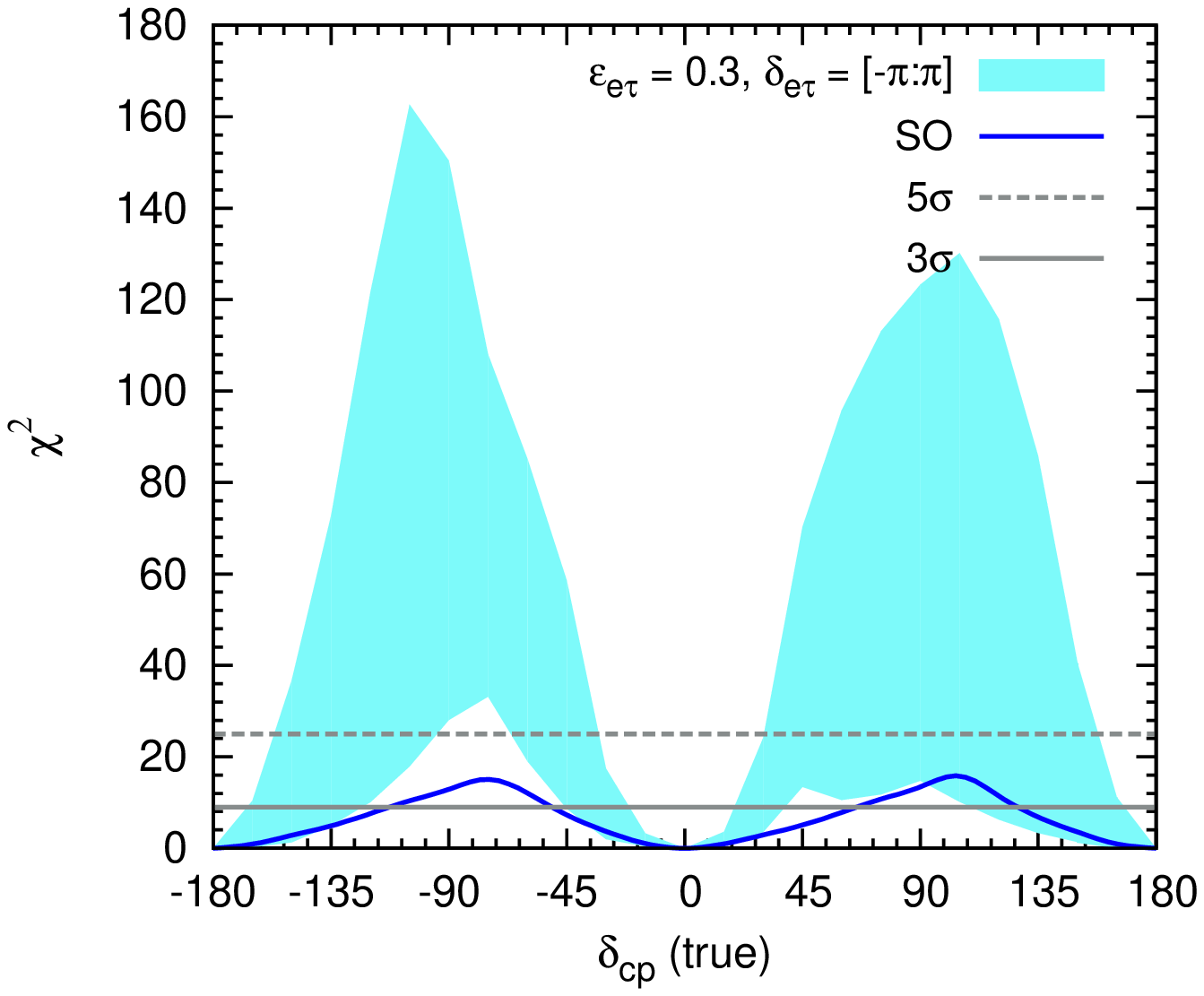}
\end{center}
\caption{The CPV potential as a function of true values of $\delta_{CP}$ for T2K (left panel), NO$\nu$A (middle panel) and DUNE (right panel) without NSI (solid blue line) and with NSI (band).}
\label{CPV-s}
\end{figure} 
 
We use the  observable so called CP asymmetry ($A_{CP}$) to quantify the effects due to CP violation and it is defined as
\begin{equation}\label{acp}
 A_{CP} = \frac{P_{\mu e} - \overline{P}_{\mu e}} {P_{\mu e} + \overline{P}_{\mu e}} 
\end{equation}
 where $P_{\mu e}$ is the $\nu_{e}$ appearance probability and $\overline{P}_{\mu e}$ is $\bar{\nu}_{e}$ appearance probability. Fig.~\ref{ACP} shows 
the CP asymmetry bands for T2K (left panel), NO$\nu$A (middle panel) and DUNE (right panel) without NSI (light coloured band) and with NSI 
(dark coloured band) by assuming both normal (top panel) and inverted (bottom panel) hierarchies. The solid black line corresponds to CP asymmetry 
for $\delta_{CP} = 0$ without NSI, whereas the dashed white line corresponds to CP asymmetry for $\delta_{CP} = 0$ with NSI ($\varepsilon_{e\tau} = 0.3$). 
The dark bands in the figure show the impact of the phase of LFV-NSI parameter on $A_{CP}$. Therefore, the dark bands  correspond to the 
fake CP signals which are coming from NSI.  From the  figures, we can see that there is not  much change in the asymmetry  with NSI and  
without NSI  in the case of T2K, whereas in the case of NO$\nu$A the bands show that there is significant change in the asymmetry with NSI and without NSI. 
Moreover, the change in the asymmetry is quite large in the case of DUNE. From the figure, it is clear that  NSI can  give fake CP signals 
even without considering contributions from the intrinsic phase ($\delta_{e\tau}$) of NSI parameter and  therefore, it is very difficult to determine 
the CP violation in the presence of NSIs.

The CP violation sensitivity as a function of true values of $\delta_{CP}$ for T2K (left panel), NO$\nu$A (middle panel) and DUNE 
(right panel) is shown  in Fig.~\ref{CPV-s}. The  CP violation sensitivity is obtained by comparing the true event spectrum and test 
event spectrum with $\delta_{CP}^{test}=0,~\pi$. We do marginalization over the SO parameter space and add a prior on $\sin^2\theta_{13}$. 
From the figure, it is clear that there is a possibility to determine CP violation above $2\sigma$, $3\sigma$ and $5\sigma$ with 30\%, 60 \% and 60 \% of $\delta_{CP}$ parameter space for T2K, NO$\nu$A and DUNE respectively.

\section{Degeneracies among oscillation parameters in presence of LFV-NSI} 

 One of the major issues in neutrino oscillation physics is the parameter degeneracy among the oscillation parameters. 
In the standard oscillation physics, there are four-fold degeneracies among the oscillation parameters and they are known as octant degeneracy 
and mass hierarchy (sign of $\Delta m^2_{31} $) degeneracy. In this section, we present a simple  way to understand the degeneracies 
among the oscillation parameters in the presence of LFV-NSI parameter $\varepsilon_{e\tau}$, by using bi-probability plots i.e., 
CP trajectory in a $P_{({\nu_{\mu}} \rightarrow {\nu_{e}})}-P_{(\bar \nu_{\mu} \rightarrow \bar \nu_{e})}$ plane and $\delta_{CP}$-$A_{CP}$ plane.

\begin{figure}
\begin{center}
\includegraphics[width=5.4cm,height=4.5cm]{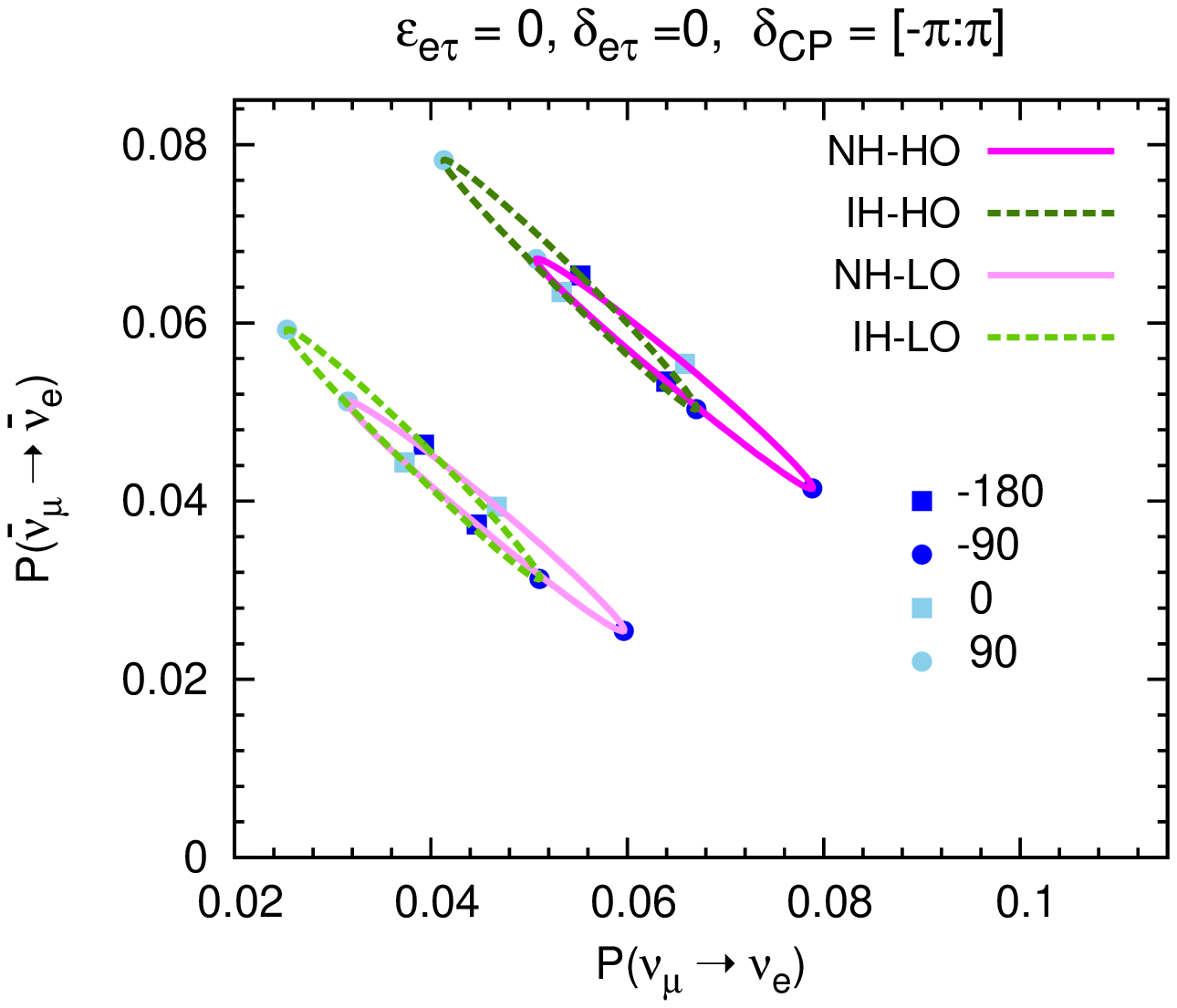}
\includegraphics[width=5.4cm,height=4.5cm]{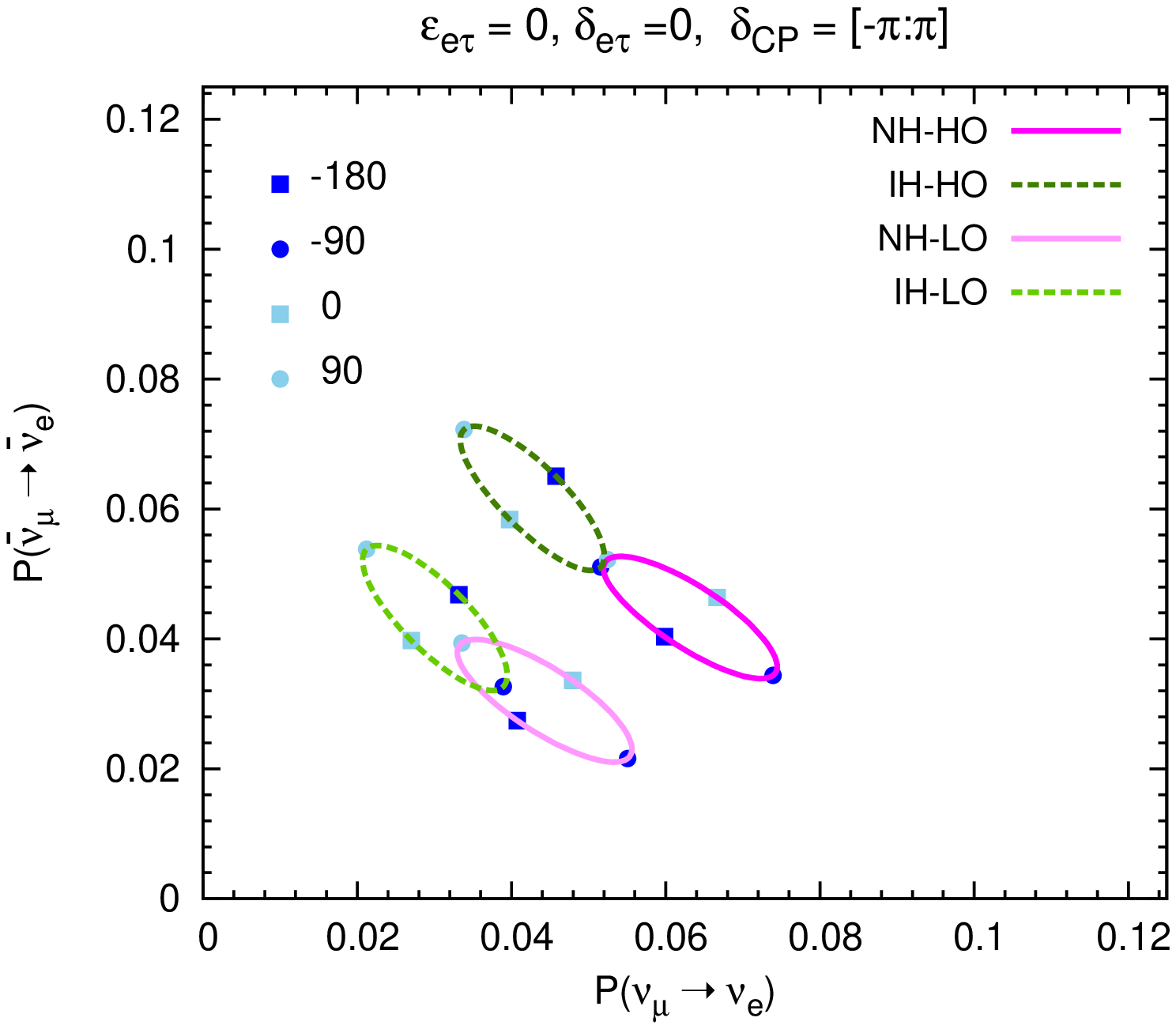}
\includegraphics[width=5.4cm,height=4.5cm]{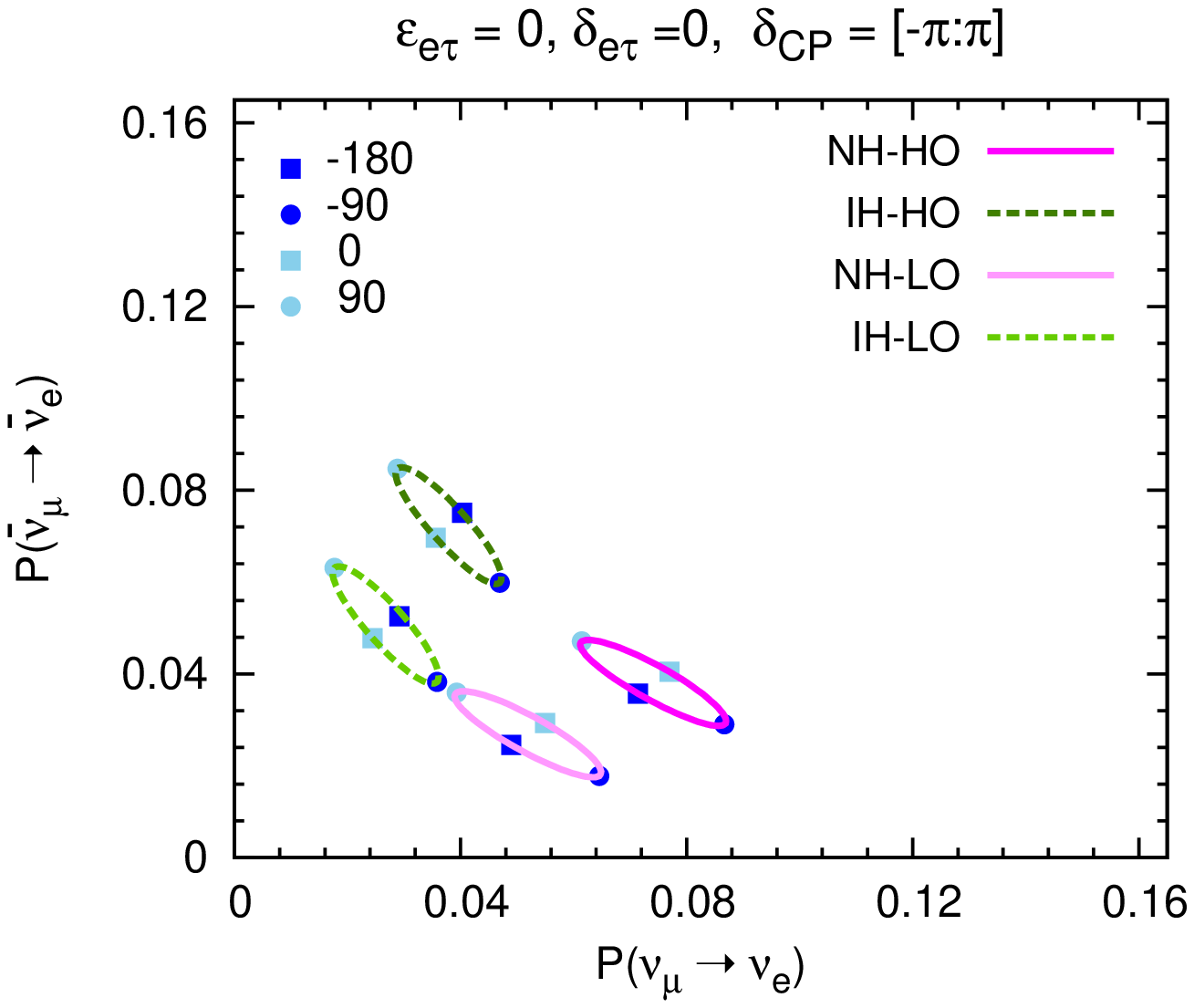}
\includegraphics[width=5.4cm,height=4.5cm]{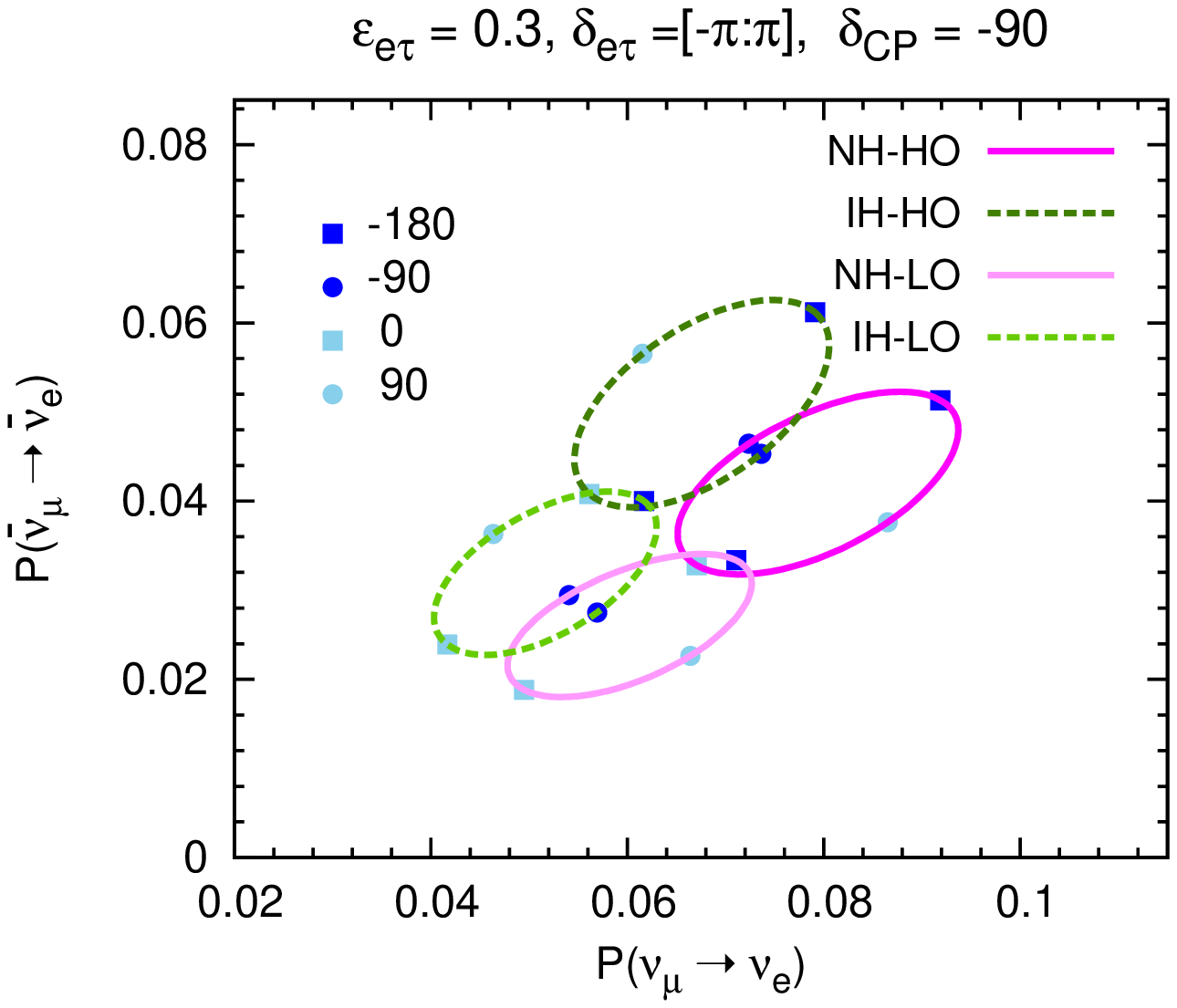}
\includegraphics[width=5.4cm,height=4.5cm]{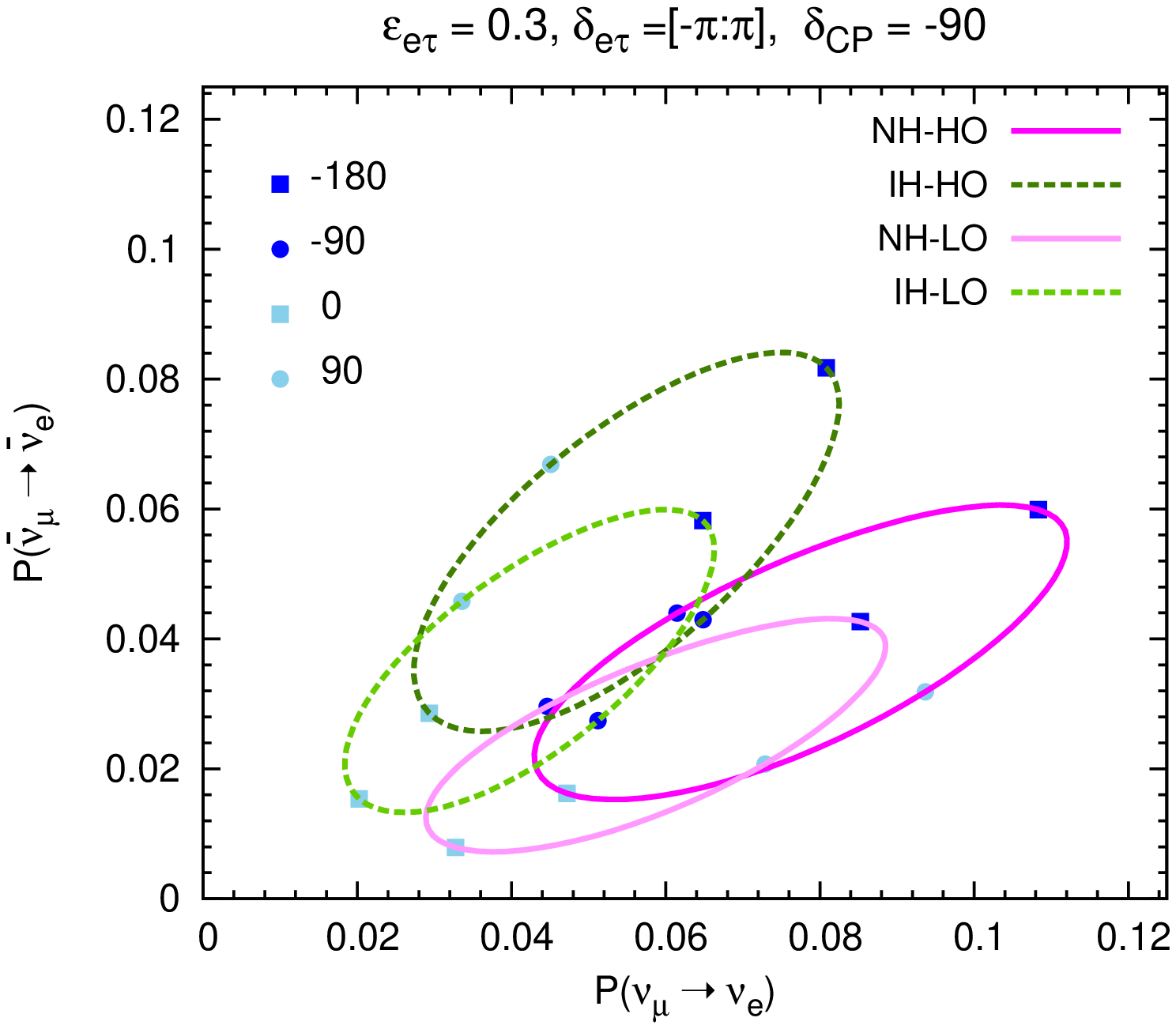}
\includegraphics[width=5.4cm,height=4.5cm]{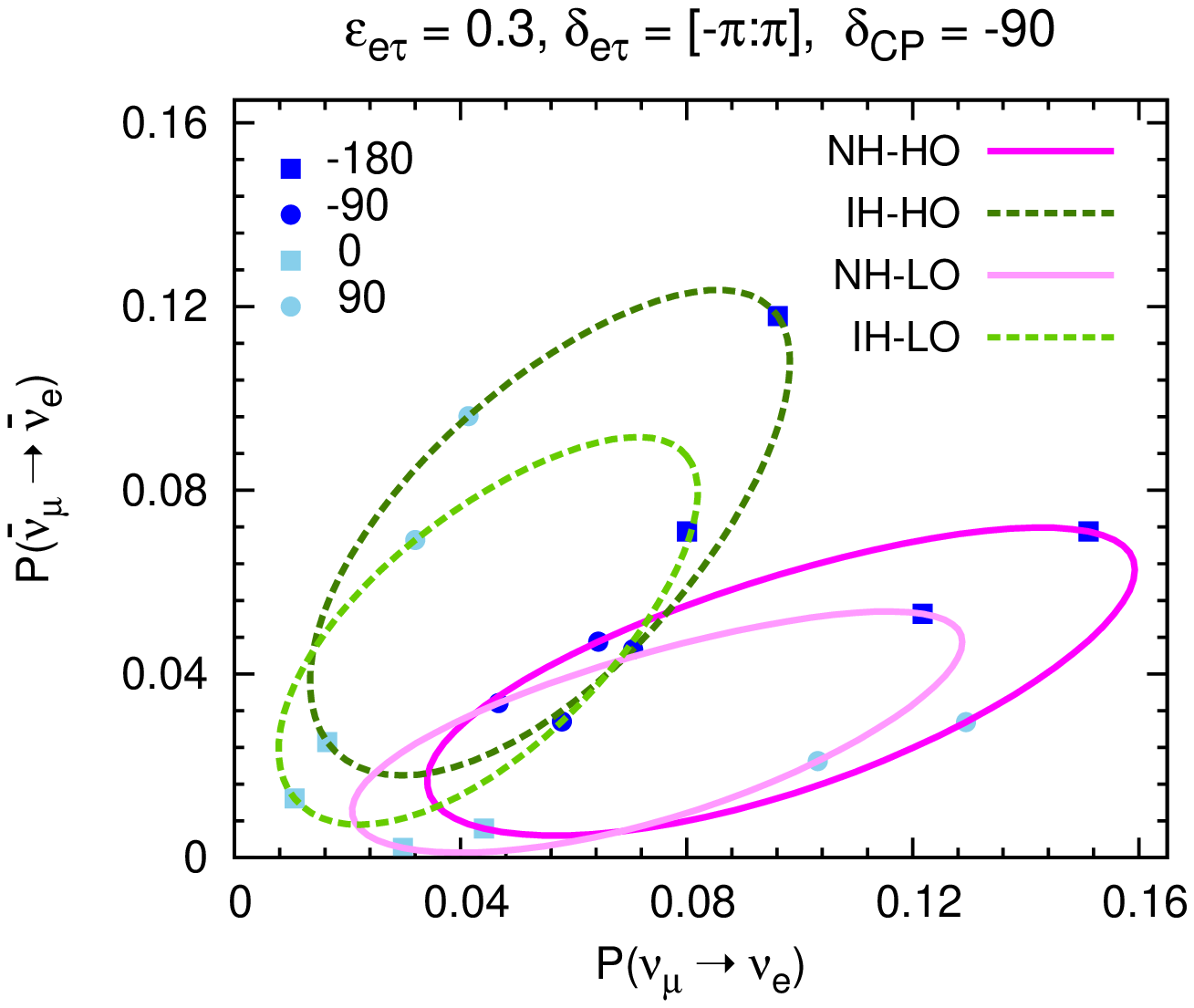}
\end{center}
\caption{The CP trajectory for T2K (left), NO$\nu$A (middle) and DUNE (right) with (bottom panel) and without 
(top panel) NSIs.}
\label{biprob}
\end{figure}

In Fig.~\ref{biprob}, we show the bi-probability plots for T2K ($E$ = 0.6 GeV, L = 295 km), NO$\nu$A ($E$ = 2 GeV, 
L = 810 km) and DUNE ($E$ = 3 GeV, L = 1300 km) for both NH (solid line) and IH (dashed line) where  dark (light) colour plot  
corresponds to HO (LO). In the figure, the upper panel corresponds to $\delta_{CP}$ trajectory without NSIs, whereas the lower panel 
corresponds to $\delta_{e\tau}$ trajectory with $\varepsilon_{e\tau}=0.3$ and $\delta_{CP}=~-90^\circ$ (it is the presently favoured value of CP phase).

In the standard oscillation paradigm, the NH and IH ellipses are  well separated in the case of DUNE experiment,  compared with 
T2K and NO$\nu$A experiments. This means that DUNE experiment has highest mass 
hierarchy determination capability. However, the ellipses in presence of LFV-NSI overlap with each other, which will  
significantly worsen the hierarchy determination capability  of DUNE experiment. It can also be seen from the figure that octant 
degeneracy can be resolved by using all the three experiments, since the light coloured ellipses are well separated from dark coloured  ellipse in the SO. 
Whereas the octant resolution capability of  NO$\nu$A and DUNE experiments become worsen in presence of LFV-NSI, 
because there is significant overlap between the CP trajectories of HO and LO  in presence of LFV-NSI.
 Moreover, there present new types of degeneracies among oscillation parameters in presents of LFV-NSI.

Now, we focus on the bi-probability plot of DUNE with NSI (bottom right panel)  of  Fig.~\ref{biprob}, for a detailed discussion 
on the resolution of parameter degeneracies among the oscillation parameters. One can see from the figure that

\begin{itemize}
\item  If  $\delta_{e\tau} =-180^\circ$, then the points in the  $P_{({\nu_{\mu}} \rightarrow {\nu_{e}})}-P_{(\bar \nu_{\mu} \rightarrow \bar \nu_{e})}$ 
plane are  well separated in the case of NH-HO and IH-HO, which is a clear indication of mass hierarchy determination even in presence of LFV-NSI. Whereas, the capability of MH is reduced in the case of IH-LO and NH-LO. 
It is also noted from the figure that, NH(IH)-HO and NH(IH)-LO are also well separated, which means that octant determination is possible in this case.
\item If  $\delta_{e\tau} =-90^\circ$, then it is extremely difficult to infer any definitive conclusion about the  determination of 
both mass hierarchy and octant, since  all the four degenerate  points in $P_{({\nu_{\mu}} \rightarrow {\nu_{e}})}-P_{(\bar \nu_{\mu} \rightarrow \bar \nu_{e})}$ plane are very close to each other .
\item If  $\delta_{e\tau} = 0$, then  all the four degenerate points are very close to each other  in $P_{({\nu_{\mu}} \rightarrow {\nu_{e}})}-P_{(\bar \nu_{\mu} \rightarrow \bar \nu_{e})}$ plane and therefore it is extremely difficult to make any 
decisive prediction  about the  determination of both mass hierarchy and octant.
\item If  $\delta_{e\tau} = 90^\circ$, then the points  correspond to NH-HO and IH-HO  in  
$P_{({\nu_{\mu}} \rightarrow {\nu_{e}})}-P_{(\bar \nu_{\mu} \rightarrow \bar \nu_{e})}$ plane are very  well separated, which is an 
indication of MH determination. However, the capability  of determination of mass hierarchy is reduced in the case of LO. 
It is also noted that  octant determination is poor in this case. 
\end{itemize}
 
All the above  predictions are made under the assumptions that the value  of 
LFV-NSI $\varepsilon_{e\tau}$ is near to its upper bound and the value of CP violating phase is near to its currently preferred 
value i.e, $\delta_{CP} = -90^\circ$. Moreover, these predictions point toward that the mass hierarchy and octant 
determinations are possible even in the presence of LFV-NSI, if $\delta_{e\tau}=-180^\circ$ or $90^\circ$.

\begin{figure}
\begin{center}
\includegraphics[width=3.5cm,height=3.5cm]{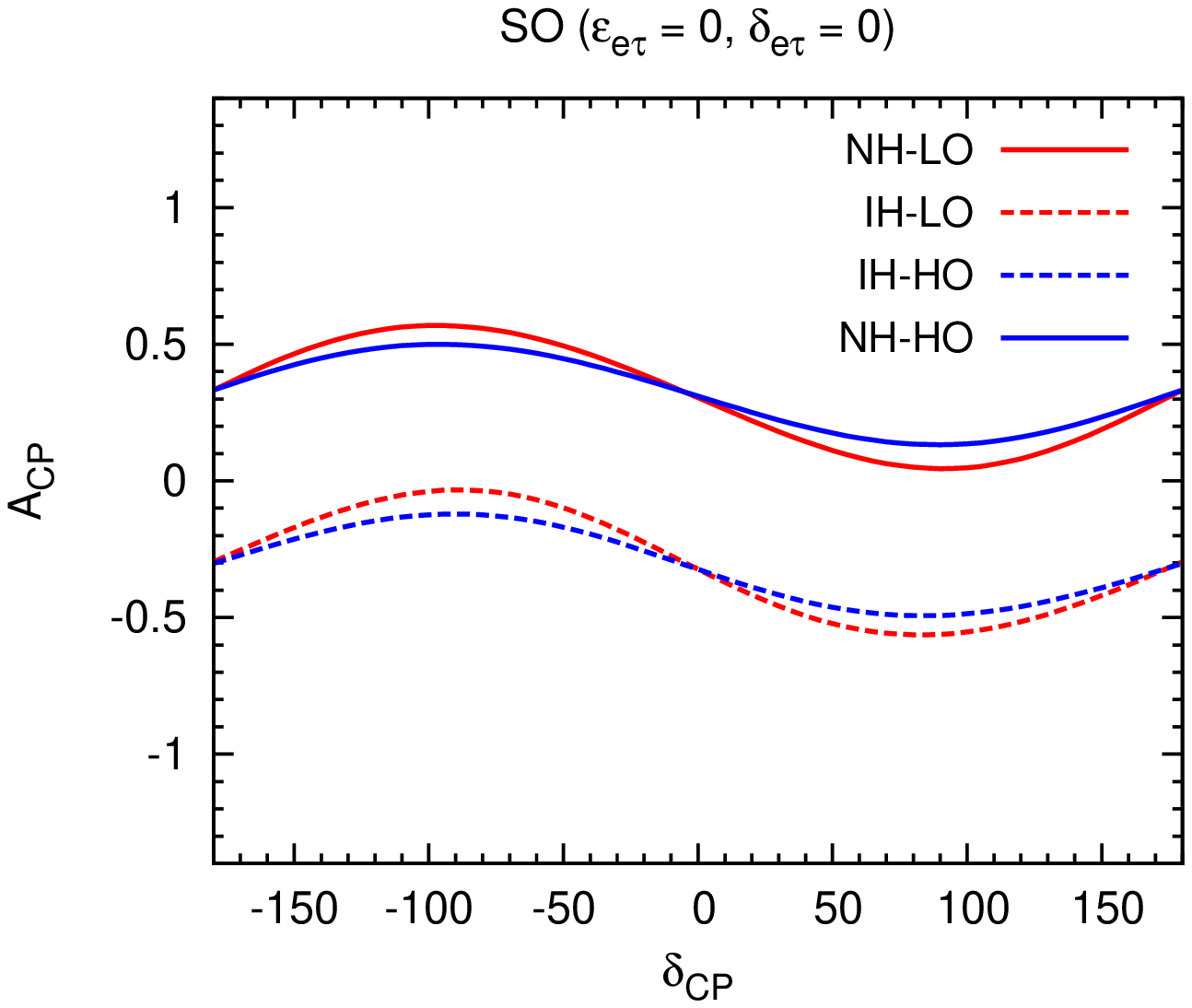}
\includegraphics[width=3.5cm,height=3.5cm]{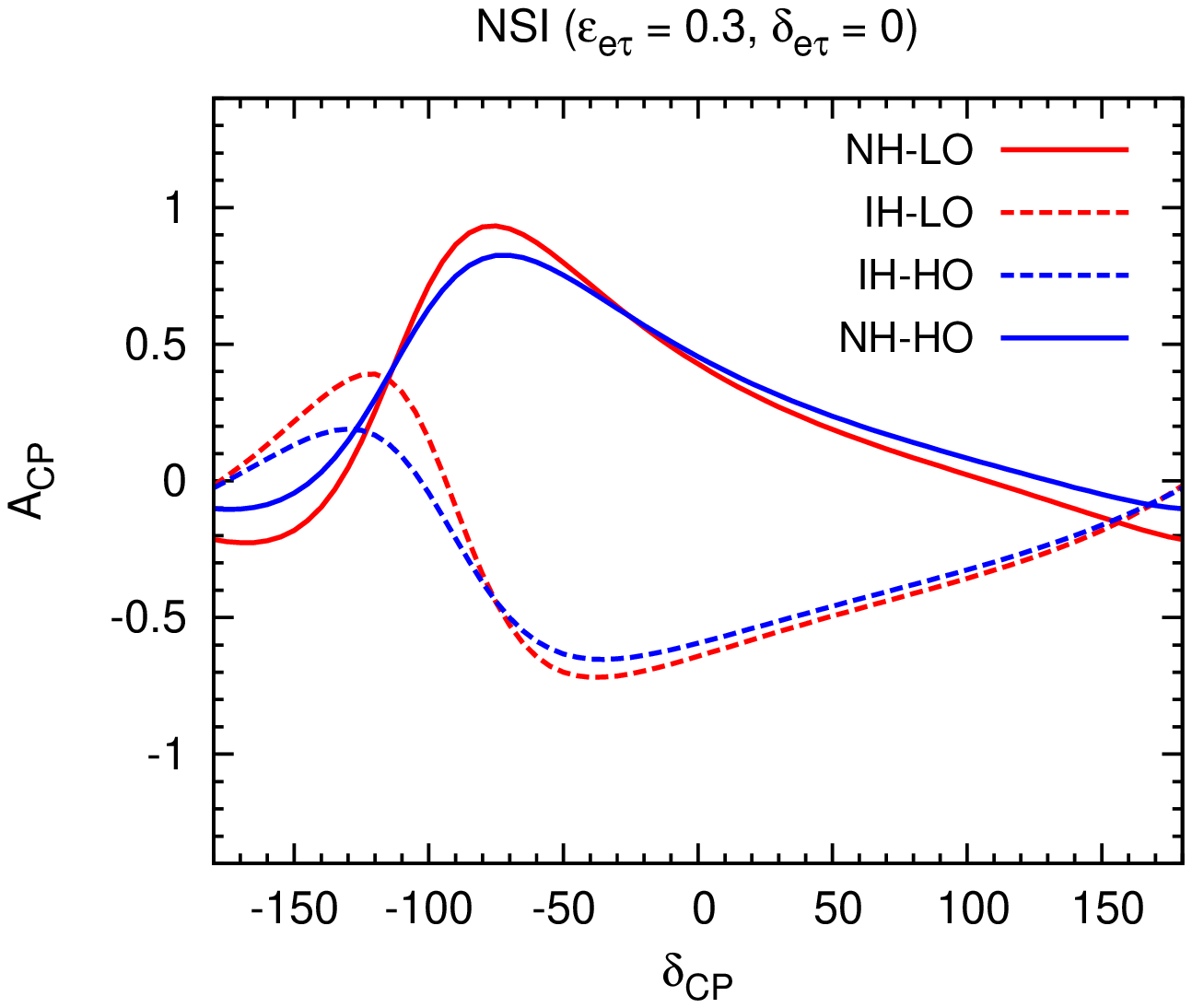}
\includegraphics[width=3.5cm,height=3.5cm]{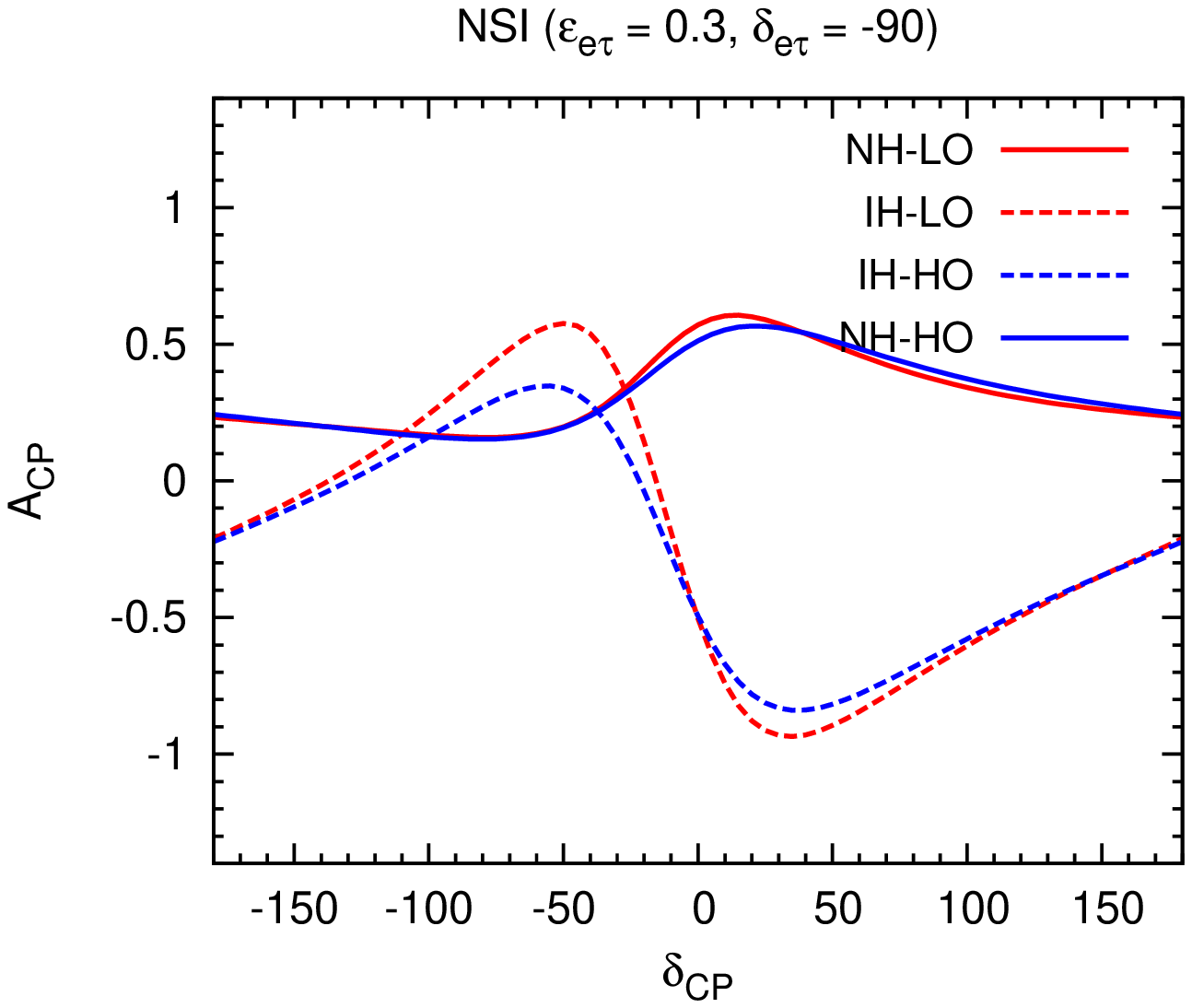}
\includegraphics[width=3.5cm,height=3.5cm]{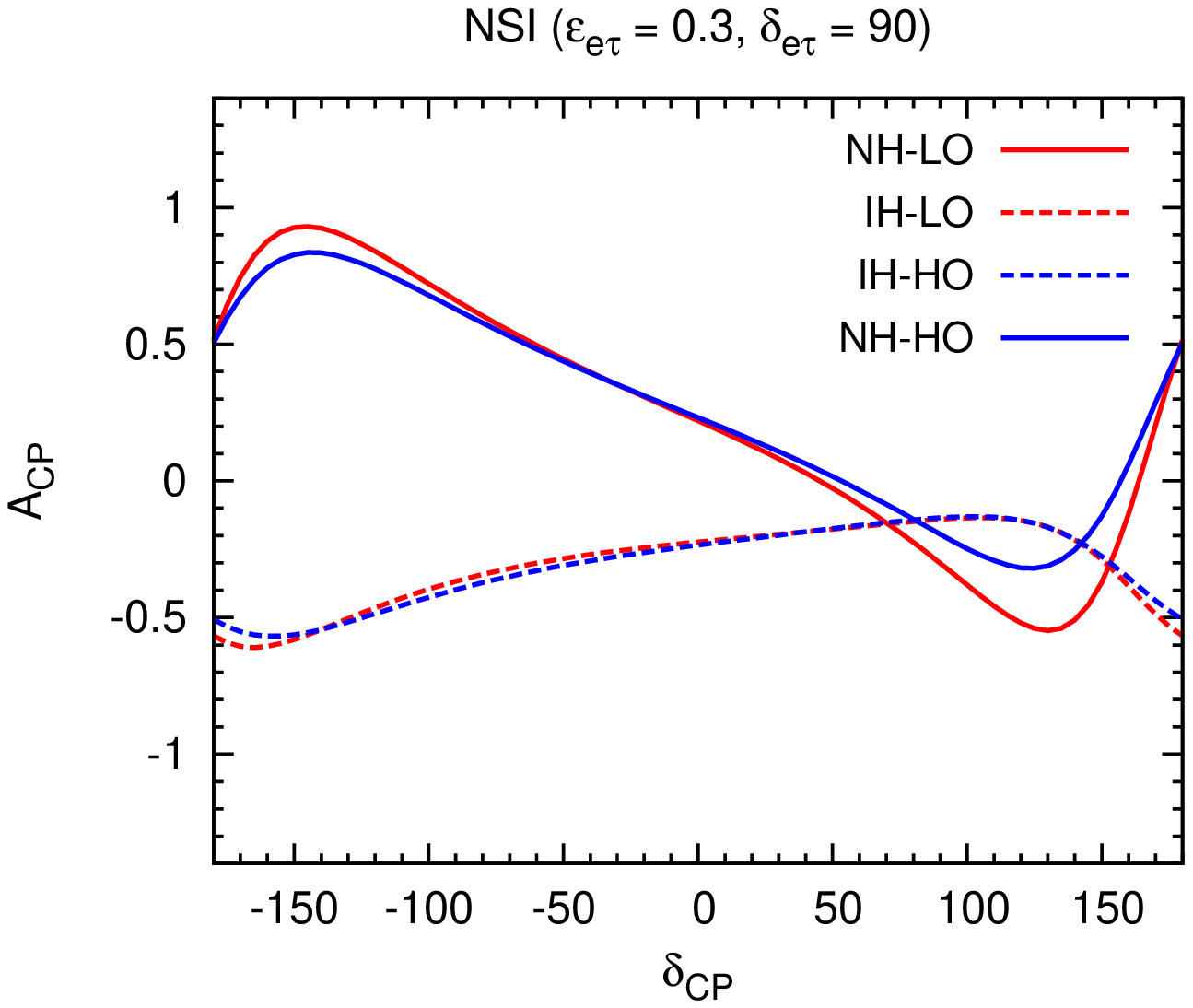}
\includegraphics[width=3.5cm,height=3.5cm]{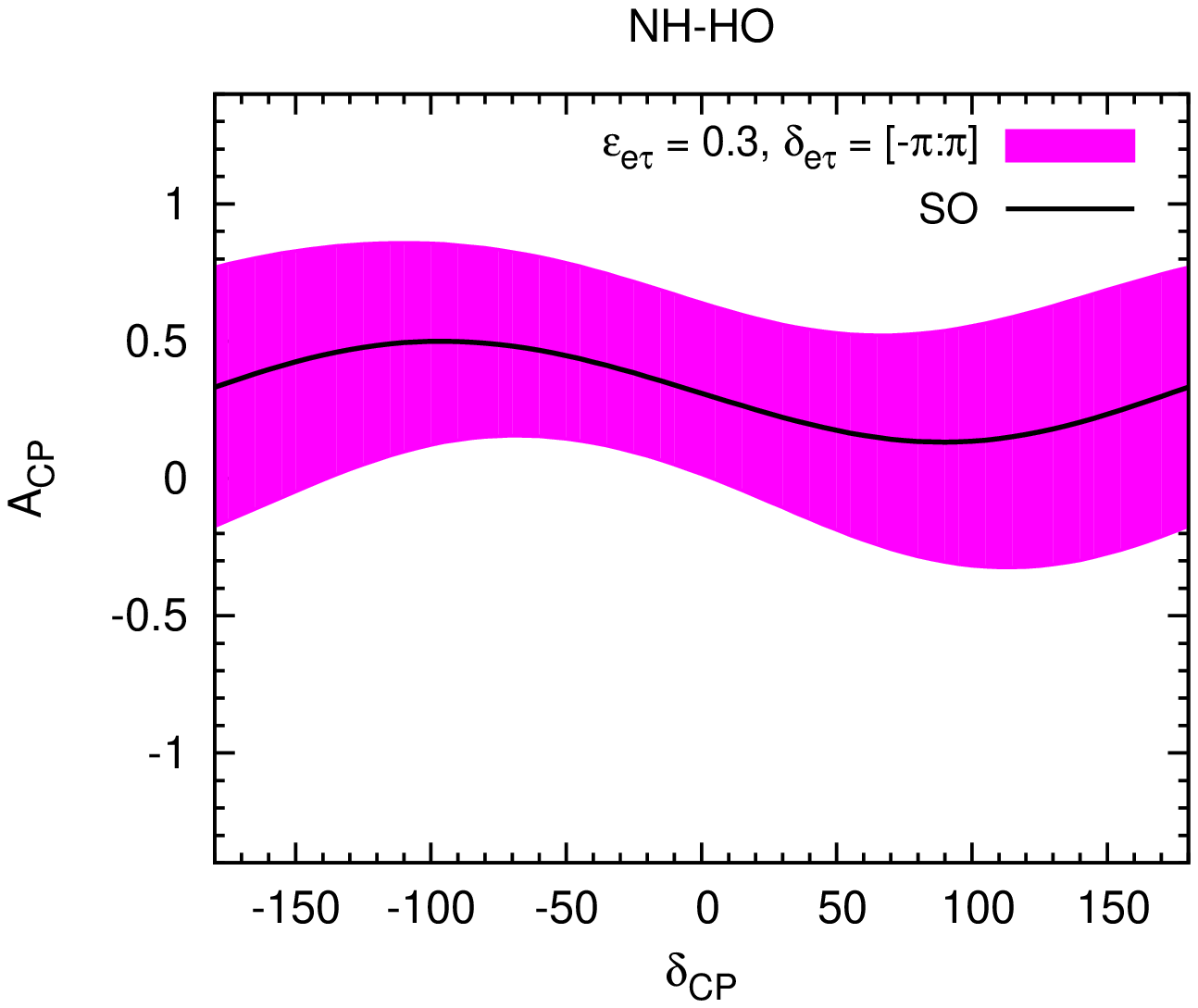}
\includegraphics[width=3.5cm,height=3.5cm]{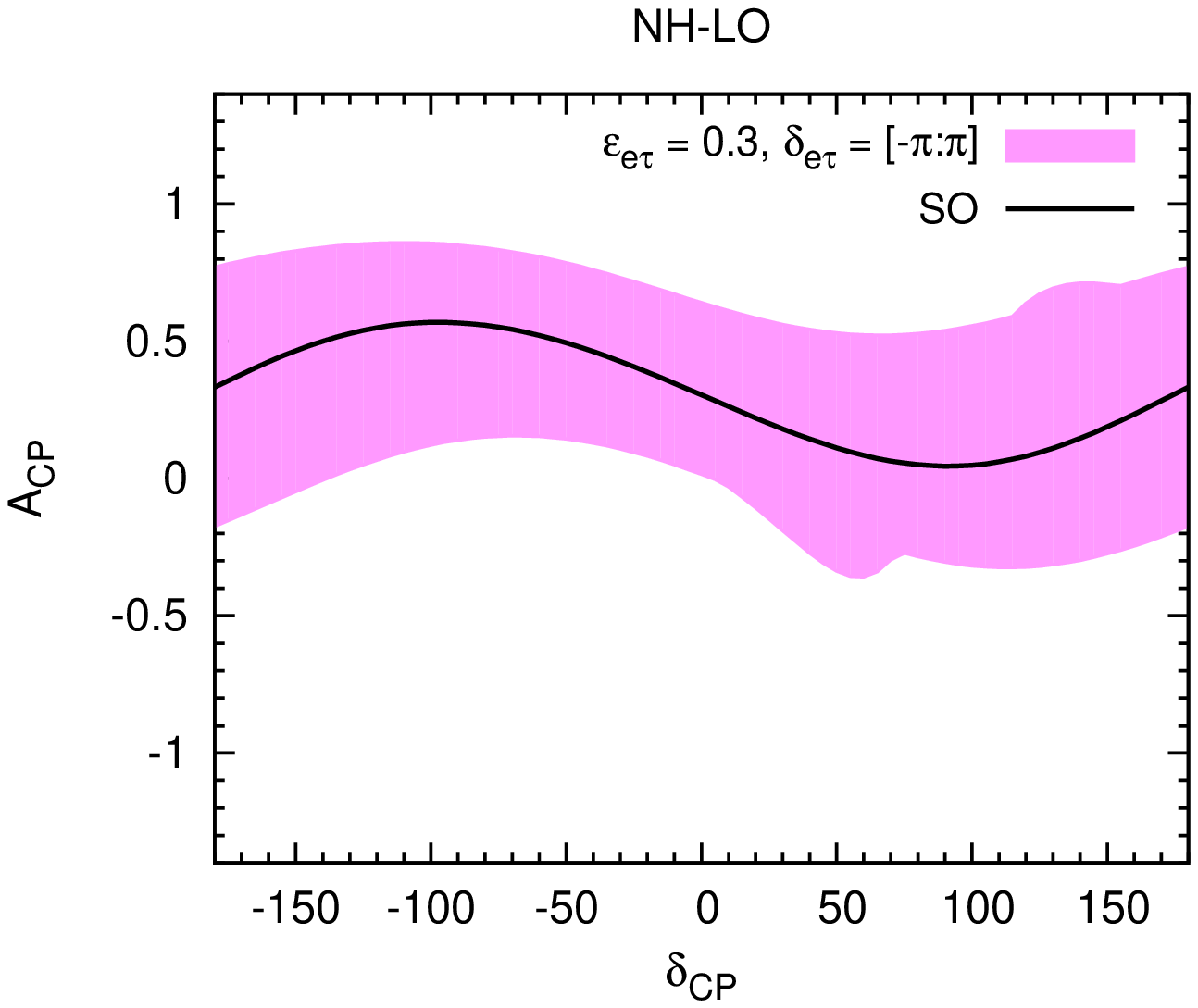}
\includegraphics[width=3.5cm,height=3.5cm]{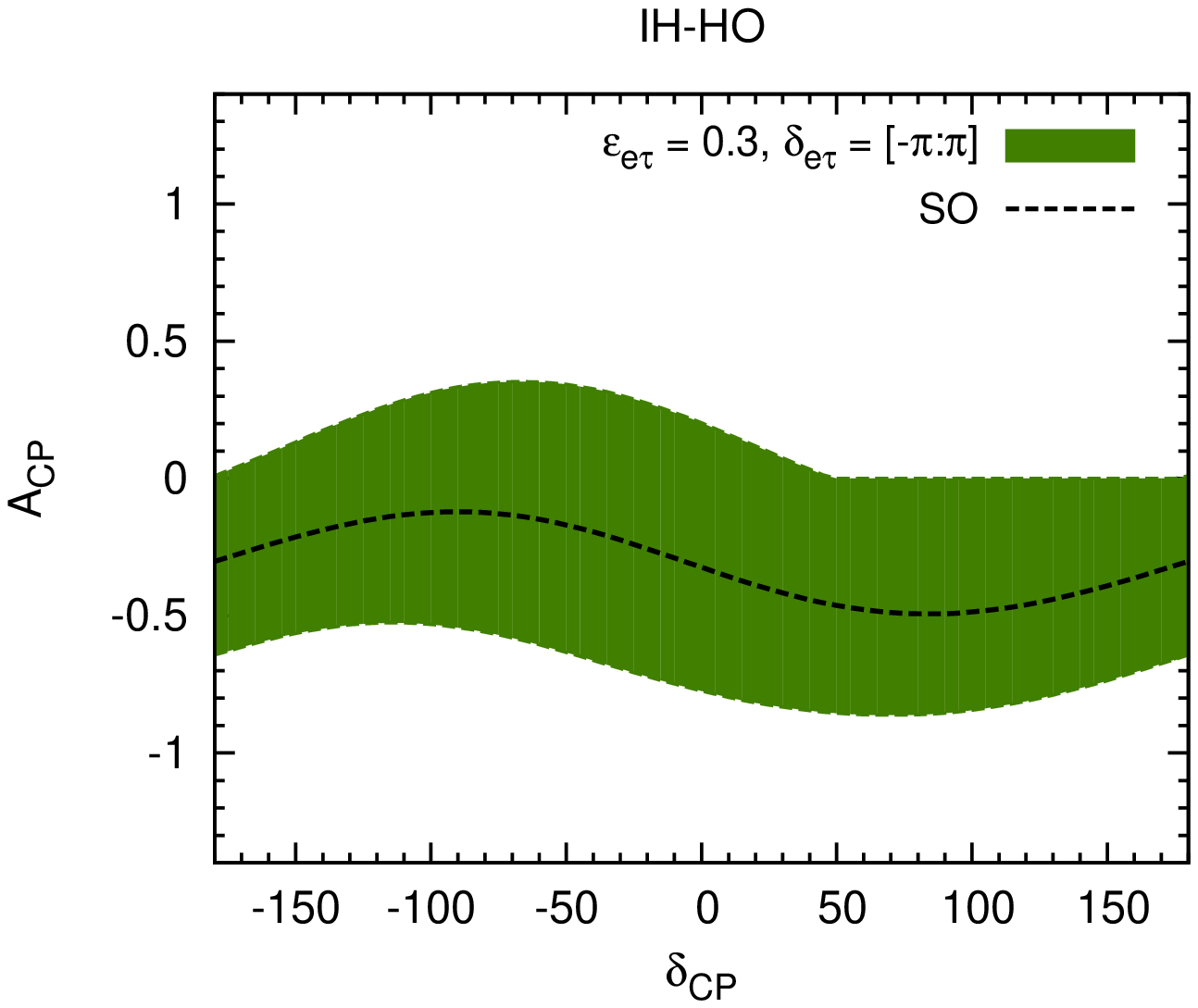}
\includegraphics[width=3.5cm,height=3.5cm]{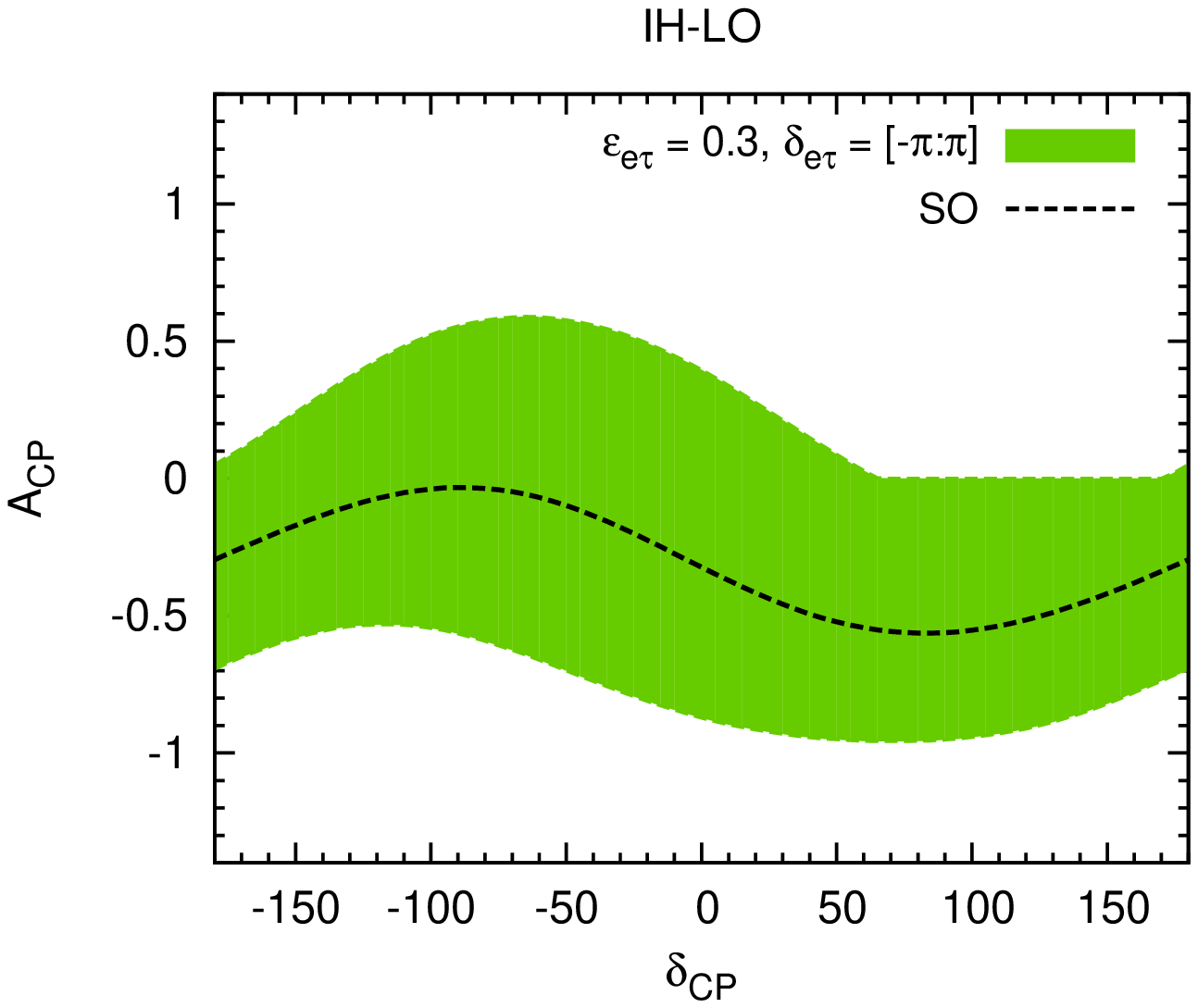}
\end{center}
\caption{The parameter degeneracy among the oscillation parameter in $\delta_{CP}$-CP asymmetry plane for  DUNE experiment. 
The top left panel shows the degeneracies in SO, whereas the other three panels show the degeneracy in presence of LFV-NSI with 
$\delta_{e\tau}$ =0, -90, and 90 respectively. The bottom panel shows the $A_{CP}$ for NH-HO, NH-LO, IH-LO and IH-LO  
in presence of NSI ($\varepsilon_{e\tau} =0.3$ and $\delta_{e\tau}=[\pi:\pi]$).}
\label{ACP-D}
\end{figure}

 Another simple way to understand the parameter degeneracies  among the oscillation parameters is by simply 
looking at the CP-asymmetry, which is defined in Eqn. (\ref{acp}). CP-asymmetry as a function of $\delta_{CP}$ for NH-LO, NH-HO, IH-LO and IH-HO for 
DUNE experiment is given in Fig.~\ref{ACP-D}. The top left panel of  the figure shows the CP asymmetry in standard oscillation and it can be 
seen from the figure that CP asymmetry is more in LO than in HO for both NH and IH. The rest of three in the top panel show the 
CP asymmetry in presence of NSI with $\delta_{e\tau} =0,-90^\circ$, and $90^\circ$ respectively. It is clear from the figure that LFV-NSI 
introduces other degeneracies among the standard oscillation parameters. Moreover, the bottom panel shows the $A_{CP}$ for NH-HO, NH-LO, 
IH-LO and IH-LO  in presence of NSI ($\varepsilon_{e\tau} =0.3$ and $\delta_{e\tau}=[\pi:\pi]$). Therefore, degeneracy resolution in presence of NSI 
extremely complicated. It also noted that degeneracy resolution capability is mainly depend 
on the value of $\delta_{e\tau}$, for instance if   $\delta_{e\tau} = 90^\circ$, then CP-asymmetry for IH-LO and IH-HO are almost same and one cannot 
distinguish between them.

\subsection{Correlation between $\delta_{CP}$ and $\theta_{23}$}
\begin{figure}
\begin{center}
\includegraphics[width=3.5cm,height=3.5cm]{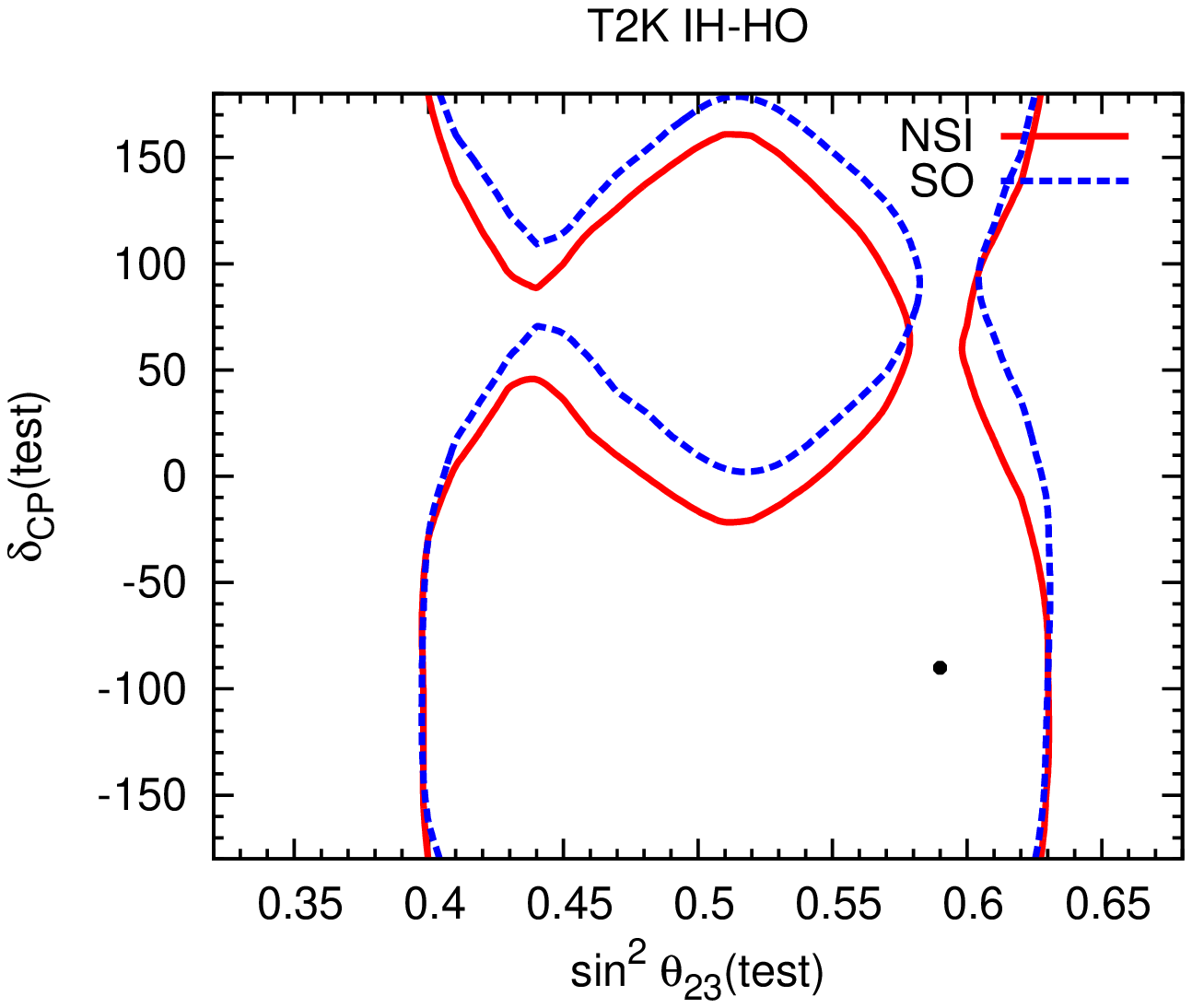}
\includegraphics[width=3.5cm,height=3.5cm]{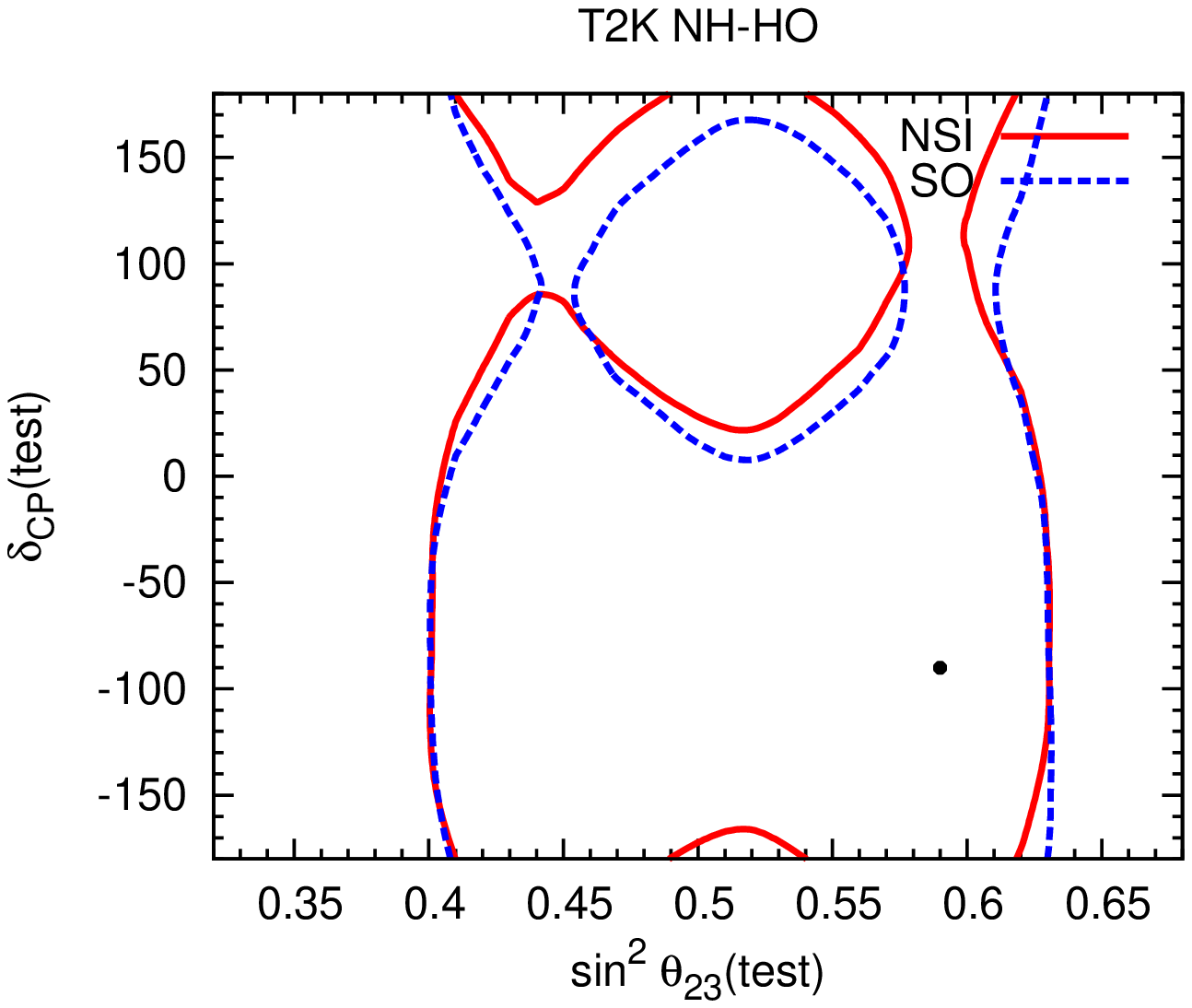}
\includegraphics[width=3.5cm,height=3.5cm]{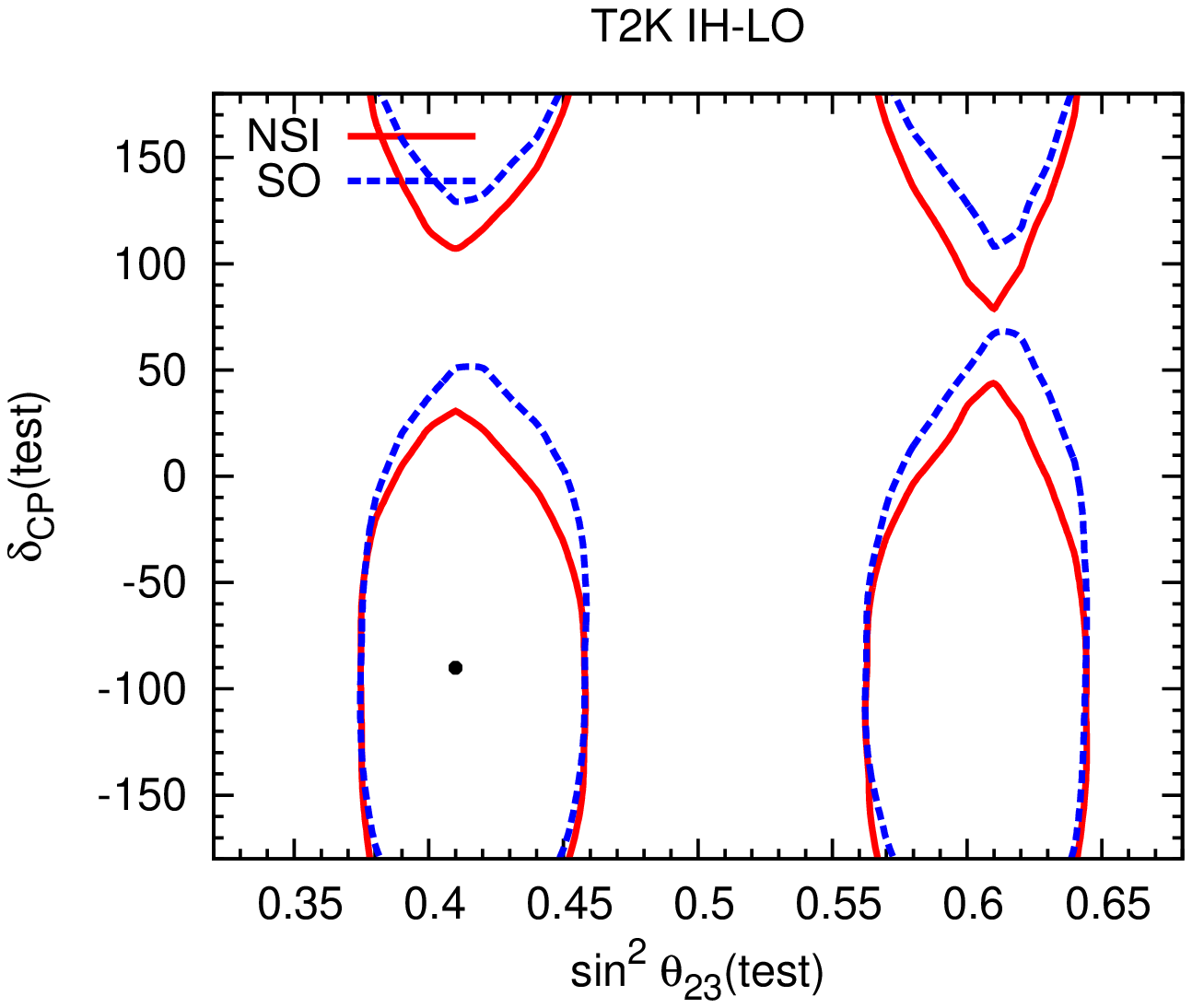}
\includegraphics[width=3.5cm,height=3.5cm]{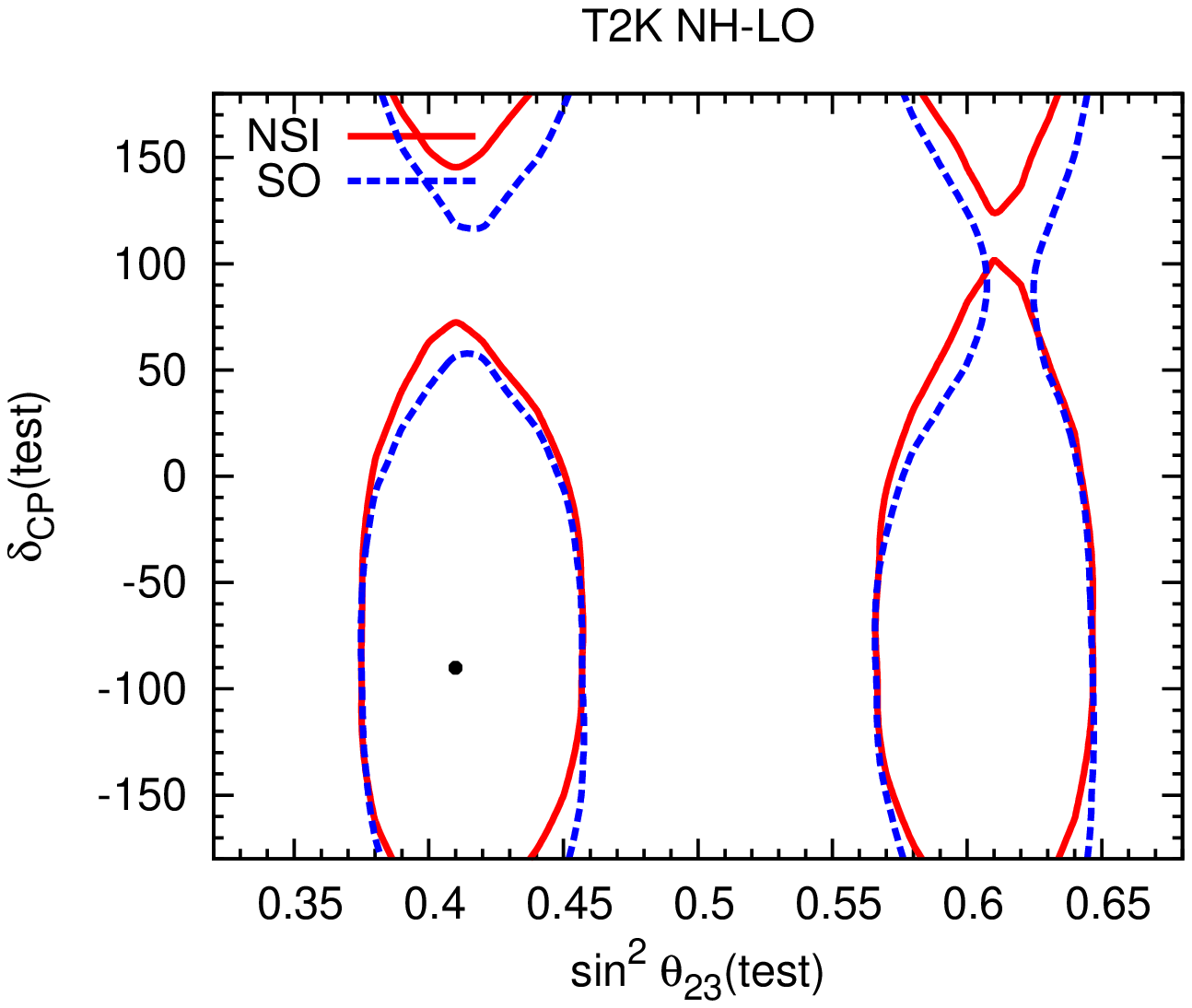}
\includegraphics[width=3.5cm,height=3.5cm]{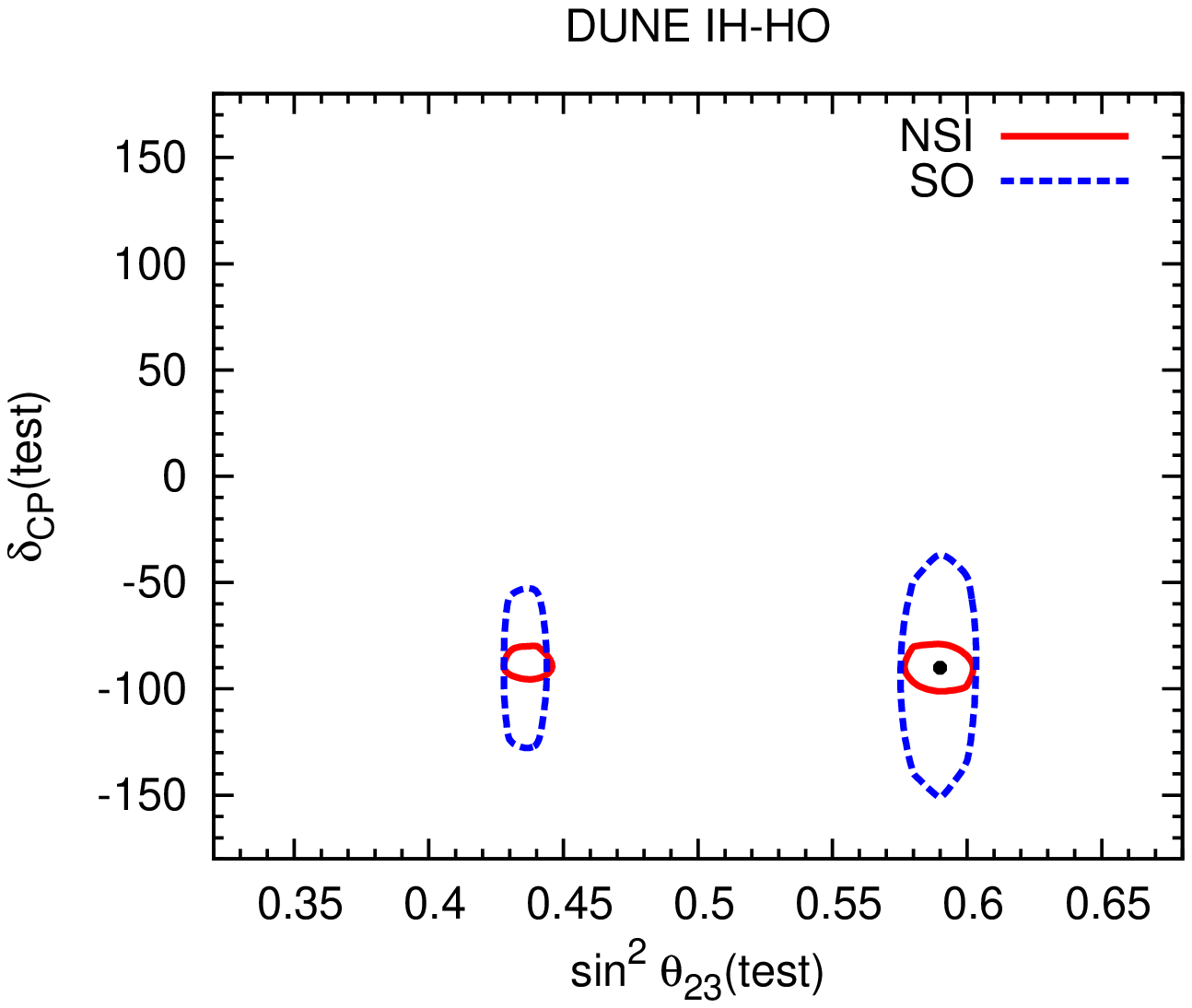}
\includegraphics[width=3.5cm,height=3.5cm]{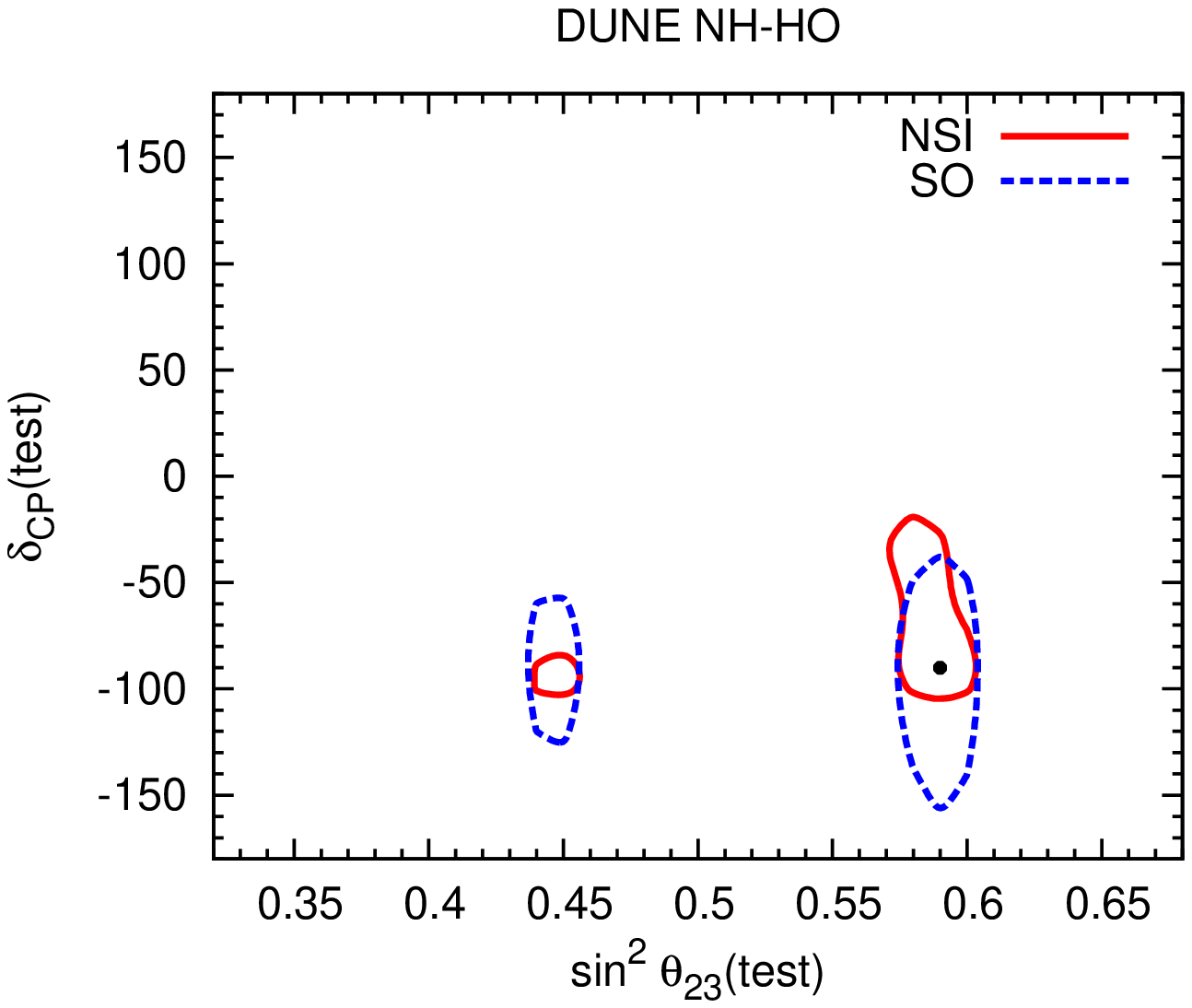}
\includegraphics[width=3.5cm,height=3.5cm]{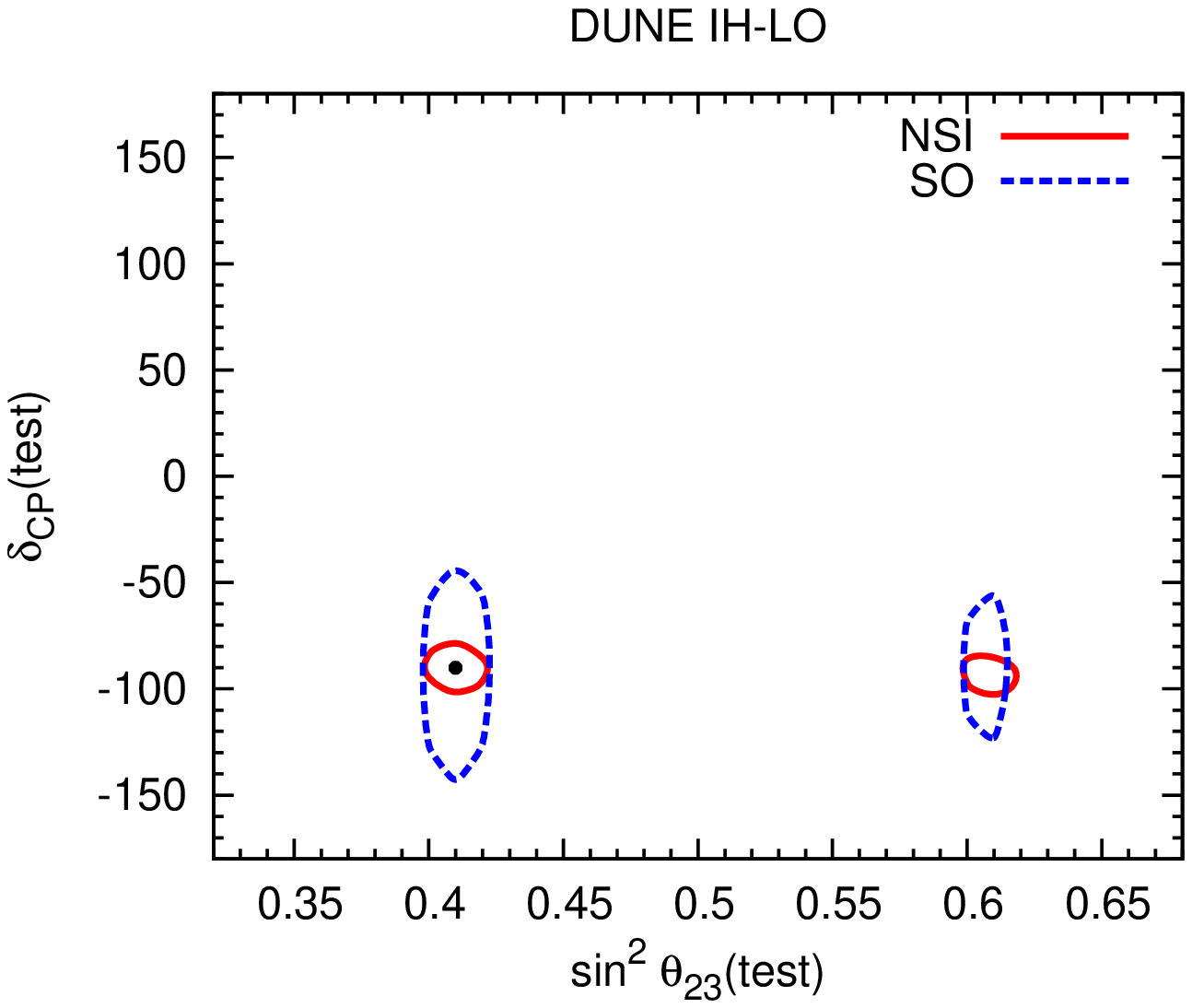}
\includegraphics[width=3.5cm,height=3.5cm]{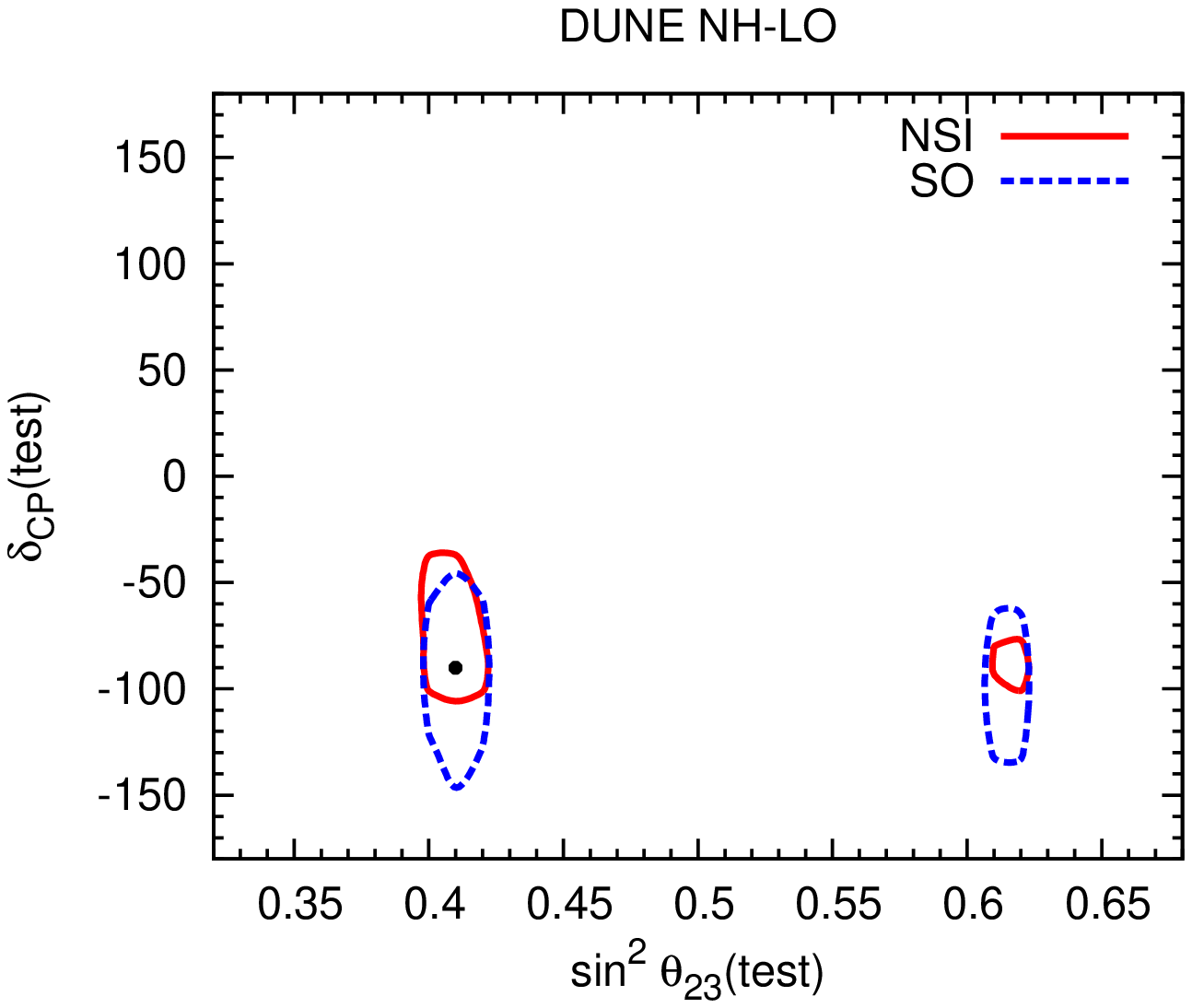}
\end{center}
\caption{The $2\sigma$ C.L. regions for  $\sin^2\theta_{23}$ vs. $\delta_{CP}$ with true $\sin^2\theta_{23}$ = 0.41 (0.59) for LO (HO) and true 
 $\delta_{CP} =-90^\circ$. The top panel corresponds to T2K and bottom panel corresponds to DUNE experiments.}
\label{DCPTH23}
\end{figure}

 In this section, we discuss the effect of LFV-NSI on the allowed parameter space of  $\sin^2\theta_{23}$ and $\delta_{CP}$. 
We show the $2\sigma$ C.L. regions for  $\sin^2\theta_{23}$ vs. $\delta_{CP}$ with true $\sin^2\theta_{23}$ = 0.41 (0.59) for LO (HO) and true  
$\delta_{CP} =-90^\circ$ in Fig.~\ref{DCPTH23}, for T2K (top panel) and DUNE (bottom panel) experiments. 
From the figure, we can see that there is significant change in the allowed  parameter space in presence of LFV-NSI for DUNE.  

\section{Summery and Conclusions}
We have investigated the  implications of LFV-NSIs on the physics potential of various neutrino oscillation experiments. 
We found that the discovery reach for the unknowns in oscillation physics by the experiments that we have considered can be altered significantly 
in the presence of LFV-NSIs.  Moreover, we found that the degeneracy 
discrimination capability of all the experiment will worsen in the presence of LFV-NSI, since  it leads to new degeneracies among the 
oscillation parameters other than the existing degeneracies in standard oscillation physics. We also found that the possibility of misinterpretation of 
oscillation data in the presence of new physics scenarios (NSIs),  give rise to wrong determination of octant of atmospheric mixing angle, 
neutrino mass hierarchy and the CP violation.

{\bf Acknowledgments}
We would like to thank Science and Engineering Research Board (SERB),
Government of India for financial support through grant No.SB/S2/HEP-017/2013.

\end{document}